\documentclass[12pt]{article}

\pdfoutput=1

\usepackage{ifpdf}
\ifpdf
\usepackage{graphicx,color}
\usepackage{hyperref}
\else
\usepackage[dvipdfmx]{graphicx,color}
\usepackage[dvipdfmx]{hyperref}
\fi
\usepackage{amssymb,amsfonts,amsmath,cancel,cite,multirow,dsfont}
\usepackage[capitalise]{cleveref}
\usepackage{slashed}
\usepackage{subcaption}
\usepackage[margin=1in]{geometry}

\newcommand{\eqs}[1]{\begin{equation}\begin{split} #1 \end{split}\end{equation}}

\begin{document}

\begin{titlepage}

\begin{flushright}
\end{flushright}

\vskip 1.35cm
\begin{center}

{\large
\textbf{
  Dark Vector Mesons at LHC Forward Detector Searches
}}
\vskip 1.2cm

Takumi Kuwahara$^{a}$ 
and Shu-Run Yuan$^{b}$

\vskip 0.4cm

\textit{$^a$
Center for High Energy Physics, Peking University, Beijing 100871, China
}

\textit{$^b$
School of Physics, Peking University, Beijing 100871, China
}

\vskip 1.5cm

\begin{abstract}
  Confining gauge dynamics in a dark sector is promising to provide dark matter with a mass in the range of sub-GeV to GeV.
  Such dark sectors consist of composite particles such as dark baryons and dark mesons, that are neutral under the standard-model charge.
  A dark photon is introduced as a portal matter between the dark sector and the standard-model sector to alleviate cosmological problems (e.g., to maintain kinetic equilibrium between two sectors or to reduce the light dark-sector particles contributing to the dark radiation), and dark hadrons are produced through the same dark photon at accelerator-based experiments. 
  As dark vector mesons and dark pions have similar masses, dark vector mesons can be long-lived particles, which will be explored by far-detector experiments.
  We study the future prospects of the LHC forward-detector experiments, FASER/FASER2 and FACET, for exploring the dark vector mesons. 
  When the dark photon is heavier than the dark pions, the LHC forward-detector searches will be comparable to DarkQuest, and the invisible decay searches of dark photons will also explore the same parameter space.
  Meanwhile, when dark photons are lightest in the dark sector, their future prospects will be comparable to the visible decay searches for dark photons at LHCb, Belle-II, and HPS.
\end{abstract}

\end{center}
\end{titlepage}

\section{Introduction}

The particle nature of dark matter (DM) remains unknown even though astrophysical observations have confirmed its existence. 
Several properties are important to explore DM and its associated particles: mass, lifetime, and interaction with the standard model (SM) particles. 
The weakly interacting massive particle (WIMP) mechanism provides the correct relic abundance via the annihilation into the SM particles.
However, both direct searches and indirect searches for DM put stringent constraints on the WIMP with a mass of $\sim 100 \,\mathrm{GeV}$. 
The DM with masses in the MeV to GeV range has recently attracted attention since the constraints from the direct detection searches weaken.
This DM would require an alternative mechanism for the correct relic abundance; indeed, these light DMs lead to the overclosure of the Universe, which is known as the Lee-Weinberg bound~\cite{Lee:1977ua} when the weak bosons mediate their annihilation.
The idea of the dark sector opens up new mechanisms for providing the correct abundance. 

A dark sector with confining dynamics would naturally provide the DM candidate with masses of order $0.1$--$10$ GeV.
Besides, the dark sector would consist of composite particles (dark hadrons) at the low-energy scale as with the SM hadrons as a consequence of confining gauge dynamics~\cite{Gudnason:2006yj, Dietrich:2006cm, Khlopov:2007ic, Khlopov:2008ty, Foadi:2008qv, Mardon:2009gw, Kribs:2009fy, Barbieri:2010mn, Blennow:2010qp, Lewis:2011zb, Appelquist:2013ms, Hietanen:2013fya, Cline:2013zca, Appelquist:2014jch, Hietanen:2014xca, Krnjaic:2014xza, Detmold:2014qqa, Detmold:2014kba, Asano:2014wra, Brod:2014loa, Antipin:2014qva, Hardy:2014mqa, Appelquist:2015yfa, Appelquist:2015zfa, Antipin:2015xia, Hardy:2015boa, Co:2016akw, Dienes:2016vei, Ishida:2016fbp, Lonsdale:2017mzg, Berryman:2017twh,  Gresham:2017zqi, Gresham:2017cvl, Mitridate:2017oky, Gresham:2018anj, Ibe:2018juk, Braaten:2018xuw, Francis:2018xjd, Bai:2018dxf, Chu:2018faw, Hall:2019rld, Tsai:2020vpi, Asadi:2021yml, Asadi:2021pwo, Zhang:2021orr, Bottaro:2021aal,Ibe:2021gil, Hall:2021zsk, Asadi:2022vkc} or of mirror-world like  dark sector~\cite{Kobzarev:1966qya,Blinnikov:1983gh,Kolb:1985bf,Khlopov:1989fj,Hodges:1993yb,Foot:2004pa} (see Ref.~\cite{Kribs:2016cew, Beylin:2020bsz} for a review).
One of the alternatives providing the correct relic abundance is known as the strongly interacting massive particle (SIMP) mechanism~\cite{Hochberg:2014dra}, where the number-changing process depletes the thermal relics. 
The SIMP mechanism can be realized by the confining dark sector \cite{Hochberg:2014kqa}.
Dark pions appear as pseudo-Nambu-Goldstone (pNG) bosons of the chiral symmetry breaking in the dark sector.
The number-changing process arises from the topological term of the low-energy effective theory, which is known as the Wess-Zumino-Witten (WZW) term~\cite{Wess:1971yu,Witten:1983tw}.
The sub-GeV dark pions are expected as the DM candidate to explain the correct relic abundance.
Meanwhile, asymmetric DM (ADM) provides another alternative, where the DM abundance is determined by particle-antiparticle asymmetry of DM.
The closeness of DM and SM baryons mass densities, which is observed in cosmological observations, would be explained by linking their asymmetries in some way and by the DM mass of order GeV-scale. 
Composite models provide key ingredients for realizing ADM: the stability of DM, the strong depletion of the symmetric component, and the DM mass of order GeV-scale.
In these models, the lightest dark baryon is the candidate for the ADM.
The DM number ensures the stability of the lightest dark baryon in analogy to the stability of SM protons.
Their annihilation into the dark pions strongly depletes the symmetric part of DM.
The dimensional transmutation would naturally provide dark baryons with masses of order GeV-scale.
The composite DM would have self-interaction as a by-product, which may solve the small-scale structure problems of the Universe~\cite{Spergel:1999mh} (for a review, see Ref.~\cite{Tulin:2017ara}).

Dark vector mesons may participate in physical processes besides the dark baryons and dark pions and may play a significant role in making these models viable in some cases.
The dark pion realization of SIMP is faced with the perturbative bound on the chiral perturbation theory in order that the number-changing process is effective in providing the correct relic abundance~\cite{Hochberg:2014kqa}.
The presence of the dark vector mesons unitarizes the number-changing process even for the pion self-coupling close to the perturbative bound~\cite{Choi:2018iit} and provides a new channel for depleting dark pions via the semi-annihilation process $\pi' \pi' \to \pi' V'$ when $m_{V'} < 2 m_{\pi'}$ with masses $m_{\pi'}$ for dark pion $\pi'$ and $m_{V'}$ for dark vector meson $V'$~\cite{Berlin:2018tvf}.
Meanwhile, the dark vector mesons would mediate the resonant scattering in theories with the dark baryon DM~\cite{Chu:2018faw}, which provides the velocity-dependent self-interaction indicated by the observations of the DM halos~\cite{Kaplinghat:2015aga}.

An additional massive vector boson, which is called the dark photon, is introduced to the dark sector and is kinematically mixed with the SM photon via a tiny mixing angle $\epsilon$. 
The dark photon may resolve several problems of the confining dark sector. 
In the dark-pion DM model, two sectors are connected with each other via the dark photon, and the dark sector is in the kinetic equilibrium with the SM sector.
Meanwhile, in the composite ADM, dark pions are largely produced by the strong depletion of the symmetric part of ADM, and the lightest dark pions are stable.
The dark pions carry huge entropy in the dark sector and lead to the overclosure of the Universe. 
In the presence of the dark photons, the entropy in the dark sector will be released into the dark photons and will be finally converted to the SM particles via the kinetic mixing.

When the mass of dark vector mesons is in the range of $m_{\pi'} < m_{V'} < 2m_{\pi'}$, invisible decay $V' \to \pi' \pi'$ is kinematically forbidden. 
The dominant decay arises from the anomalous interactions among $V'\,, \pi'$, and $A'$ for the dark-charged $V'$ and from the kinetic mixing between $V'$ and $A'$ for the dark-neutral $V'$, and then $A'$ decay into the SM particles leaves the visible signals.
The decay width of $V'$ into SM is suppressed by $\epsilon^2$, and the dark vector mesons may have a long lifetime.
In particular, their proper lifetime can be of $\mathcal{O}(1)\,\mathrm{[m]}$ when $\epsilon$ is in the range of $10^{-4}\text{--}10^{-3}$.
The dark photon plays a significant role in producing the dark hadrons at terrestrial experiments besides resolving cosmological problems.
There are two ways to produce the dark vector mesons at the experiments; via the on-shell dark photon and via the off-shell dark photon.
For the former case, Ref.~\cite{Berlin:2018tvf} has studied the visible decay searches at the fixed-target experiment: such as E137~\cite{Bjorken:1988as}, HPS~\cite{Celentano:2014wya}, LDMX~\cite{Izaguirre:2014bca}, and DarkQuest~\cite{SeaQuest:2017kjt}.
Meanwhile, as for the dark baryon DM (composite ADM), the dark hadrons are produced at the accelerator-based experiments only through the off-shell dark photon.
In this paper, we discuss the dark vector meson searches at the forward detector experiments at the Large Hadron Collider (LHC): FASER/FASER2~\cite{Feng:2017uoz} and FACET~\cite{Cerci:2021nlb}.

Various experiments are proposed to explore the long-lived particles with a proper decay length of $10^0 \text{--} 10^2 \, \mathrm{[m]}$ at the LHC: CODEX-b~\cite{Gligorov:2017nwh}, MATHUSLA~\cite{Chou:2016lxi}, FASER/FASER2~\cite{Feng:2017uoz}, and FACET~\cite{Cerci:2021nlb}.
For the former two experiments, the detectors are located at the off-axis from the beamline of the LHC. 
FASER/FASER2 (FACET) is located at about $10^2 \, \mathrm{m}$ away from the collision point of the ATLAS (CMS) along the beam axis. 
The particles with (sub-)GeV mass radiated in the forward direction tend to be highly boosted due to the collision energy of the LHC. 
Therefore, they would have the potential to explore the dark vector mesons with masses in the sub-GeV to GeV range and with the proper lifetime of meters.

This paper is organized as follows. 
In \cref{sec:models}, we briefly review the dark vector mesons in dark-pion DM models and in composite ADM. 
We describe FASER/FASER2 and FACET in detail and introduce our simulation setups in \cref{sec:detectors}. 
In \cref{sec:results}, we show the prospective sensitivity of FASER2 and FACET to the dark vector mesons produced through on-shell or off-shell dark photons. 
We devote \cref{sec:summary} to summarizing our study.

\section{Models\label{sec:models}}

In this section, we review the dark hadrons in the dark-pion DM models and in the composite ADM. 
First, we discuss common features of dark hadrons among these models.
We consider a dark sector with dark quantum chromodynamics (QCD) and dark quantum electrodynamics (QED).
The dark QCD is described by a gauge symmetry $SU(N_c)$ with $N_c = 3$ colors. 
We also introduce $N_f = 3$ flavors of Dirac fermions, called dark quarks, in the fundamental representation.
This theory possesses a (approximate) global chiral symmetry $SU(N_f)_L \times SU(N_f)_R$, which is expected to be spontaneously broken to $SU(N_f)_V$ by the dark quark condensate, giving $N_f^2 -1$ pNG bosons called dark pions.
Chiral perturbation theory describes the dark pions at the low energy, and the chiral Lagrangian at the leading order is given by
\eqs{
  \mathcal{L} = \frac{f_{\pi'}^2}{4} \mathrm{Tr} (D_\mu \Sigma D^\mu \Sigma^\dag) 
  + \frac{B f_{\pi'}^2}{4} \mathrm{Tr} \left[ M^\dag \Sigma + M \Sigma^\dag \right]\,.
  \label{eq:chiralLagrangian}
}
Here, the first term gives the kinetic term of the dark pions, and the second term provides the masses of the dark pion.
$f_{\pi'}$ denotes the decay constant, and $M$ denotes the mass matrix of dark quarks.
A dimensionful parameter $B$ is proportional to the quark condensate, $B f_{\pi'}^2 \propto - \langle \overline q' q' \rangle$ (for a review, see Ref.~\cite{Scherer:2002tk}).
The non-linear sigma field $\Sigma$ is defined by
\eqs{
  \Sigma \equiv \exp \left(  \frac{2 i \pi'^a T^a}{f_{\pi'}}\right) \,, \qquad
  \mathrm{Tr}(T^a T^b) = \frac12 \delta^{ab} \,.
}
The dark quarks are charged under the dark QED, described by $U(1)_D$, and hence the dark pions are also charged.
We will discuss the charge matrix in each model individually. 
The dark photon of $U(1)_D$ is kinematically mixed with the SM photon:
\eqs{
  \mathcal{L} = \frac{\epsilon}{2} F'_{\mu\nu} F^{\mu\nu} \,,
}
where $F_{\mu\nu}$ and $F'_{\mu\nu}$ are the field strength of the SM photon and the dark photon, respectively.

As with the hadron spectrum in the SM, dark vector mesons $V'$ are expected to appear in the dark hadron spectrum at a scale close to the cutoff scale of the chiral perturbation theory.
There are several ways to incorporate vector mesons into the theory: the matter field model~\cite{Ecker:1989yg}, the massive Yang-Mills model~\cite{Schwinger:1967tc,Gasiorowicz:1969kn,Kaymakcalan:1983qq,Meissner:1987ge}, the anti-symmetric tensor field model~\cite{Gasser:1983yg,Ecker:1989yg}, the hidden local symmetry (HLS)~\cite{Bando:1984ej} (for review see \cite{Harada:2003jx}), and so on.
We use the HLS to introduce the dark vector mesons into the chiral Lagrangian, discussed in \cref{app:vector_meson} in detail. 
The original global symmetry of the chiral Lagrangian may be enhanced to the local symmetry, known as the HLS.
The dark vector mesons are realized as dynamical gauge bosons in the HLS.
The Lagrangian for dark mesons is given by
\eqs{
  \mathcal{L} & =
  \mathrm{Tr} (\partial_\mu \pi' \partial^\mu \pi')
  - m_{\pi'}^2 \mathrm{Tr} (\pi'^2)
  - \frac12 \mathrm{Tr} (V'_{\mu\nu} V'^{\mu\nu}) 
  + m_{V'}^2 \mathrm{Tr}(V'_\mu V'^\mu) \\
  & \qquad 
  - 2 i g \mathrm{Tr} (V'_\mu [\pi', \partial^\mu \pi'])
  + \cdots \,.
  \label{eq:VpiLagrangian}
}
Here, $V'_\mu$ and $V'_{\mu\nu}$ are the dark vector meson field and its field strength tensor.
We use the matrix notations for the dark pion, $\pi' = \pi'^a T^a$, and for the dark vector mesons, $V_\mu' = V_\mu'^a T^a$.
We assume that the mass terms and the interaction terms respect the residual global symmetry $SU(N_f)_V$: in other words, we assume that any corrections to them violating the global symmetry can be negligible, such as the dark quark mass difference and the dark QED corrections.
The mass of the dark vector mesons and the dark vector meson coupling to the dark pions are related to each other through the Kawarabayashi-Suzuki-Riazuddin-Fayyazuddi (KSRF) relation~\cite{Kawarabayashi:1966kd,Riazuddin:1966sw}:
\eqs{
  m_{V'}^2 = 2 g^2 f_{\pi'}^2 \,.
  \label{eq:KSRF_relation}
}
The KSRF relation has been established by the use of the current algebra and vector meson dominance and has been phenomenologically confirmed in the SM.
The dark meson coupling $g$ originates from the gauge coupling of the hidden local gauge dynamics in the HLS. 
The gauged WZW Lagrangian provides the interactions between dark photon and dark hadrons.
\eqs{
  \mathcal{L}_\mathrm{WZW} \supset 
  - \frac{3 e' g}{8 \pi^2 f_{\pi'} }\epsilon^{\mu\nu\rho\sigma} \partial_\mu A'_\nu \mathrm{Tr} \left( 
    \{Q, V'_\rho\} \partial_\sigma \pi' 
  \right) \,, 
  \label{eq:ApiVcoupling}
}
where $e'$ is the $U(1)_D$ coupling; $Q$ is the $U(1)_D$ charge matrix of the dark quarks.

\subsection{Dark-Pion Dark Matter Models}

Dark QCD can realize the SIMP models~\cite{Hochberg:2014kqa}; dark pions are the SIMP DM, and the number-changing processes arise from the topological term of the low-energy effective theory, called WZW term~\cite{Wess:1971yu,Witten:1983tw}.
The stability can be ensured by the unbroken flavor symmetry, $SU(N_f)_V$.
The dark sector should be in kinetic equilibrium with the SM sector until the freeze-out of the number-changing process in order that the dark sector does not stay hot~\cite{Hochberg:2014dra}.
The dark photon is one of the candidates for keeping two sectors in thermal equilibrium.
Meanwhile, the presence of dark photon leads to the anomalous decay of dark pions as $\pi^0 \to \gamma\gamma$ in the SM; we assume that the charge matrix of dark quarks $Q$ should satisfy the condition $Q^2 \propto \mathds{1}$ to avoid the anomalous decay.
Following Refs.~\cite{Hochberg:2015vrg,Berlin:2018tvf}, we take the charge matrix in the SIMP models to be 
\eqs{
  Q_\mathrm{SIMP} = \mathrm{diag}(1, -1, -1) \,.
}
Even if there is no anomalous decay of dark pions to dark photons, the dark neutral pion can decay via $\mathcal{O}(p^6)$ terms of the chiral Lagrangian by taking charge and mass matrices as spurion fields.
\eqs{
  \mathcal{L}_{p^6} & \supset
  \frac{i \alpha'}{(4\pi)^2 f_{\pi'}} \epsilon^{\mu\nu\rho\sigma} F'_{\mu\nu} F'_{\rho\sigma} \mathrm{Tr}(Q) \mathrm{Tr}(Q M \Sigma^\dag) \\
  & \qquad + 
  \frac{i \alpha'}{(4\pi)^2 f_{\pi'}} \epsilon^{\mu\nu\rho\sigma} F'_{\mu\nu} F'_{\rho\sigma} \mathrm{Tr}(M^\dag \Sigma Q \Sigma^\dag Q \Sigma^\dag) + \mathrm{h.c.} 
  \,,
  \label{eq:piondecay_operators}
}
where the coefficients are determined by na\"ive dimensional analysis (NDA)~\cite{Manohar:1983md,Georgi:1992dw} with $\alpha' \equiv e'^2/4\pi$.
A detailed study of the decay of SIMP and the phenomenological consequence of the late-time decay of the dark pion has been discussed in Ref.~\cite{Katz:2020ywn}.
A specific spectrum $m_{A'} \gtrsim m_{\pi'}$ is required to avoid the late-time annihilation of DM, $\pi' \pi' \to A' A'$, impacting on the distortion of the CMB via $A' \to \ell^+ \ell^-$.

The cross section for the number-changing process is proportional to $(m_{\pi'}/f_{\pi'})^{10}$, while the self-scattering cross section of the dark pions is proportional to $(m_{\pi'}/f_{\pi'})^{4}$.
A large pion self-coupling $m_{\pi'}/f_{\pi'}$ is required to obtain the correct relic abundance.
On the other hand, the observations of merging galaxy clusters put an upper bound on the self-scattering cross section, $\sigma_\mathrm{scatter}/m_{\pi'} \lesssim 1 \, \mathrm{cm}^2/\mathrm{g}$~\cite{Clowe:2003tk,Markevitch:2003at,Randall:2008ppe}.
The sub-GeV dark pions with self-coupling close to the perturbative bound, $m_{\pi'}/f_{\pi'} \simeq 4 \pi/\sqrt{N_c}$, would be consistent with these requirements~\cite{Hochberg:2014kqa}.%
\footnote{
  We take into account the large-$N_c$ scaling of the decay constant $f^2_{\pi'} \sim \mathcal{O}(N_c)$~\cite{tHooft:1973alw,tHooft:1974pnl,Witten:1979kh,Witten:1979vv,Coleman:1980mx,Witten:1980sp} for the NDA bound on the pion self-coupling~\cite{Manohar:1983md,Georgi:1992dw}.
}

The presence of dark vector mesons alleviates the bound from correct relic abundance on $m_{\pi'}/f_{\pi'}$ in two ways.
There will be resonant contributions to the $3 \to 2$ annihilation cross section through $V'\pi'\pi'$ and $V'\pi'\pi'\pi'$ interactions from the gauged WZW terms, and the required $m_{\pi'}/f_{\pi'}$ is reduced for $m_{V'} \simeq 2 m_{\pi'}$ and $m_{V'} \simeq 3 m_{\pi'}$~\cite{Choi:2018iit}.
The resultant self-scattering cross section is also reduced.
Meanwhile, for the dark vector mesons $m_{V'} < 2 m_{\pi'}$, there is a new channel, via semi-annihilation process $\pi' \pi' \to \pi' V'$, to deplete the DM number density.
The invisible decay $V' \to \pi' \pi'$ is kinematically forbidden for this spectrum.
Since we also have the inverse process, $\pi' V' \to \pi' \pi'$, the DM density is determined by the rate of the dark vector meson decay to the SM particles, which depends on the kinetic mixing $\epsilon$.
The dark pion annihilation to the SM particles, $\pi'\pi' \to \mathrm{SM}$ via dark photons, is efficient and determines the DM abundance for $\epsilon \gtrsim \mathcal{O}(10^{-3})$.
The semi-annihilation and the $V'$ decay to the SM determine the abundance of the dark pions in the mass range of sub-GeV to GeV for the parameter range of $10^{-6} \lesssim \epsilon \lesssim 10^{-3}$~\cite{Berlin:2018tvf}. 
Thus, the presence of $V'$ is more significant for the dark-pion DM scenarios within the range of the kinetic mixing, $10^{-6} \lesssim \epsilon \lesssim 10^{-3}$, for the dark pion mass of sub-GeV to GeV, which will be covered at the lifetime frontier experiments as we will discuss in \cref{sec:results}.

\subsection{Composite Asymmetric Dark Matter}

The composite ADM model is another alternative that provides the dark vector mesons in the low-energy spectrum.
The ADM scenario is motivated by the observation that the mass density of DM is about five times larger than that of the SM baryons, $\Omega_\mathrm{DM} \simeq 5\Omega_{b}$.
This fact suggests that there is a connection between the cosmological evolution of the dark sector and that of the SM sector.
Analogous to the present baryon density originating from the baryon asymmetry in the early universe, the ADM scenarios postulate that the present DM density is due to the DM particle-antiparticle asymmetry. 
Once the asymmetries are shared with each other in some way during the early universe epoch, the number densities of DM and baryons are related to each other.

In the composite ADM models, the lightest dark baryon is the DM, the stability of DM is ensured by the accidental dark baryon number conservation, and the symmetric component of the relics is significantly reduced by their annihilation into dark pions as with the SM nucleons. 
The dimensional transmutation of the dark confining dynamics would explain the GeV-scale DM mass, and hence the coincidence of the energy densities is naturally realized~\cite{Berryman:2017twh,Lonsdale:2017mzg,Ibe:2018juk,Chu:2018faw,Tsai:2020vpi,Asadi:2021yml,Bottaro:2021aal,Zhang:2021orr,Ibe:2021gil,Hall:2021zsk}.%
\footnote{
  There is no need for the dynamical scale in the dark sector to appear at the GeV-scale. 
  However, once the dark strong coupling is close to the SM strong coupling at a certain scale, the dark dynamical scale can be naturally close to the QCD scale. 
  The coincidence of the energy densities is explained in the context of mirror sector setup with grand unification~\cite{Lonsdale:2018xwd,Ibe:2018tex,Ibe:2019ena}.
}
A portal interaction sharing the asymmetry connects the dark sector to the SM sector in the early universe, and hence the temperature is shared in both sectors. 
After the decoupling of the portal interaction, entropy is conserved separately; 
the strong annihilation of the symmetric component will lead to a significant contribution to the dark radiation~\cite{Blennow:2012de}.
The dark photon is introduced in the dark sector as a low-energy portal interaction in order to release the superfluous entropy of the dark sector into the SM sector.
The number of dark pions is reduced via the annihilation or decay into the dark photons, and then the dark photons finally decay into the SM particles. 
In contrast to the dark-pion DM scenarios, the dark-neutral pions should decay into dark photons through the chiral anomaly in order to avoid the additional contribution to dark radiation.
We assume $m_{A'} \lesssim m_{\pi'}$ for the composite ADM models and assume that the charge matrix of dark quarks takes the same form as the charge matrix of the SM quarks:
\eqs{
  Q_\mathrm{ADM} = \frac{1}{3} \mathrm{diag}(2, -1, -1) \,.
  \label{eq:Qmat_ADM}
}

The dark vector meson would play a role in providing velocity-dependent self-scattering of DMs~\cite{Chu:2018faw}.
One of the dark quarks may have a mass greater than the dark confining scale, while the other remains nearly massless. 
Dark baryons consisting of the heavier dark quark are bound in a nuclei-like state by mediating dark $\eta$ meson consisting of light dark quarks; the dark vector mesons (dark $\rho$ meson) consisting of light dark quarks provide coherent spin-independent scattering of dark ``nuclei''.
The dark nucleus radius is enhanced by the nuclei mass number, $\sim A^{1/3} m_{\eta'}^{-1}$, but the range of the scattering force is $\sim m_{\rho'}^{-1}$.
A mass spectrum, $m_{\rho'} \simeq m_{\eta'}$, is expected in order to realize the velocity-dependent self-scattering.

\section{Forward Searches for Dark Vector Mesons \label{sec:detectors}}

In this section, we discuss the visible decay searches of the dark vector mesons at the LHC forward detectors, FASER/FASER2 and FACET. 

\subsection{Dark Photon Searches}

Before going to discuss the decay of dark vector mesons, we summarize the dark photon searches and introduce the basic concept of searching for long-lived particles.
The search strategy for dark photons depends on the spectrum of dark particles. 
Concerning the dark-pion DM models, the dark photon is required to be heavier than dark pions.
Thus, dark photons produced at accelerator-based experiments mostly decay into the dark-sector particles, and then dark photons leave only invisible signals at detectors.
The mono-photon signal searches at $e^+ e^-$ colliders are available for constraining the invisible decay of the dark photons ($e^+ e^- \to \gamma A'$) (for the existing upper bound from BaBar, see Ref.~\cite{BaBar:2017tiz}; and for the future prospect at Belle II, see Ref.~\cite{Belle-II:2018jsg}).
The missing-energy measurements are also available and most powerful to constrain invisible signals when it is possible to precisely measure beam energy and the energy of final-state particles at the detector.
The dark-sector particles produced from the decay of dark photons carry away some fraction of the primary beam energy.
The missing-energy measurements at the fixed-target experiments have been proposed for searching for the light DM models: e.g., NA64~\cite{Banerjee:2019pds} and LDMX~\cite{Izaguirre:2014bca,LDMX:2018cma}.

Concerning the composite ADM models, dark photons tend to be the lightest particle in the dark sector. 
Dark photons decay only into the SM particles through the kinetic mixing with the SM photons, and hence the decay leaves visible signals.
The existing beam-dump experiments have already constrained the visible decay of dark photons: electron beam-dump experiments, E137~\cite{Bjorken:1988as}, E141~\cite{Riordan:1987aw}, E774~\cite{Bross:1989mp}, Orsay~\cite{Davier:1989wz}, KEK~\cite{Konaka:1986cb}, and NA64~\cite{NA64:2018lsq},; proton beam-dump experiments, CHARM~\cite{CHARM:1985nku,Gninenko:2012eq}, NA48/2~\cite{NA482:2015wmo}, LSND~\cite{LSND:1997vqj,Batell:2009di}, U70/$\nu$Cal~\cite{Blumlein:2011mv,Blumlein:2013cua}.
A fixed-target experiment, SeaQuest~\cite{SeaQuest:2017kjt}, will be upgraded by inserting a decontaminated electromagnetic calorimeter~\cite{Gardner:2015wea,Berlin:2018pwi,Apyan:2022tsd}, which is called DarkQuest experiment.
The visible decay search at DarkQuest will cover a broad range of unexplored parameters for dark photons. 
In addition to the fixed-target experiments, several experiments have been proposed to search for the dark-sector particles at the LHC, collectively called the lifetime frontier at the LHC.
These are located outside the ATLAS/CMS detector systems and $\mathcal{O}(1) \text{--} \mathcal{O}(10^2) \, \mathrm{m}$ away from the collision points, but there is a difference among them: ones will be located in a very forward direction (along the beam axis of the LHC), FASER/FASER2~\cite{Feng:2017uoz} and FACET~\cite{Cerci:2021nlb}, while others will be located at the off-axis region, e.g., CODEX-b~\cite{Gligorov:2017nwh} and MATHUSLA~\cite{Chou:2016lxi}.
The forward detector experiments are suitable for searching for the dark-sector particles with a mass in the range of sub-GeV to GeV since these light particles tend to be highly boosted at the LHC.
We focus on the forward detector experiments at the LHC, namely FASER/FASER2 and FACET, in this study.

\subsection{Visible Decay of Dark Vector Mesons}

Dark vector mesons promptly decay into two dark pions, $V' \to \pi'\pi'$, as with the SM when $m_{V'} > 2 m_{\pi'}$.
In contrast to the SM, the dark hadron spectrum is model-dependent, and the dark vector mesons can have the mass similar to the dark pion mass. 
The dark vector mesons are possible to be long-lived in such cases, in particular for $m_{V'} < 2 m_{\pi'}$.
We give the detailed formulae for the decay of dark vector mesons in \cref{app:Decay_Rate}.

There are two kinds of dark vector mesons in the presence of the dark QED: one is neutral under the dark QED, and another is charged.
The dark-neutral vector mesons will mix with dark photons through the kinetic mixing:
\eqs{
  \mathcal{L} \supset - \frac{e'}{g} \mathrm{Tr} (Q V'_{\mu \nu}) F'^{\mu\nu} \,.
}
The physical gauge boson contains the vector meson components via the kinetic mixing, which is known as the vector meson dominance in the SM.
Therefore, the dark-neutral vector mesons decay to the SM particles through the mixing.
The decay rate into SM leptons takes the form:
\eqs{
  \Gamma(V'^{a} \to \ell^+ \ell^-) \simeq \frac{16 \pi \alpha' \alpha \epsilon^2}{3 g^2} \frac{m_{V'}^5}{m_{A'}^4} \left(\frac{m_{A'}^2}{m_{A'}^2 - m_{V'}^2}\right)^2 [\mathrm{Tr}(Q T^a)]^2 \,.
}
in the massless limit of SM leptons.
Here, $\alpha$ denotes the electromagnetic structure constant.
The dark-neutral vector mesons also decay into the SM hadrons; we incorporate the hadronic decays by the use of the data-driven $R$-ratio taken from Ref.~\cite{Zyla:2020zbs}.
\eqs{
  \Gamma(V'^{a} \to \text{SM hadrons}) & = R(\sqrt{s} = m_{V'}) \Gamma(V'^{a} \to \mu^+ \mu^-) \,, \\
  R(\sqrt{s}) & = \frac{\sigma(e^+ e^- \to \text{hadrons})}{\sigma(e^+ e^- \to \mu^+ \mu^-)} \,.
}
Meanwhile, the dark-charged vector mesons have a coupling to the dark photon and the dark pion, which is given in \cref{eq:ApiVcoupling}.
When the on-shell dark photon channel $V' \to \pi' A'$ does not open, the dark vector mesons decay into the SM leptons via off-shell dark photon by emitting the dark pion: $V' \to \pi' A'^\ast \to \pi' \ell^+ \ell^-$.
Thus, the dark-charged vector mesons tend to have a longer lifetime than the dark-neutral vector mesons due to the phase space of three-body decay.
The three-body decay rate approximately takes the form
\eqs{
  \Gamma(V'^{a} \to \pi'^b \ell^+ \ell^-) \simeq \frac{1}{256 \pi^3} \frac{\alpha' \alpha g^2 \epsilon^2}{3 \pi^2} \frac{m_{V'}^7}{f_{\pi'}^2 m_{A'}^4} [\mathrm{Tr}(\{Q, T^a\} T^b)]^2 \,.
}
The proper decay lengths of dark vector mesons are approximately given by 
\eqs{
  c\tau(V'^{a} \to \text{SM}) 
  & \simeq \mathcal{O}(10^{-5})\, \mathrm{[m]} \left(\frac{0.01}{\alpha'}\right) \left(\frac{10^{-3}}{\epsilon}\right)^2 \left(\frac{m_{\pi'}/f_{\pi'}}{3}\right)^2 \left(\frac{500\, \mathrm{MeV}}{m_{V'}}\right) \,, \\
  c\tau(V'^{a} \to \pi'^b + \text{SM}) 
  & \simeq \mathcal{O}(1)\, \mathrm{[m]} \left(\frac{0.01}{\alpha'}\right) \left(\frac{10^{-3}}{\epsilon}\right)^2 \left(\frac{3}{m_{\pi'}/f_{\pi'}}\right)^4 \left(\frac{500\, \mathrm{MeV}}{m_{V'}}\right) \,.
  \label{eq:decay_of_V}
}
Here, we assume the mass spectrum as $m_{\pi'}/m_{A'}=1/3$ and $m_{V'}/m_{A'}=0.6$.
We take the dark QED coupling as $\alpha' = 0.01$ in the following: it is expected that there will be no Landau pole up until an ultraviolet scale such as the grand unification scale or the Planck scale.
For the parameter space providing proper decay lengths of $\mathcal{O} (1) \, [\mathrm{m}]$, they do not cause the late-time energy injection to the electromagnetic channels, which is stringently constrained by big-bang nucleosynthesis (BBN) or cosmic microwave background (CMB) anisotropy~\cite{Poulin:2016anj}.
On the other hand, their proper decay lengths are suitable to the accelerator-based searches for the long-lived particles.

The fixed-target experiments can be utilized for searching the dark vector mesons in the dark-pion DM models~\cite{Berlin:2018tvf}.
No visible/invisible signal observation at existing (electron) beam-dump experiments, such as E137, has placed constraints on a parameter region.
There are several proposals for searching for long-lived particles with the fixed-target experiments: LDMX, HPS, and DarkQuest. 
LDMX is designed to search for the missing momentum with precisely measuring both momentum of the beam electrons and the energy deposition in a calorimeter. 
This facility will be able to search for two distinct channels: one is the invisible decay of dark photons, $A' \to \pi' \pi'$, and another is the visible decay of the dark vector mesons inside of the calorimeter. 
The others are designed to search for the visible decay signals and use different beam particles: an electron beam for HPS and a proton beam for DarkQuest.
Besides, the invisible decay of dark photons, $A' \to \pi' \pi'$, can be explored at the collider experiments, such as BaBar and Belle(-II) experiments using $e^+e^-$ colliders. 
Meanwhile, the visible decay searches at the LHC lifetime frontier have been studied in the context of the composite ADM models in Ref.~\cite{Kamada:2021cow}: in particular, dark nucleon transition via $U(1)_D$ breaking, $N'_1 \to N'_2 \ell^+ \ell^-$, and the anomalous decay of dark pions $\pi' \to A' \ell^+ \ell^-$.
Two distinct searches are available for exploring the models at the visible-decay search experiments: signals from dark photon decay $A' \to \mathrm{SM}$ and from decay/transition of dark hadrons.

\subsection{Dark Vector Mesons at the LHC}

The forward detector searches at the LHC have been proposed: one is the FASER/FASER2 experiments~\cite{Feng:2017uoz} and another is the FACET experiment~\cite{Cerci:2021nlb}.
FASER/FASER2 (FACET) detector will be located far from the collision points in the forward direction to the ATLAS (CMS) beam collision point.
FASER has started taking data during Run 3 of the LHC and will be upgraded to FASER2 during the High-luminosity LHC (HL-LHC). 
There exists the LHC infrastructure upstream of the detector, and thus the FASER/FASER2 detector is shielded by $\sim \mathcal{O}(100) \, \mathrm{m}$ of concrete and rock.
FASER/FASER2 is located $\sim 480 \, \mathrm{m}$ downstream of ATLAS just after the main LHC tunnel starts to curve.
The FASER/FASER2 detector has a cylindrical shape.
Meanwhile, FACET is a proposed experimental subsystem for the long-lived particle searches during the HL-LHC.
FACET detector is shielded by $\sim \mathcal{O}(40) \, \mathrm{m}$ of magnetized iron, which is located upstream of the detector.
The FACET experiment is proposed to be located on the LHC beamline, and hence it is required to replace a part of the LHC beam pipe with a circular pipe of a 50\,cm radius. 
The detector design of these experiments is summarized in \cref{tab:detectors}.

\begin{table}[t]
    \centering
    \begin{tabular}{c|ccccccc}
    \hline
    Detector & Distance & Length &Radius & Threshold & Luminosity \\
    \hline
    FASER& 480\,m & 1.5\,m & 10\,cm & 100\,GeV & $150\,\mathrm{fb}^{-1}$ \\
    FASER2& 480\,m & 5\,m & 1\,m & 100\,GeV & $3\,\mathrm{ab}^{-1}$ \\ \hline
    FACET & 120\,m & 18\,m & $0.18 \mathrm{m} < R < 0.5 \mathrm{m}$ & 10\,GeV & $3\,\mathrm{ab}^{-1}$ \\
    \hline
    \end{tabular}
    \caption{
      Shown is the detector design of FASER/FASER2 and FACET. 
      FASER is running during the LHC run-3 (with integrated luminosity of $150\,\mathrm{fb}^{-1}$), while FASER2 and FACET are planed to run at HL-LHC (with the total luminosity of $3 \, \mathrm{ab}^{-1}$ and the collision energy of $14 \, \mathrm{TeV}$). \
      ``Distance'' denotes distances from the collision point to the front of detectors.
      ``Length'' denotes the length of the detector decay volume, and ``Radius'' is the radius of the detectors. 
      ``Threshold'' denotes the minimum energy of visible particles to be deposited at the detectors, which is assumed in our numerical simulations.
    }
    \label{tab:detectors}
\end{table}

We impose the minimum energy of the visible particles to be deposited at the detectors in order to reduce the trigger rate and the background events. 
Following Refs.~\cite{FASER:2018eoc,Cerci:2021nlb}, we apply $E_\mathrm{vis} > 100\,\mathrm{GeV}$ for FASER/FASER2 and $E_\mathrm{vis} > 10\,\mathrm{GeV}$ for FACET.%
\footnote{
  The future sensitivities have a strong dependence on the minimum energy of the visible particles, as shown in Ref.~\cite{Berlin:2018jbm}. 
  Further studies of the background events are required in order to determine the future sensitivities precisely. 
}
Both experiments have the shielding upstream of the detectors in order to remove most of the SM-charged particles.
Muons can reach the detectors from the interaction points, but putting an additional system upstream of the detectors reduces background events: in particular, a system consisting of scintillation layers and a lead layer vetoes the high-energy muons at FASER, and hodoscopes tag the charged particles at FACET.
The background events are estimated at zero in FASER/FASER2 detector, and the first results of the FASER experiment support the assumption of no background event~\cite{MoriondFASER:2023,Petersen:2023hgm}.
Meanwhile, there would remain enormous background events from the decay of neutral hadrons in the FACET~\cite{Cerci:2021nlb}: these background events would not be negligible for the decay into a charged-hadron pair of the long-lived particle with a mass below $0.8\,\mathrm{GeV}$, and for the decay into a leptonic-pair with a mass below $0.6\,\mathrm{GeV}$.
The dark-sector particles produced at the interaction points travel in the forward directions, penetrate the shielding and veto/tagging system, and get into the decay volume.
The dark-sector particles with a proper decay length of $\mathcal{O}(1) \,\mathrm{m}$ decay into the SM particles in the decay volume, leaving signals at the tracking system and calorimeters. 

\begin{table}[t]
  \centering
  \begin{tabular}{c|cc}
  \hline
  & Dark Pion DM & Composite ADM \\
  \hline
  Mass spectrum & $m_{\pi'} \lesssim m_{V'} < m_{A'}$ & $m_{A'} < m_{\pi'} \lesssim m_{V'}$ \\
  & $r_{\pi'}=1/3\,, r_{V'}=3/5$ 
  & $r_{\pi'}=1.2\,, r_{V'}=2$  \\
  Charge Matrix $Q$ & $\mathrm{diag}(+1,-1,-1)$ & $\frac13 \mathrm{diag}(+2,-1,-1)$ \\
  $V'$ production & $A'$ decay & off-shell $A'$ \\
  $A'$ signal & invisible decay & visible decay \\
  \hline
  \end{tabular}
  \caption{
    Summary of two dark hadron models.
    The mass hierarchy is shown in the second row, and we show the specific ratios assumed in the text: the ratios defined by $r_{\pi'} = m_{\pi'}/m_{A'}$ and $r_{V'} = m_{V'}/m_{A'}$.
    The charge matrices are different: there is no chiral anomaly in the dark pion DM models, and the dark-neutral pion should decay into dark photons in the composite ADM. 
    In the last row, we show the dark photon signals that are compatible to the dark vector meson searches.
  }
  \label{tab:models}
\end{table}

Now, we consider two different production mechanisms of dark hadrons: via on-shell dark photon (dark-pion DM scenarios) and via off-shell dark photon (composite ADM scenarios).
We summarize the differences of these models in \cref{tab:models}: mass spectra, production processes of $V'$, and signals of $A'$ that can be explored simultaneously.
Concerning the dark-pion DM scenarios, we adopt the same parameters chosen in Ref.~\cite{Berlin:2018tvf}: the dark QED coupling is fixed to be $\alpha'=0.01$; and the mass spectrum among dark photon, dark pion, and dark vector meson is fixed to be $r_{\pi'} = m_{\pi'}/m_{A'}=1/3$ and $r_{V'} = m_{V'}/m_{A'}=3/5$. 
This spectrum allows the decay $A' \to \pi' V'$, but the decays $V' \to A' \pi'$ and $V' \to \pi'\pi'$ are kinematically forbidden.
The dark photons are produced at the accelerator-based experiments, and then dark vector mesons are predominantly produced via the decay of $A'$ in the dark-pion DM scenarios. 
Therefore, the number of dark vector mesons produced at the accelerator-based experiments is given by
\eqs{
  N_{V'} \simeq N_{A'} \mathrm{Br}(A' \to \pi' V') \,.
}
Here, $N_{A'}$ is the number of dark photons produced at the accelerator-based experiments, which we will discuss later. 
Meanwhile, as for the off-shell dark photon case, we take the mass spectrum to be $r_{\pi'} = m_{\pi'}/m_{A'}=1.2$, $r_{V'} = m_{V'}/m_{A'}=2$; this spectrum does not allow $A' \to \pi' V'$, $V' \to A' \pi'$, and $V' \to \pi'\pi'$.
Here, the dark photons are the lightest particle in the dark sector, and hence the dark photon predominantly decays into the SM particles through the kinetic mixing. 
Dark vector mesons can be produced only through mediating off-shell dark photons at the accelerator-based experiments.

We use a package \texttt{FORESEE}~\cite{Kling:2021fwx} to get the expected sensitivity of forward-detector experiments with proton-proton colliders, in particular the LHC.
We consider three different production processes of the dark vector mesons at the collider experiments: on/off-shell production of dark photons from meson decays, proton bremsstrahlung, and the Drell-Yan process. 
It is important for the sensitivity to know the kinetic distribution of dark vector mesons produced at the interaction point at the LHC.
The kinetic distribution of dark photons, momentum $p_{A'}$ and angle $\theta_{A'}$ with respect to the beam axis, gives that of dark vector mesons, momentum $p_{V'}$ and angle $\theta_{V'}$, via the decay $A' \to V' \pi'$ in the case of the on-shell dark photon.
We generate the kinetic distribution of dark photons by the use of \texttt{FORESEE}, and then we recast it as that of dark vector mesons.

Concerning the composite ADM scenarios, we directly compute the distribution of dark vector mesons. 
We briefly explain our estimation of the dark vector meson production via the off-shell dark photon.
We discuss the dark vector meson production in \cref{app:production} in detail.
The SM mesons are enormously produced in the forward direction at the LHC: in particular, the SM pions, $\eta$ mesons, and $\phi$ mesons.
These mesons, denoted by $\varphi$, decay into dark vector mesons as $\varphi \to \gamma V' \pi'$ via anomalous decay.
For the production via meson decay, it is important to provide the momenta of the SM meson and its angle with respect to the beam axis.
The kinetic distributions of the SM mesons are obtained with \texttt{EPOS-LHC}~\cite{Pierog:2013ria}, which is incorporated in \texttt{FORESEE}.
The meson decay is a dominant process to produce dark hadrons with a mass below a few hundred MeV. 

Above the meson mass threshold, the proton bremsstrahlung dominantly produces dark vector mesons. 
The differential cross section for the dark vector meson production is approximately given by%
\footnote{
  We may have to take into account the effect from the spin-density matrix for the precise simulation since the mediator particle has a spin one. 
  However, the kinetic distribution strongly depends on the kinetic mixing, and the effect provides a subdominant correction to the distribution. 
  Therefore, we ignore the effect in this study.
} 
\eqs{
  \frac{d \sigma}{dx d m_{A'}^{\ast 2}}
  & \simeq \frac{d \sigma_{A'}}{dx} \times \frac{1}{\pi} \frac{m_{A'}^\ast \Gamma_{A' \to V' \pi'}}{(m_{A'}^{\ast 2} - m_{A'}^2)^2 + (m_{A'} \Gamma_{A'})^2} \,.
  \label{eq:offshell_productionxsec}
}
Here, we divide the full cross section into two subprocesses: the production of the dark photon with the fictitious mass $m^\ast_{A'}$ and its decay into $V' \pi'$.
$d \sigma_{A'}/dx$ denotes the production cross section of dark photon with $m_{A'}^\ast$ and $x$ collectively denoting the kinetic variables for the dark photon production, and we compute the cross section by the use of the Fermi-William-Weizs\"acker (FWW) approximation~\cite{Blumlein:2013cua}.
We incorporate the electromagnetic form factor of the proton, and hence there is enhanced production of dark photons near $m_{A'}^{(\ast)} \simeq 0.7 \, \mathrm{GeV}$ due to hadronic resonances $\rho$ and $\omega$.
The production cross section via proton bremsstrahlung drastically decreases for $m_{A'}^{(\ast)} \gtrsim 1 \, \mathrm{GeV}$ since the internal structure of nucleons is probed.

At the large mass region of dark vector mesons, above $\mathcal{O}(1) \, \mathrm{GeV}$, the dominant contribution to the production is the Drell-Yan process $q \bar q \to A'^{\ast} \to V' \pi'$.
The kinetic distribution of dark hadrons is obtained using \texttt{FeynRules}~\cite{Alloul:2013bka}, \texttt{MadGraph5}~\cite{Alwall:2014hca}, and \texttt{PYTHIA 8}~\cite{Sjostrand:2014zea}, and using the \texttt{NNPDF 3.1} parton distribution function (PDF) set at the leading order (LO)~\cite{NNPDF:2017mvq}.
The dominant uncertainty arises from the choice of the scale where the PDFs are evaluated: we choose the scale to be $\mu = m_{A'}$ for the on-shell production case, and $\mu = m_{A'}^{\ast}$ for off-shell production case. 

In the above discussion, we assume that the dark vector mesons are produced via the direct coupling of $A'$-$V'$-$\pi'$ shown in \cref{eq:ApiVcoupling}, which arises from the gauged WZW Lagrangian.
Vector mesons have masses of order of the dynamical scale $\Lambda_\mathrm{QCD'} \simeq 4\pi f_{\pi'}/\sqrt{N_C}$. 
It is uncertain that whether the production of dark hadrons with a mass of the dynamical scale is governed by the perturbative coupling in the effective models for hadrons, such as the HLS model of dark vector mesons.
In this study, we also consider dark hadronization in the dark vector meson production. 
However, contrary to the SM hadronization, we do not have any data for dark hadronization, and we are agnostic about hadronization.
Hence, we model the production cross section and kinematic distribution of the dark vector mesons via dark hadronization under na\"ive assumptions.
As with the production through the perturbative couplings, the differential cross section for the production of dark vector mesons is written as 
\eqs{
  \frac{d\sigma}{d x d m_{A'}^{\ast 2}} & = \frac{ d\sigma_{A'}(m_{A'}^\ast)}{dx} \frac{1}{\pi} \frac{m_{A'}^\ast \Gamma_{A'}(A'\to V'+X)}{(m_{A'}^{\ast 2}-m_{A'}^2)^2 + (m_{A'}^\ast \Gamma_{A'})^2}\,.
  \label{eq:diffxsec_had}
}
Here, we model the decay rate of $A'$ into $V'$ with the associated hadrons denoted by $X$, given by 
\eqs{
  \Gamma_{A'}(A'\to V'+X)
  & = n_{V'} \frac{\alpha' m_{A'}^\ast}{3} \mathrm{tr}(Q^2_\mathrm{ADM})\,, 
}
where $\alpha' m_{A'}^\ast \mathrm{tr}(Q^2_\mathrm{ADM})/3$ denotes the sum of $A'$ decay rate into a pair of dark quarks in the massless dark quark limit. 
$n_{V'}$ denotes the $V'$ multiplicity assumed to be constant; in our simulation, we assume $n_{V'} = 1.0$ and $n_{V'} = 0.01$.
$\mathrm{tr}(Q^2_\mathrm{ADM}) = 2$ in our choice of the charge matrix (\ref{eq:Qmat_ADM}).
We also assume the energy equipartition among dark hadrons: a dark hadron carries the energy of $E_{A'}/N_h$ with $E_{A'}$ being the energy of the dark photon and $N_h$ being the total numbers of dark hadrons. 
Since all dark hadrons have a similar mass, we na\"ively estimate the total numbers as $N_h = m_{A'}^\ast/m_{h'}$ where $m_{h'}$ denotes the averaged dark hadron mass, $m_{h'} = (m_{\pi'} + m_{V'})/2$.
This factorization of the hadronization part implies that, at a certain energy injection $m_{A'}^\ast$, the dark vector mesons follow the same distribution as the dark photons with a mass of $m_{A'}^\ast$ except for the overall normalization. 

The signal number from dark meson decay is given by
\eqs{
  N_\mathrm{signal} = 
  \int dx \frac{d N_{V'}}{d x} \mathcal{P}_\mathrm{dec} (x) \,, \qquad
  \mathcal{P}_\mathrm{dec} (x) = 
  \left( e^{-L_\mathrm{min}/d} - e^{-L_\mathrm{max}/d} \right) A_\mathrm{geo} \,.
  \label{eq:signalnumber}
}
Here, $x$ collectively denotes the kinetic variables of dark vector mesons produced at the interaction point, such as its angle $\theta_{V'}$ with respect to the beam axis. 
$\mathcal{P}_\mathrm{dec}$ denotes the probability of a dark vector meson to decay inside the detector.
$L_\mathrm{min}$ is the distance from the interaction point to the starting point of the detector, and $L_\mathrm{max}-L_\mathrm{min}$ denotes the fiducial decay range of the detector. 
$d = c \tau \beta \gamma$ is the boosted decay length of the produced dark vector mesons.
$A_\mathrm{geo}$ denotes the geometric acceptance, which is determined by requiring that the dark vector mesons decay inside the detector volume.
The geometric acceptance for FASER/FASER2 and FACET is 
\begin{align}
  A_\mathrm{geo} & = \Theta(R - L_\mathrm{max} \tan \theta_{V'}) \,, & \text{(FASER/FASER2)} \,, \\
  A_\mathrm{geo} & = \Theta(R_\mathrm{max} - L_\mathrm{max} \tan \theta_{V'}) \Theta(L_\mathrm{max} \tan \theta_{V'}-R_\mathrm{min}) \,, & \text{(FACET)} \,.
\end{align}
Here, $R\,, R_\mathrm{max}$, and $R_\mathrm{min}$ are listed in the Radius column of \cref{tab:detectors}.

\section{Results\label{sec:results}}

In this section, we show the prospective sensitivities of forward detectors at the LHC to the dark vector mesons.
First, we discuss the dark vector mesons produced via on-shell dark photons. 
Since a dark photon promptly decays into a dark vector meson and a dark pion, we assume that dark vector mesons come from the interaction point of the LHC.
As discussed in the previous section, we compute the kinetic distribution of the dark photon, which is incorporated in \texttt{FORESEE}, and we recast it as that of the dark vector mesons. 

\begin{figure}[t]
  \begin{minipage}[t]{0.50\linewidth}
    \centering
    \includegraphics[width=1\textwidth]{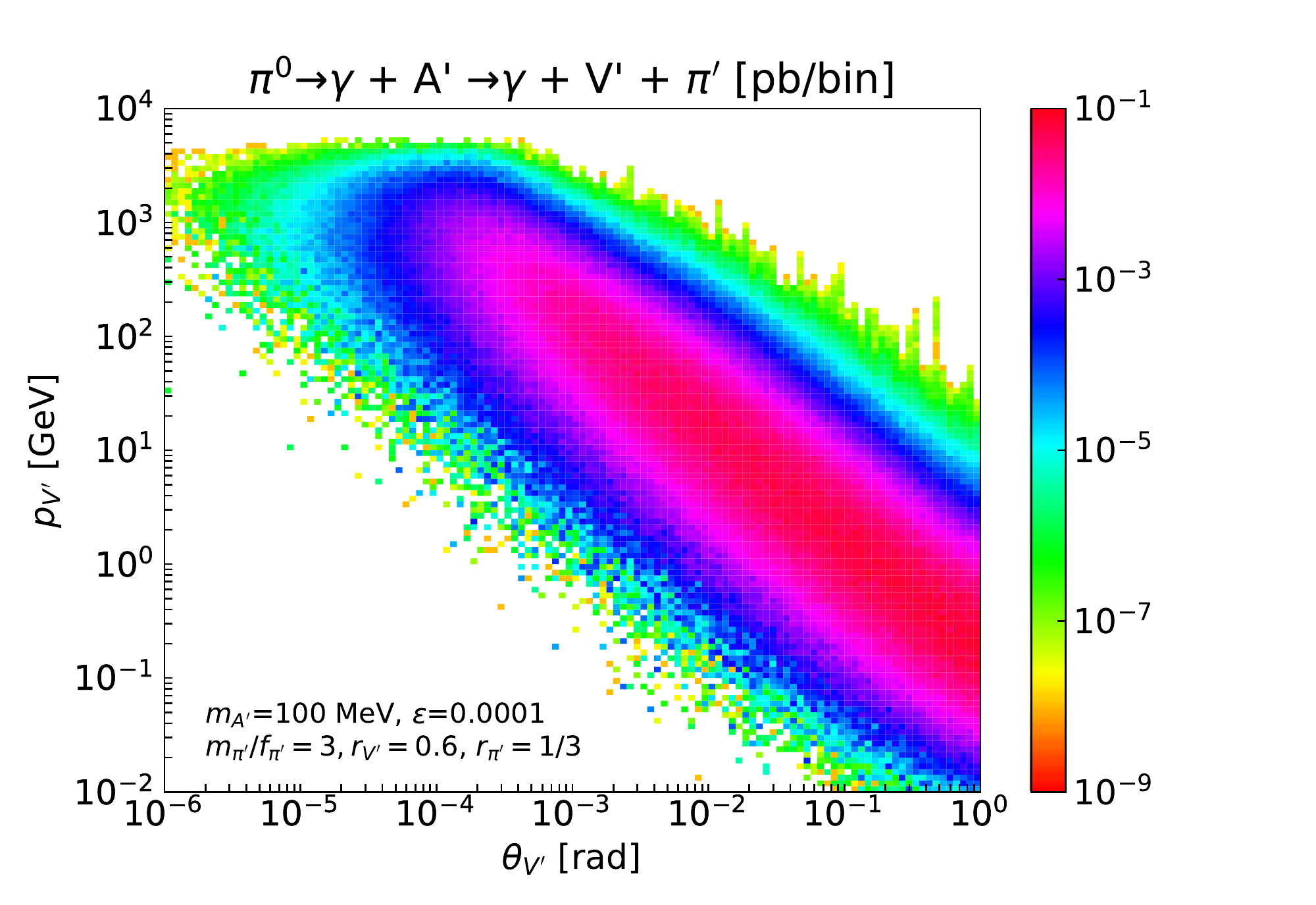}
  \end{minipage}
  \begin{minipage}[t]{0.50\linewidth}
    \centering
    \includegraphics[width=1\textwidth]{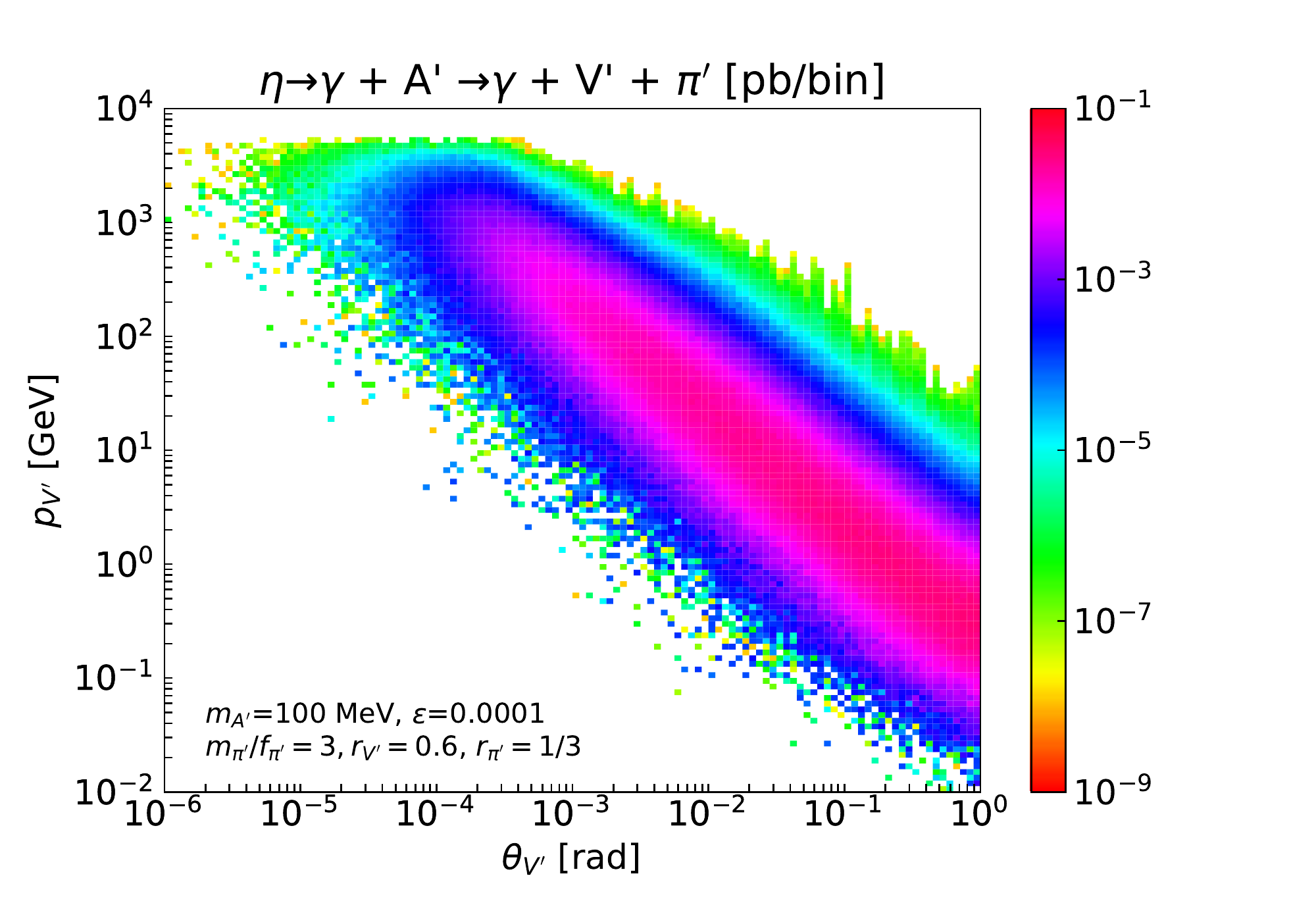}
  \end{minipage} \\
  \begin{minipage}[t]{0.50\linewidth}
    \centering
    \includegraphics[width=1\textwidth]{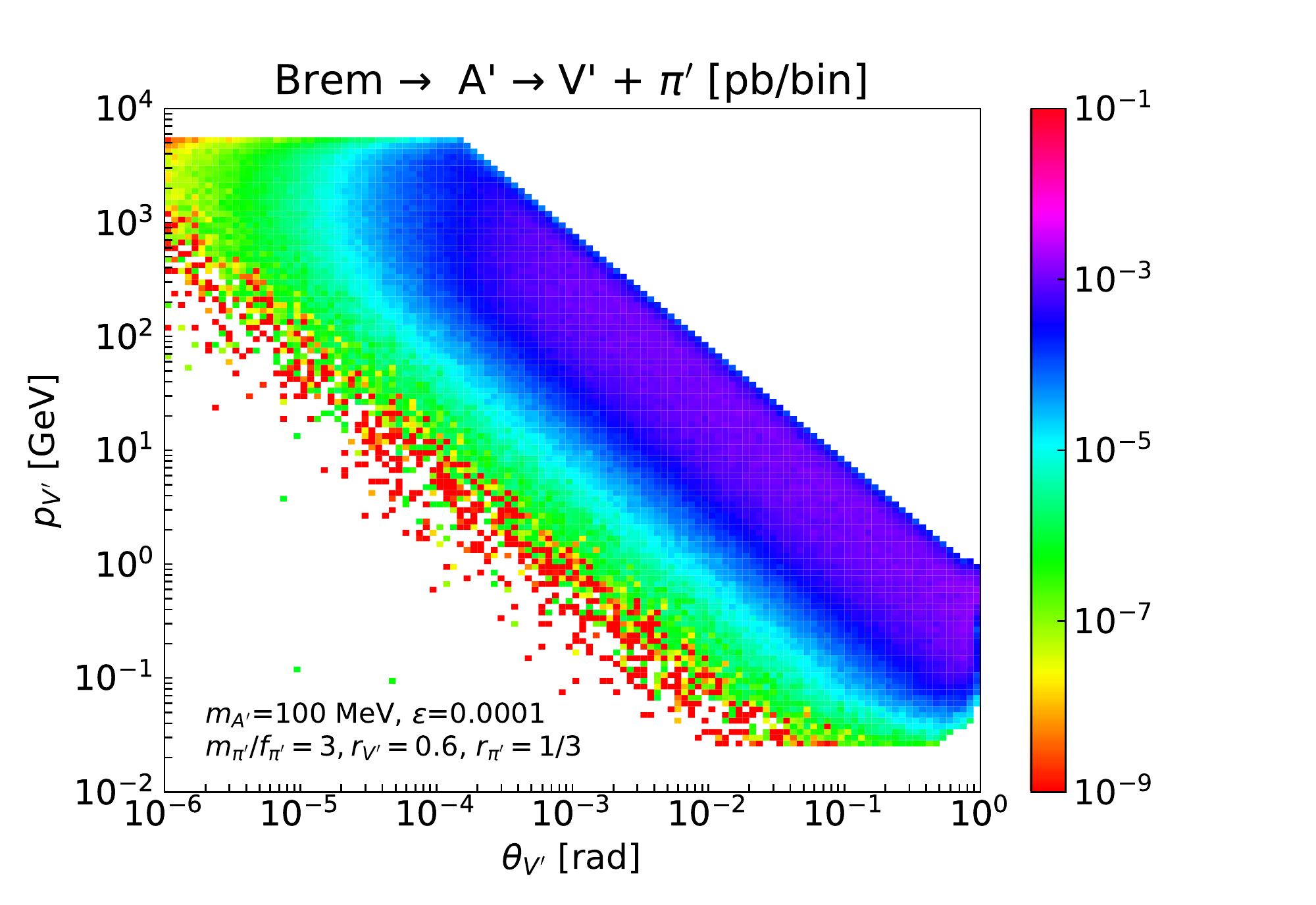}
  \end{minipage}
  \begin{minipage}[t]{0.50\linewidth}
    \centering
    \includegraphics[width=1\textwidth]{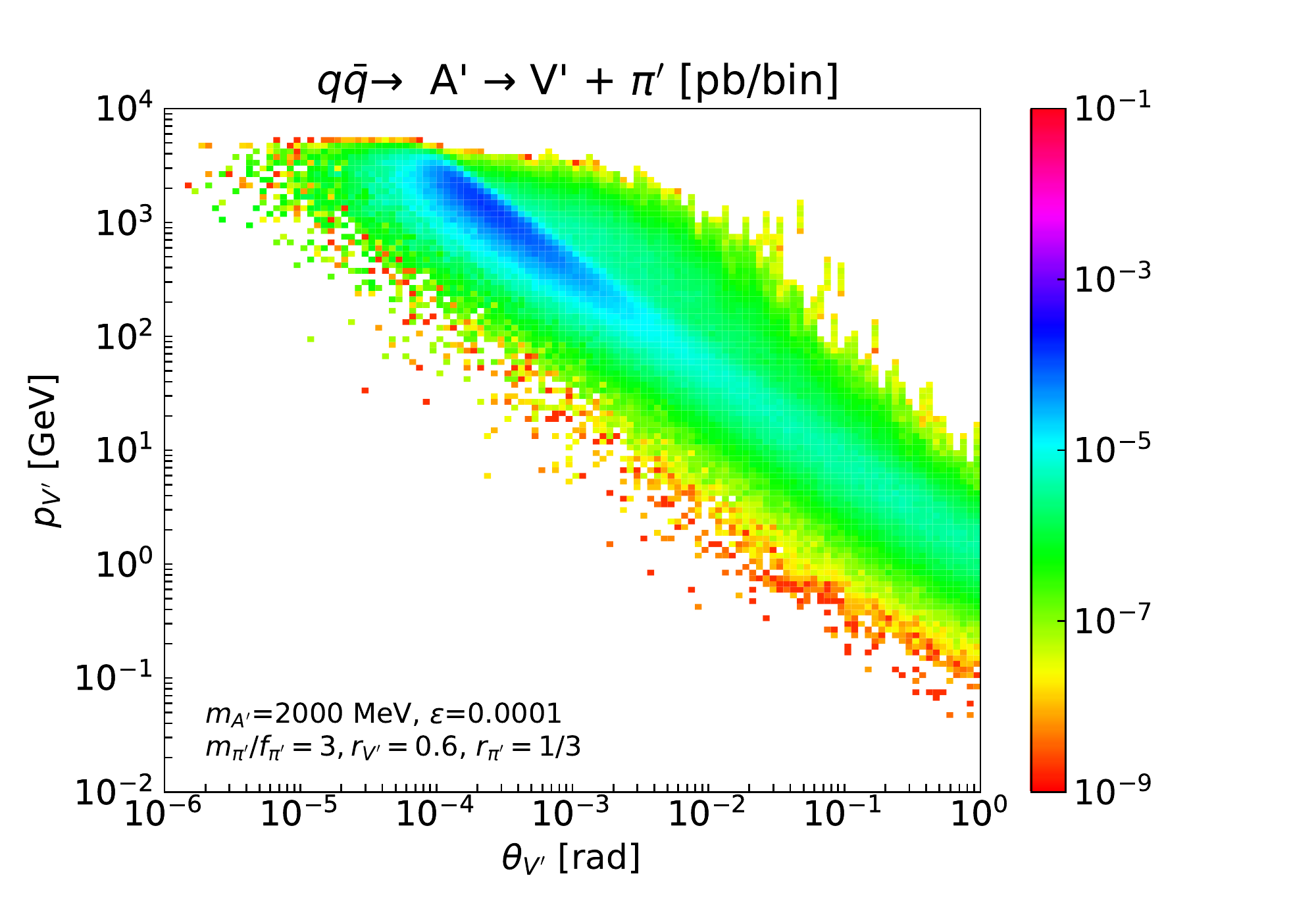}
  \end{minipage}
  \caption{
    Dark vector meson distribution from on-shell dark photon in the $(\theta_{V'},p_{V'})$ plane. 
    $\theta_{V'}$ is the angle with respect to the beam axis and $p_{V'}$ is the momentum for the dark vector meson. 
    Each panel corresponds to the kinetic distribution from the different production mechanisms: pion decay (left top), $\eta$-meson decay (right top), proton bremsstrahlung (left bottom), and Drell-Yan process (right bottom).
    }
\label{fig:spectrum_SIMP}
\end{figure}

We show the kinetic distribution of a dark vector meson in \cref{fig:spectrum_SIMP}.
The dark vector mesons produced with the angle of $\theta_{V'} \lesssim 2 \times 10^{-3}$ ($1.5 \times 10^{-3} \lesssim \theta_{V'} \lesssim 4 \times 10^{-3}$) would arrive in the detector area of FASER2 (FACET).
We also show the masses and the couplings that we use in order to compute each kinetic distribution in \cref{fig:spectrum_SIMP}. 
The kinetic distribution of the dark vector mesons depends on the production mechanisms. 
Each panel in \cref{fig:spectrum_SIMP} shows one of the four dominant production mechanisms: pion decay (left top), $\eta$-meson decay (right top), proton bremsstrahlung (left bottom), and the Drell-Yan production (right bottom).
For top panels, since decay products are lighter than these mesons, the distribution is maximized at $p_{V'} \simeq \Lambda_\mathrm{QCD}/\theta_{V'}$ with $\Lambda_\mathrm{QCD} \simeq 250\,\mathrm{MeV}$ being the QCD scale.
As for the production via the proton bremsstrahlung, kinematical conditions assumed in the FWW approximation are summarized as $E_\mathrm{beam} \,, E_{A'} \,, E_\mathrm{beam} - E_{A'} \gg m_p \,, m_{A'} \,, p_T$, which are required in order that the virtuality of the intermediate proton can be negligibly small~\cite{Kim:1972gw,Kim:1973he} (see also \cite{Blumlein:2013cua}).
Here, $E_\mathrm{beam}$ is the proton-beam energy, $(E_{A'},m_{A'})$ denotes the energy and the mass of the dark photon, and $p_T$ is the transverse momentum of produced dark photon.
We require the transverse momentum to be $p_T < 1\,\mathrm{GeV}$ following Ref.~\cite{Blumlein:2013cua}, which is shown as the distinct upper boundary in the left-bottom panel of \cref{fig:spectrum_SIMP}.%
\footnote{
  Since the beam energy $E_\mathrm{beam}$ is $7\,\mathrm{TeV}$ and the typical upper bound of $p_T$ is given by experimental setup as $p_T/E_p \lesssim \mathcal{O}(10^{-3})$ determined by detector radius and distance, we may choose $p_T$ as taken in Ref.~\cite{Feng:2017uoz}, $p_T < 10\,\mathrm{GeV}$.
  However, the intermediate proton has sizable virtuality for the large transverse momentum. 
  We took the value $p_T < 1\,\mathrm{GeV}$ instead of taking account of the off-shell form factor as discussed in \cite{Foroughi-Abari:2021zbm}.
}
Compared to the other three production mechanisms, the Drell-Yan production requires large $m_{A'}$, $m_{A'} \gtrsim 2\,\mathrm{GeV}$ due to the validity of the calculation with parton description, and hence the production cross section is suppressed by the mass. 
However, as for this mass range, the Drell-Yan production is the dominant channel to produce dark photons.

\begin{figure}[t]
\begin{minipage}[t]{0.5\linewidth}
    \centering
    \includegraphics[width=1\textwidth]{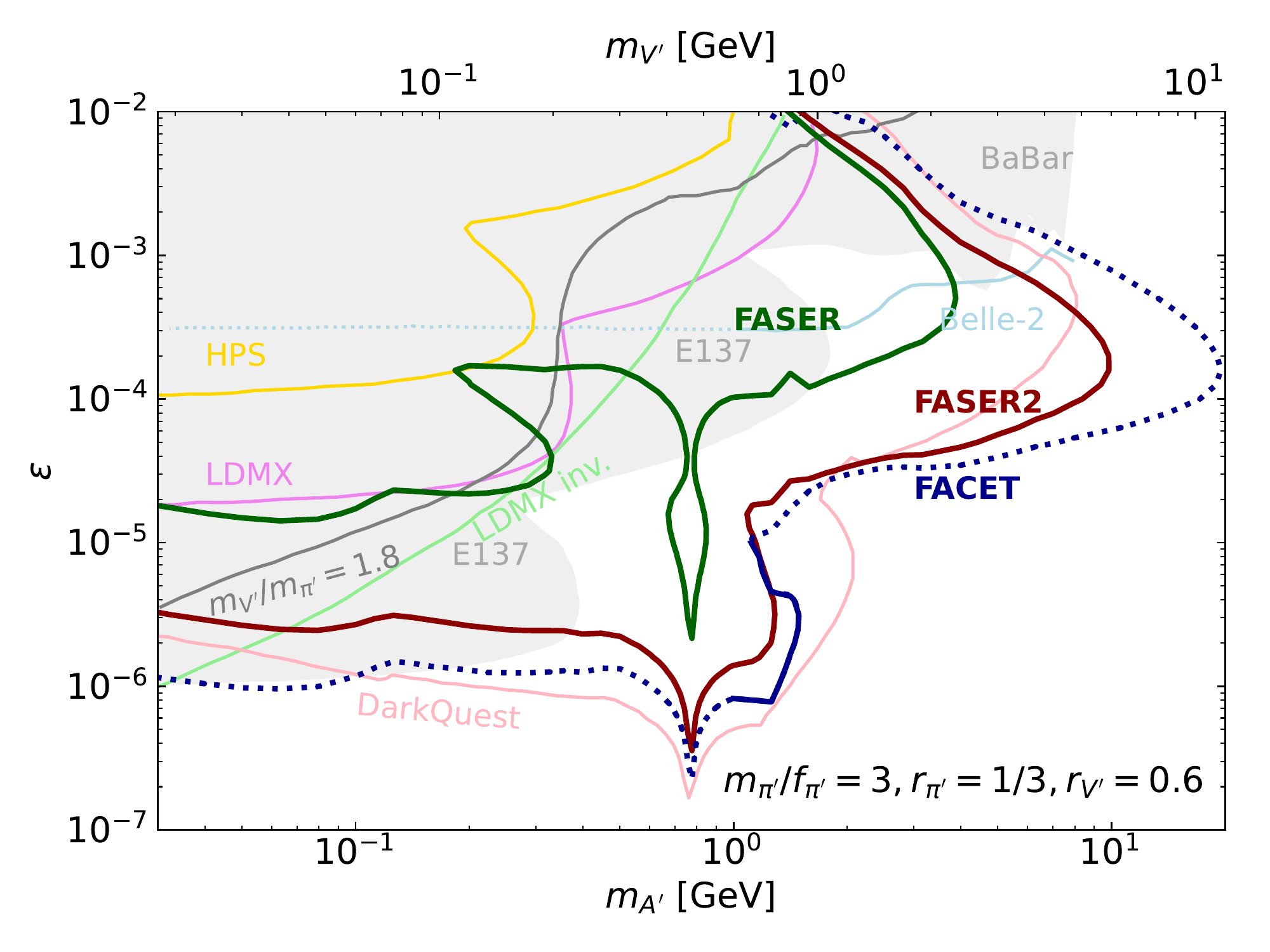}
\end{minipage}
\begin{minipage}[t]{0.5\linewidth}
    \centering
    \includegraphics[width=1\textwidth]{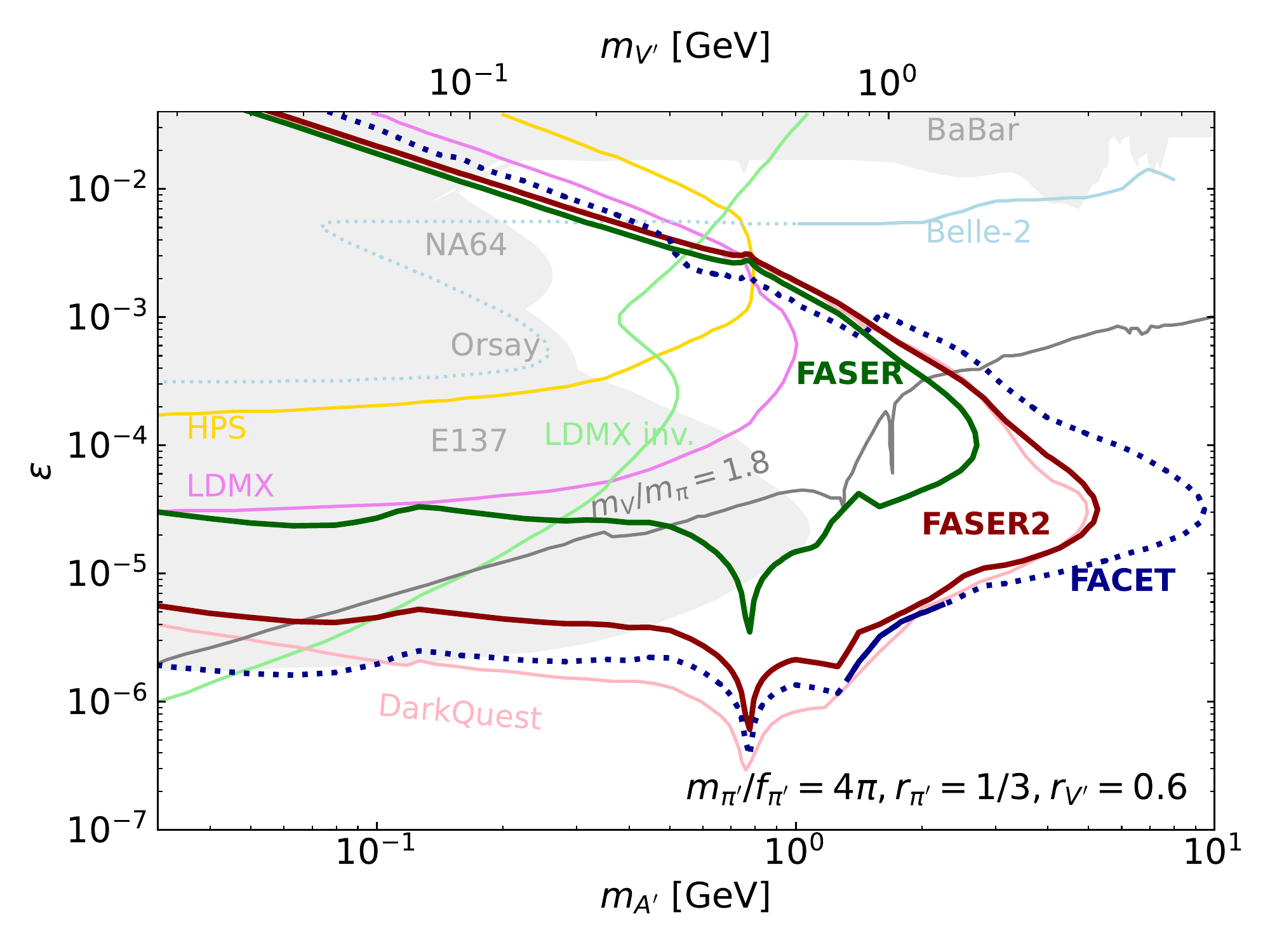}
\end{minipage}
\caption{
  Existing constraints and sensitivity of future searches to signals of dark vector mesons. 
  The shaded regions are excluded by mono-photon search at BaBar~\cite{BaBar:2017tiz}; invisible decay search at NA64~\cite{Banerjee:2019pds}; and visible decay signals at E137~\cite{Bjorken:1988as,Bjorken:2009mm}, Orsay~\cite{Davier:1989wz,Andreas:2012mt}. 
  The light-color contours correspond to the projected reach of Belle-II~\cite{Belle-II:2018jsg}, HPS~\cite{Celentano:2014wya}, DarkQuest~\cite{SeaQuest:2017kjt}, LDMX~\cite{Izaguirre:2014bca}.
  The future sensitivity of dark vector mesons at the LHC forward-detector searches during HL-LHC are shown as thick lines: FASER/FASER2~\cite{Feng:2017uoz} (dark green/dark red) and FACET~\cite{Cerci:2021nlb} (dark blue). 
  We use different line-type for the FACET sensitivity from the background consideration.
  The parameters are fixed to be $\alpha_{D}=0.01$, $r_{\pi'} = m_{\pi'}/m_{A'}=1/3$, and $r_{V'} = m_{V'}/m_{A'}=3/5$. 
  We take the pion self-coupling to be $m_{\pi'}/f_{\pi'} = 3 \, (4\pi)$ in the left (right) panel. 
  The correct relic abundance is obtained on the black-solid contour for the mass ratio $m_{V'}/m_{\pi'}=1.8$.
}
\label{fig:simp_ons_detect}
\end{figure}

\cref{fig:simp_ons_detect} shows the future sensitivities of the dark vector meson searches for the dark-pion DM scenarios at the forward-detecter experiments, FASER (dark green) during the LHC Run-3, FASER2 (dark red) and FACET (dark blue) during the HL-LHC.
We also place existing constraints and future prospects of fixed-target experiments in these figures. 
Since the dark photon dominantly decays into the dark-sector particles, we have invisible signals once dark photons are produced.
The invisible signals have been explored (gray-shaded region) and will be further explored (light-colored lines) by various experiments: mono-photon searches at the $e^+e^-$ collider experiments, BaBar~\cite{BaBar:2017tiz} and Belle-II~\cite{Belle-II:2018jsg}, and the missing-energy searches at the fixed-target experiments, NA64~\cite{Banerjee:2019pds} and LDMX~\cite{Izaguirre:2014bca}.
These are taken from Ref.~\cite{Berlin:2018tvf}, but we update the existing constraints from the invisible signal searches by including the latest result by NA64, which is discussed in detail in \cref{app:invisibleNA64}.
In the figure, we also include the visible signal searches at the fixed-target experiments: existing constraints from E137~\cite{Bjorken:1988as,Bjorken:2009mm} and Orsay~\cite{Davier:1989wz,Andreas:2012mt}, and future prospects at HPS~\cite{Celentano:2014wya}, DarkQuest~\cite{SeaQuest:2017kjt}, and LDMX~\cite{Berlin:2018tvf}.
For sufficiently large $\epsilon$, the produced dark pions can scatter with the SM particles, which is constrained by searches for the recoil energy from the DM-SM scattering at LSND~\cite{deNiverville:2011it}, E137~\cite{Batell:2014mga}, and MiniBooNE~\cite{Aguilar-Arevalo:2017mqx}. 
However, the latest result of NA64 for the invisible signals has already covered the excluded region by the dark-pion scattering.
On the black-solid line, we obtain the correct relic abundance of dark pions; in other words, the dark pions are overabundant below the black-solid line.

The dark-neutral pions can decay into SM particles for odd-$N_f$ flavors of light dark quarks through the higher-order terms of chiral Lagrangian as shown in \cref{eq:piondecay_operators}.
The late-time decay of dark pions via the operators leads to the energy injection to the electromagnetic channels, as we mentioned.
The constraints by BBN or CMB anisotropy strongly depend on the lifetime and the fraction of decaying dark pions~\cite{Poulin:2016anj}.
These constraints would appear in the bottom-right part of the figures. 
However, a dedicated study would be necessary for determining the fraction, which depends on the dark-pion mass difference arising from the isospin symmetry violation, such as quark masses and $U(1)_D$ interactions. 
A dedicated study has been performed in Ref.~\cite{Katz:2020ywn} in the absence of the dark vector mesons.

Let us discuss the sensitivity of the LHC forward detectors, FASER/FASER2 and FACET, which is comparable to that of DarkQuest~\cite{SeaQuest:2017kjt}.
The sensitivity contours consist of two parts: one comes from visible signals of dark-charged vector mesons, $V' \to \pi' \ell^+ \ell^-$, (located at the top part of the contours) while another comes from the signals of dark-neutral vector mesons, $V' \to \ell^+ \ell^-$ (located at the bottom part of the contours).
The two-body proper decay length increases, and the three-body proper decay length decreases as $m_{\pi'}/f_{\pi'}$ gets larger, as shown in \cref{eq:decay_of_V}.
Therefore, these parts overlap with each other as for $m_{\pi'}/f_{\pi'} \simeq 4 \pi$.

We assume that there is no background event, and three events are required for the sensitivity lines. 
However, as we have discussed in the previous section, there remain enormous background events from the neutral SM hadrons at the FACET: such as the neutral hadron decay to charged hadrons and to a lepton-pair.
These background events will be negligible for the long-lived particles with a mass above $0.8\,\mathrm{GeV}$.
The mass of the parent particle corresponds to the invariant mass of the final-state charged particles in the two-body decay but not in the three-body decay since the dark pion carries a portion of energy. 
In our simulation, we take into account the angle acceptance in terms of the angle $\theta_{V'}$ of the dark vector mesons, not the kinetic information of the final-state particles.
The assumption of a background-free environment would be available only for the visible signals from the dark-neutral vector mesons with a mass above $0.8\,\mathrm{GeV}$.
For more solid sensitivities of the FACET experiment, they require further study of the background estimate at low mass and the detailed information of the final-state particles for the three-body decay.
In the following figures, we use dotted lines for the three-body decay regions and the regions where no-background assumption is not good enough.

A similar sharp dip appears around $m_{A'} \simeq 0.77\,\mathrm{GeV}$ in FASER/FASER2 and FACET sensitivity curves, which is from the resonant production of the dark photon via vector meson dominance of the form factor of the proton.
In the short lifetime limit, in other words $d = c \tau \beta \gamma \ll L_\mathrm{min}$, the probability of a dark vector meson to decay inside the detector takes the approximate form,
\eqs{
  \mathcal{P}_\mathrm{dec} \simeq e^{- L_\mathrm{min}/d} A_\mathrm{geo} \,.
  \label{eq:prob_topline}
}
The upper boundary of the sensitivity is mostly determined by the lifetime, the beam energy, and the distance from the interaction point.
The FACET detector will be located closer to the interaction point than the FASER2 detector, and hence FACET is more sensitive to a shorter lifetime than FASER2.

Meanwhile, the fiducial decay region along the beam axis of FACET will be longer than that of FASER2, and the minimum energy deposit required for the detection at FACET is lower compared to that at FASER2.
In the long lifetime limit, $L_\mathrm{max} \ll d$, the probability of a dark vector meson to decay inside the detector takes the form 
\eqs{
  \mathcal{P}_\mathrm{dec} \simeq \frac{\Delta L}{c \tau \beta \gamma} A_\mathrm{geo} \,.
  \label{eq:prob_bottomline}
}
Here, $\Delta L$ defines the fiducial decay region along the beam axis, and the boost factor $\beta \gamma$ is given by $\beta \gamma \simeq p_{V'}/m_{V'}$.
The lower boundary of the sensitivity is determined by the lifetime, the typical energy of the long-lived particles, and the length of the detector volume.
In this study, we assume that the minimum energy deposit at FACET is lower than that at FASER. 
It allows us to have the low boost factor at FACET, which implies that we can explore a longer lifetime of the dark vector mesons.
However, as discussed in the previous section, the minimum energy deposit will be determined by the background consideration of these experiments.
For sufficient background reduction, it may require a larger minimum energy deposit even at FACET, and FACET is no longer sensitive to the longer lifetime for such a case.

We note that there is a dip on the upper boundary of the FACET contour in the range of $500\,\mathrm{MeV} \lesssim m_{A'} \lesssim 1.5\,\mathrm{GeV}$.
The proton bremsstrahlung process dominantly contributes to the $V'$ production in this range. 
The FACET detector has a tube-like shape, hence it cannot detect the particles that are very close to the beam axis.
This detector design restricts the angle to the range of $1.5 \times 10^{-3} \lesssim \theta \lesssim 4 \times 10^{-3}$.
Combining the detector design and the $p_T$ cut ($p_T \leq 1\,\mathrm{GeV}$) for the proton bremsstrahlung, we impose the maximum momentum as $p_{V'} \lesssim 1\,\mathrm{GeV}/\theta_\mathrm{min} \simeq 600\mathrm{GeV}$ at FACET.
On the other hand, we find $V'$ with the energy of beam particles at FASER2 since the detector has a cylindrical shape and there is no restriction on the small angle.
This maximum momentum is an order of magnitude smaller than that of $V'$ produced via other processes.
Hence, FACET loses its sensitivity to the short lifetime in the mass range where the proton bremsstrahlung dominates the production mechanism.

\begin{figure}[t]
\begin{minipage}[t]{0.5\linewidth}
    \centering
    \includegraphics[width=1\textwidth]{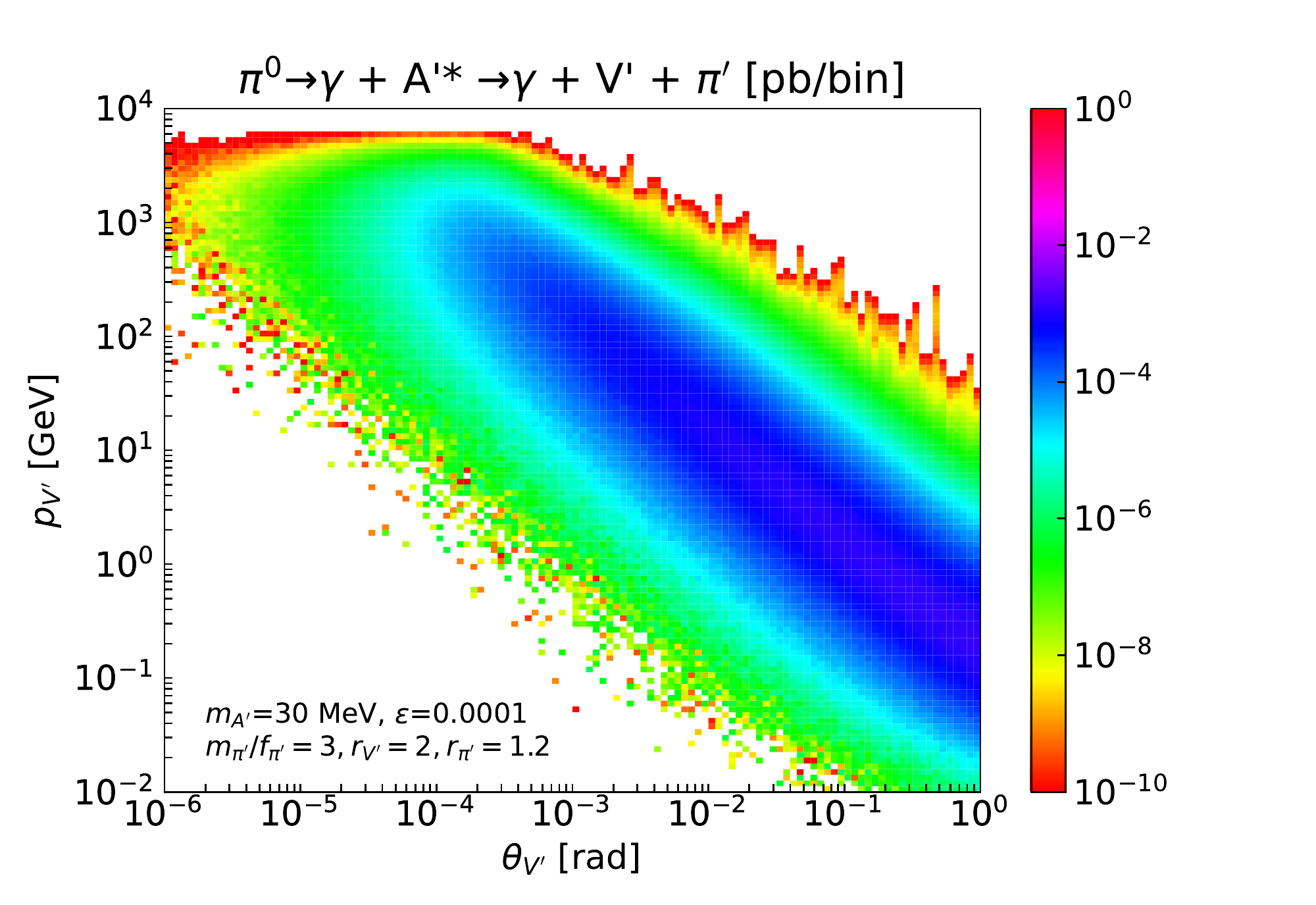}
\end{minipage}
\begin{minipage}[t]{0.5\linewidth}
    \centering
    \includegraphics[width=1\textwidth]{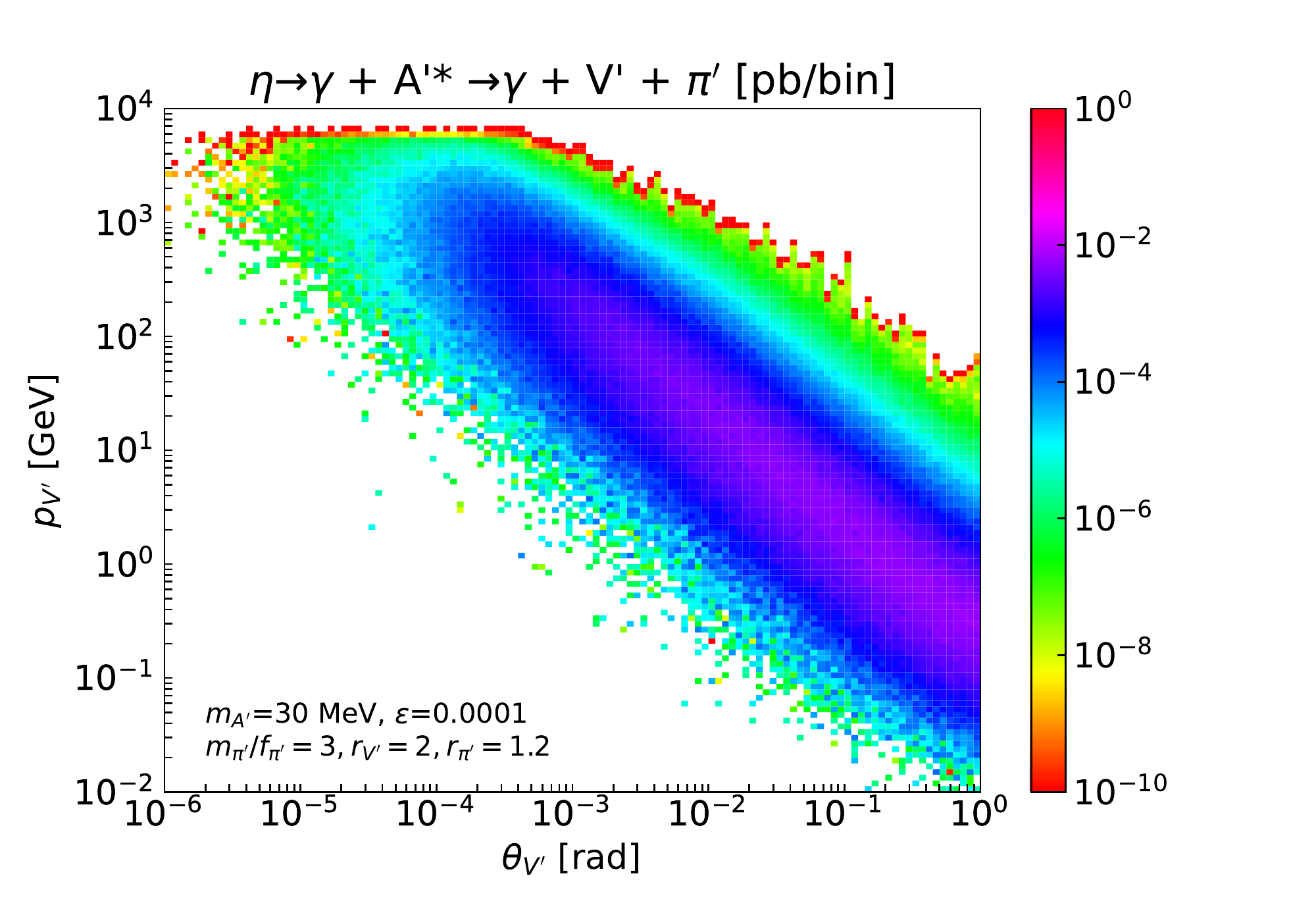}
\end{minipage}

\begin{minipage}[t]{0.5\linewidth}
    \centering
    \includegraphics[width=1\textwidth]{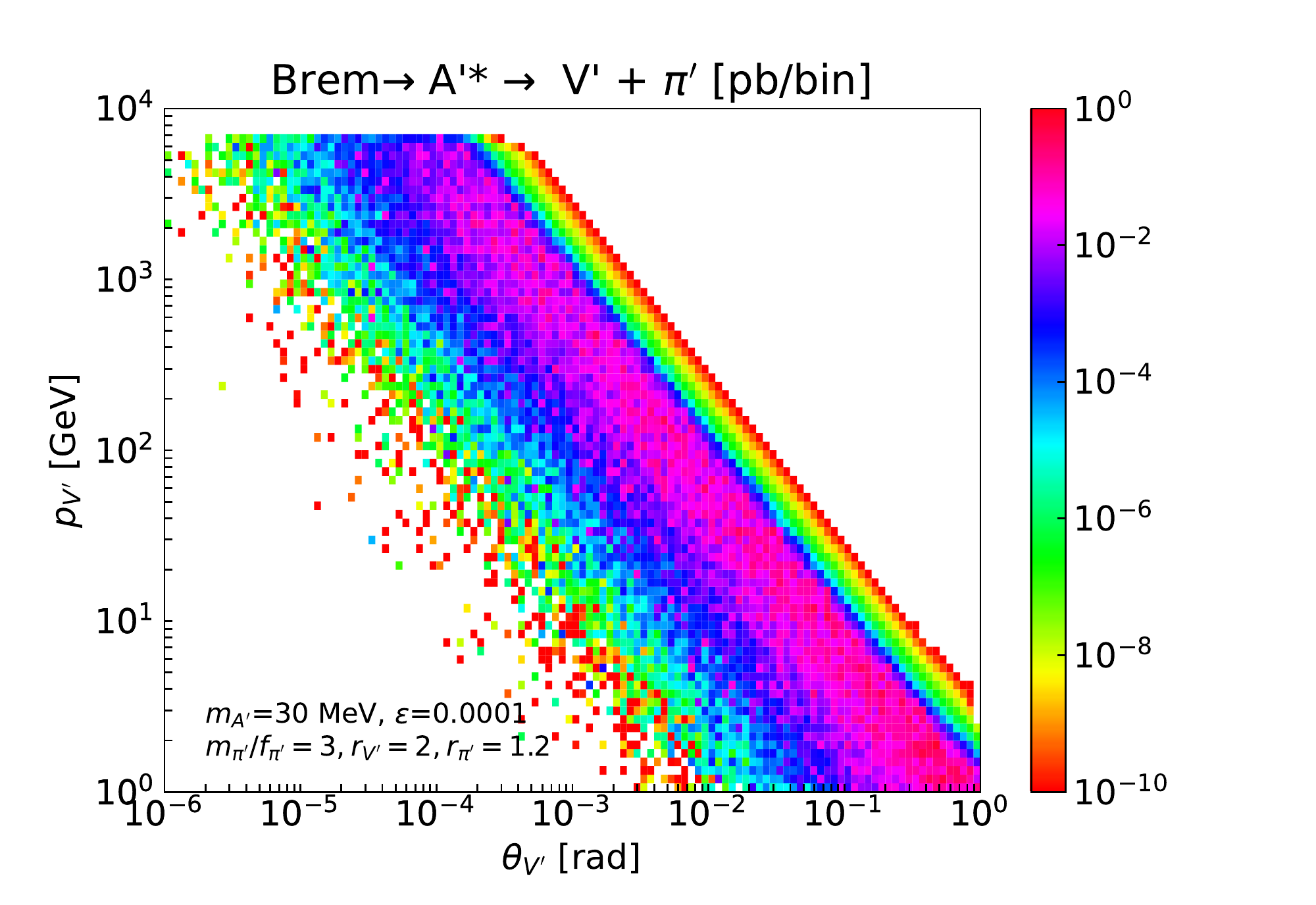}
\end{minipage}
\begin{minipage}[t]{0.5\linewidth}
    \centering
    \includegraphics[width=1\textwidth]{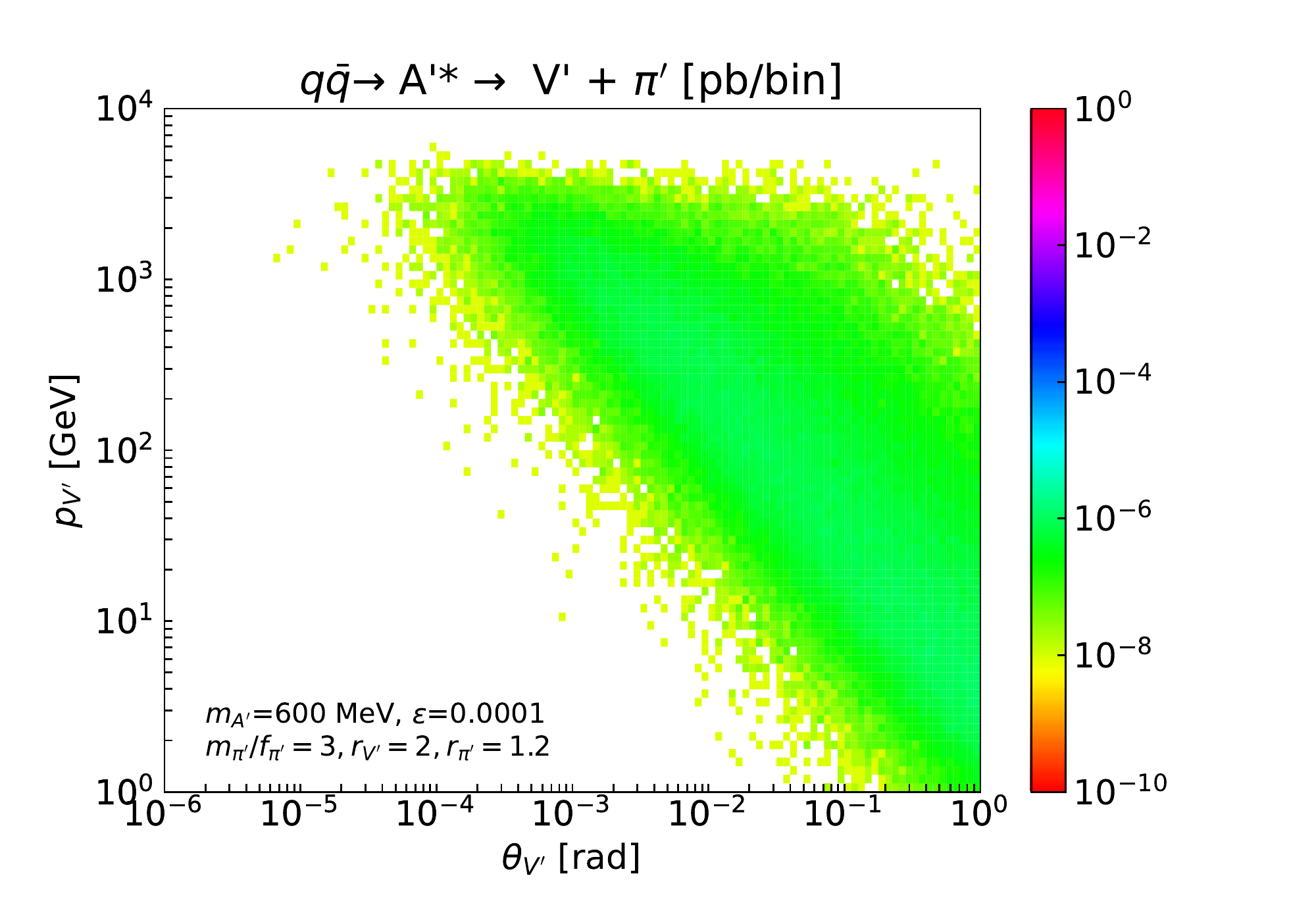}
\end{minipage}
\caption{
  Same as \cref{fig:spectrum_SIMP}, but for dark vector meson distribution from off-shell dark photon. 
  }
\label{fig:spectrum_ADM_set2}
\end{figure}

Next, we discuss the dark vector mesons produced via off-shell dark photon in the case that the production is governed by couplings in the effective theory, given by \cref{eq:ApiVcoupling}. 
The kinetic distributions are shown in \cref{fig:spectrum_ADM_set2}.
We take the same value of $m_{V'}$ as corresponding panels in \cref{fig:spectrum_SIMP}, and we list the masses and the couplings that we use in order to compute the kinetic distribution. 
The production cross section of the dark vector mesons is suppressed by a factor of $\alpha'/\pi$ due to the off-shell production. 
However, we find that the production cross section via proton bremsstrahlung can be larger than that via on-shell dark photon (see the left-bottom panel in \cref{fig:spectrum_SIMP}). 
As for the proton bremsstrahlung, since the production cross section is obtained by integrating over the fictitious mass $m_{A'}^\ast$, the dark vector mesons are resonantly produced at $m_{A'}^\ast \simeq m_\rho$ as far as the kinematic threshold is below $m_\rho$.
We note that the upper boundary is determined by the $p_T$ cut of the off-shell dark photon, and the distribution gradually decreases above $p_{V'} \gtrsim 1\,\mathrm{GeV}/\theta_{V'}$.

\begin{figure}
\begin{minipage}[t]{0.5\linewidth}
    \centering
    \includegraphics[width=1\textwidth]{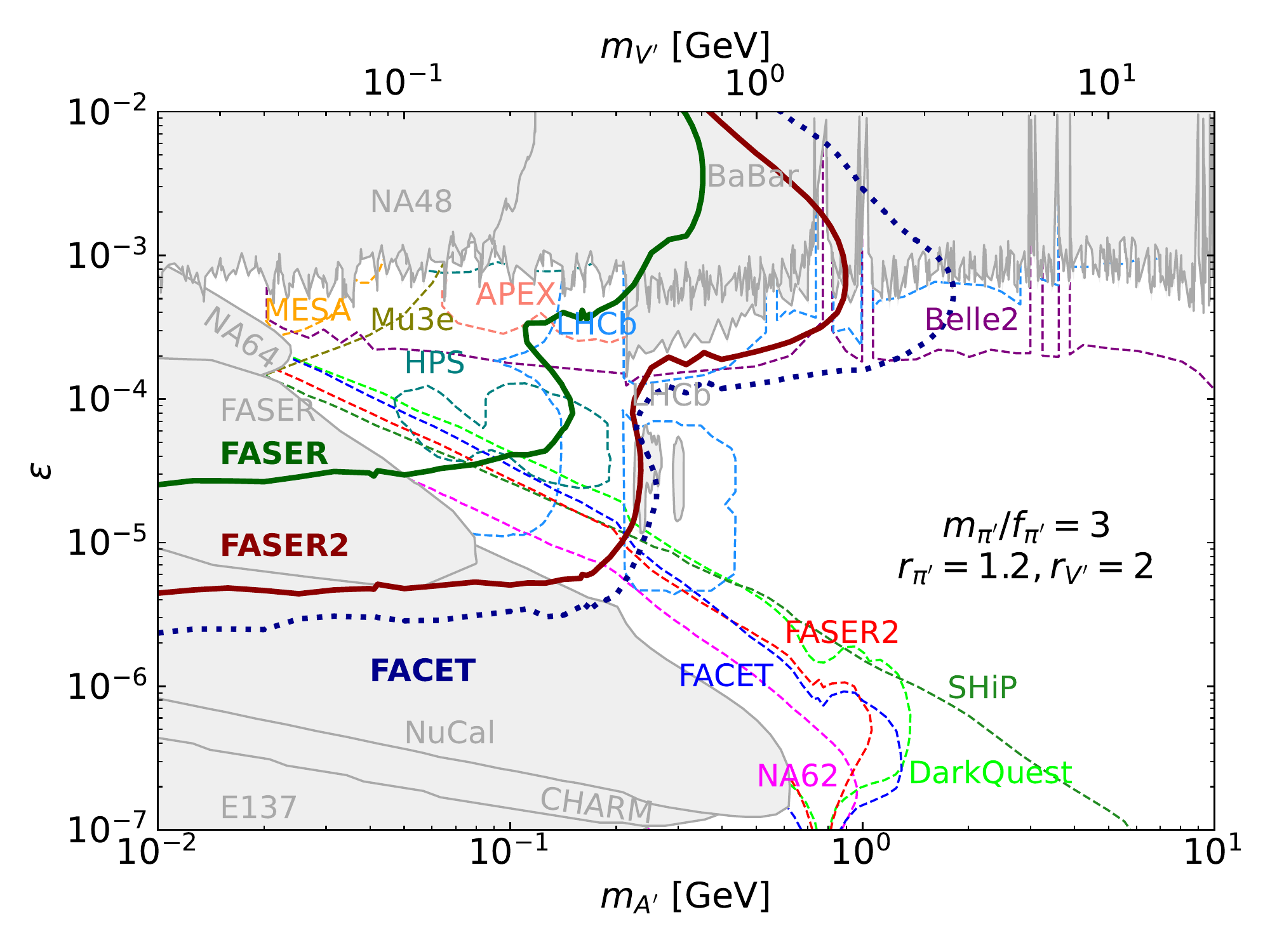}
\end{minipage}
\begin{minipage}[t]{0.5\linewidth}
    \centering
    \includegraphics[width=1\textwidth]{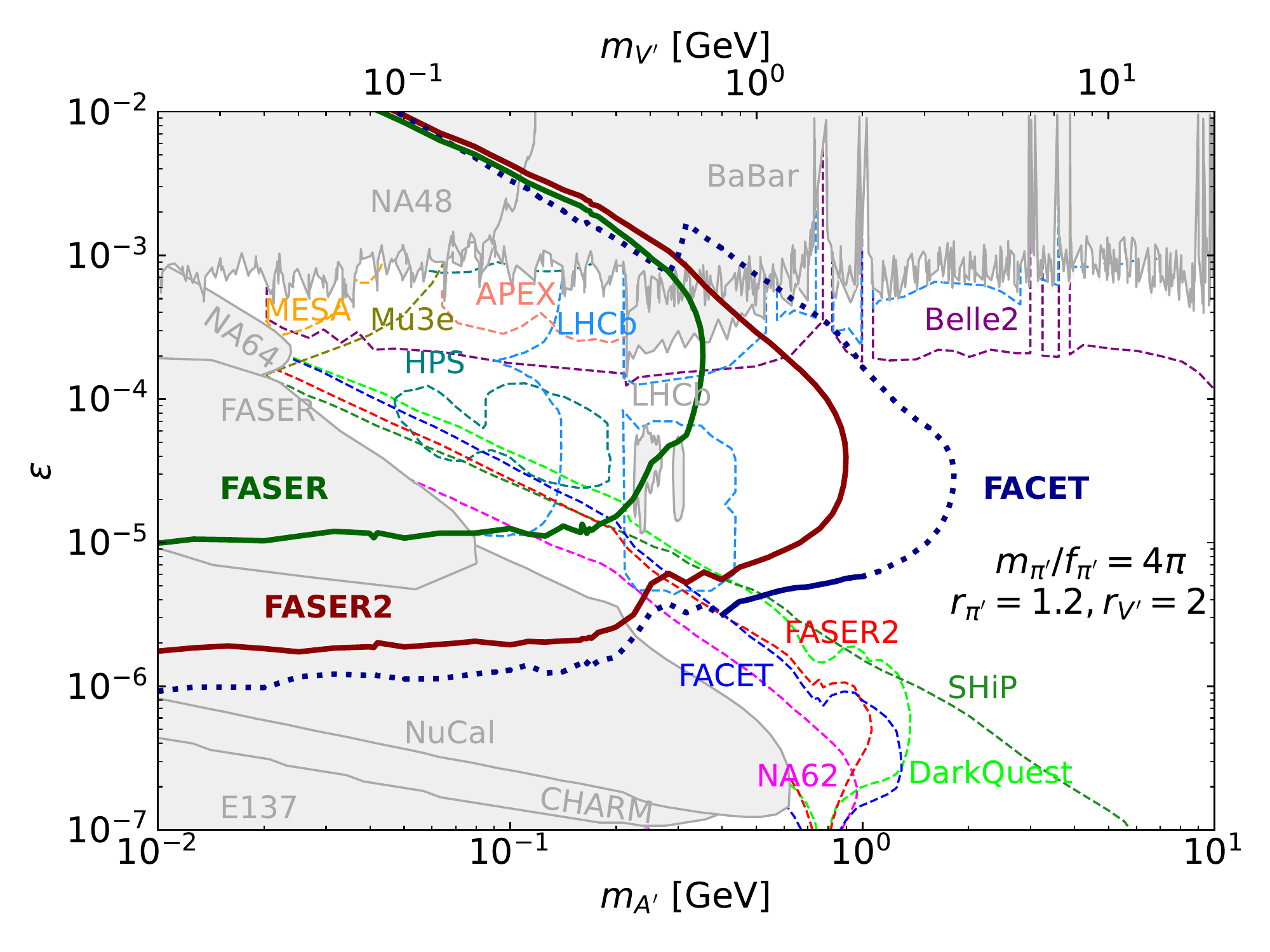}
\end{minipage}
\\
\begin{minipage}[t]{0.5\linewidth}
    \centering
    \includegraphics[width=1\textwidth]{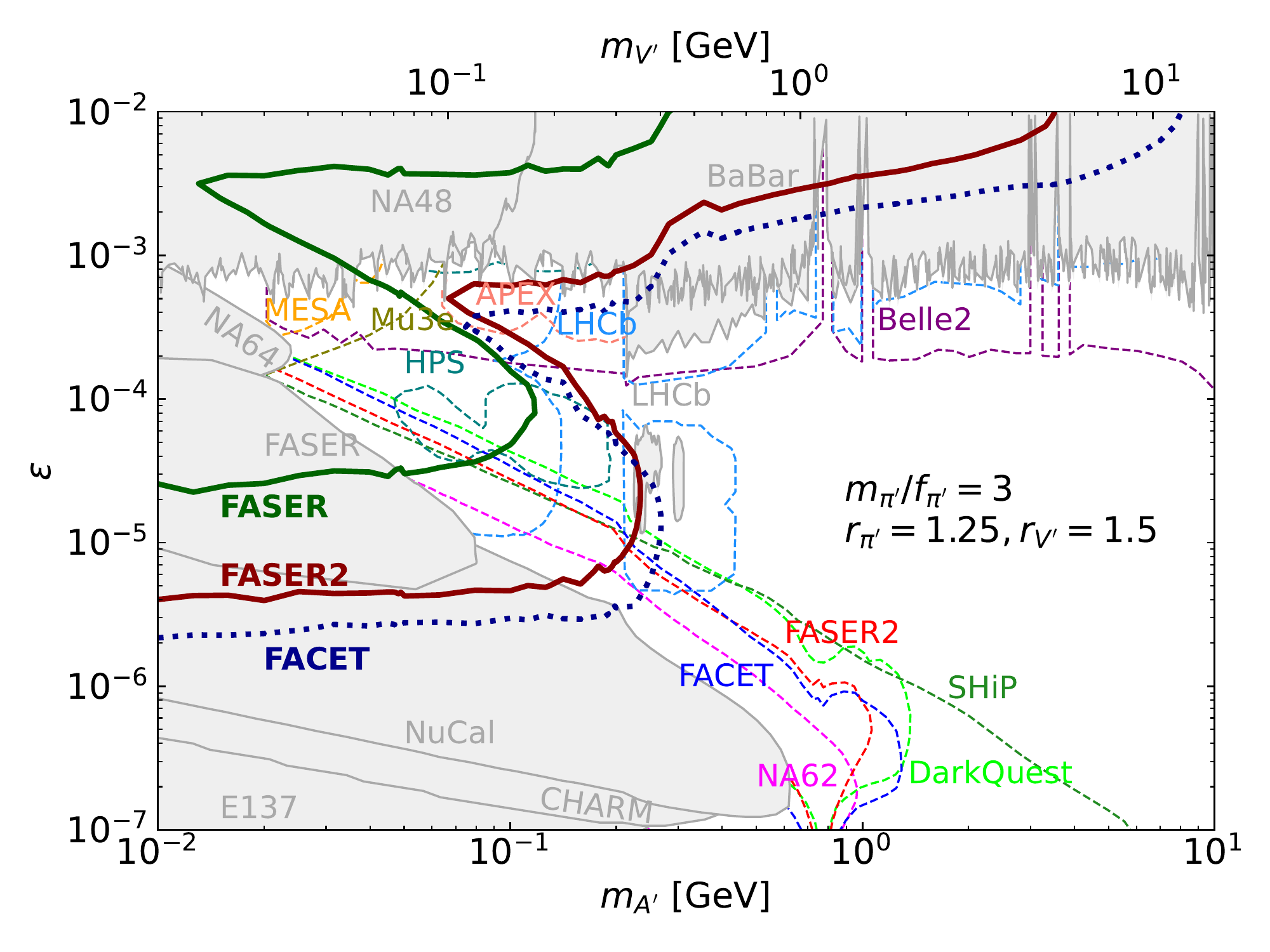}
\end{minipage} 
\begin{minipage}[t]{0.5\linewidth}
    \centering
    \includegraphics[width=1\textwidth]{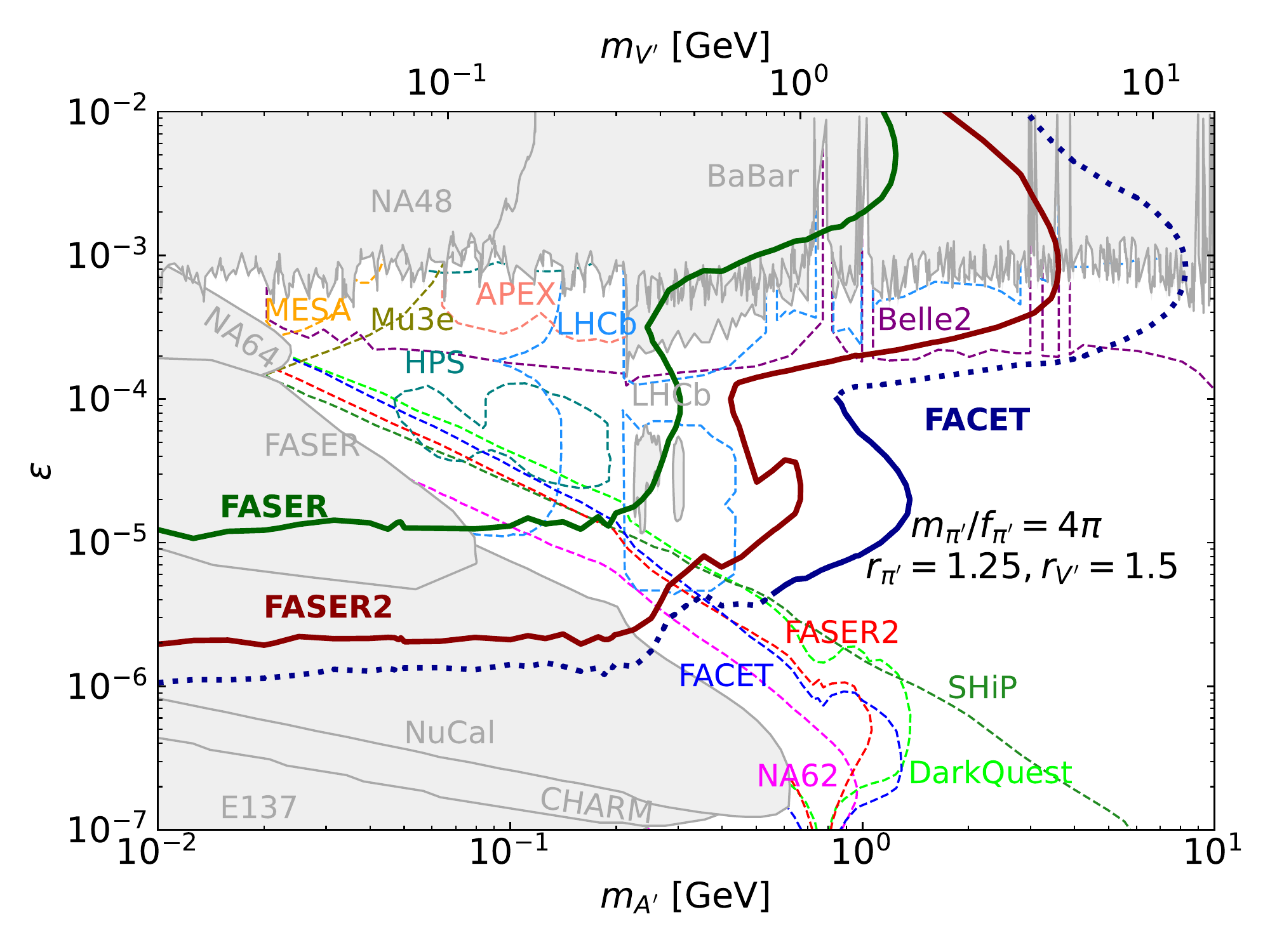}
\end{minipage}
\caption{
  Sensitivities of FASER/FASER2 and FACET to dark vector meson produced through off-shell dark photon. 
  We choose the mass spectrum to be $r_{\pi'} = m_{\pi'}/m_{A'}=1.2$, $r_{V'} = m_{V'}/m_{A'}=2$ in the top panels, while we choose it to be $r_{\pi'} = 1.25$, $r_{V'} = 1.5$ in the bottom panels. 
  We take $m_{\pi'}/f_{\pi'} = 3 \, (4 \pi)$ in the left (right) panels.
  The dark photon visible decay searches are also depicted: existing constraints (gray shaded region) and future sensitivities (thin-dashed lines).
  The existing constraints are given by
  electron beam-dump experiments, E137~\cite{Bjorken:1988as}, NA64~\cite{NA64:2018lsq}, and E141~\cite{Riordan:1987aw}; 
  proton beam-dump experiments, CHARM~\cite{CHARM:1985nku,Gninenko:2012eq}, NA48/2~\cite{NA482:2015wmo}, and U70/$\nu$Cal~\cite{Blumlein:2011mv,Blumlein:2013cua}; 
  and collider experiments, BaBar~\cite{BaBar:2014zli} and LHCb~\cite{LHCb:2019vmc}.
  We also include the latest result from the dark photon search at the FASER~\cite{MoriondFASER:2023,Petersen:2023hgm}.
  The future prospects include 
  electron (muon) beam-dump experiments, MESA~\cite{Doria:2018sfx,Doria:2019sux}, Mu3e~\cite{Echenard:2014lma}, APEX~\cite{APEX:2011dww}, and HPS~\cite{Celentano:2014wya}; 
  proton beam-dump experiments, SHiP~\cite{SHiP:2015vad,Alekhin:2015byh} and DarkQuest~\cite{Berlin:2018pwi,Apyan:2022tsd};
  and collider experiments, Belle-II~\cite{Belle-II:2018jsg} and LHCb~\cite{Ilten:2015hya,Ilten:2016tkc,LHCb:2017trq}.
  }
\label{fig:adm_ofs_set2_detect}
\end{figure}

\cref{fig:adm_ofs_set2_detect} shows the sensitivity reach on dark vector mesons produced via off-shell dark photon at FASER (dark green), FASER2 (dark red) and FACET (dark blue).
We also depict the visible decay searches of dark photons since the dark photons are the lightest particle in the dark sector and decay into the SM particles: future sensitivities are shown as the thin-dashed lines, and the existing constraints are shown as the shaded regions.%
\footnote{
  We include the latest result from the visible decay search for the dark photon at the FASER~\cite{MoriondFASER:2023,Petersen:2023hgm}, and it partly covers the existing constraints from the beam-dump experiments, such as E141 and NA64. 
}
Similar to the on-shell production, sensitivity reaches at the LHC forward detector experiments are composed of two components: two-body decay of dark neutral vector mesons and three-body decay of dark-charged vector mesons.
The lower boundary of $V'$ sensitivity for $m_{A'} \lesssim \mathcal{O}(100)\,\mathrm{MeV}$ is flat compared to the case with the production via on-shell dark photon.
On the lower boundary, the signal number is about 3, which is determined by the product of $V'$ number produced at the collider and the probability to decay inside the decay volume. 
In the mass range, $V'$ production cross section is dominated by the proton bremsstrahlung on the $\rho$-meson resonance, $m_{A'}^\ast \simeq m_\rho$, since we integrate over the fictitious mass $m_{A'}^\ast$ above the kinematic threshold.
For fixed $m_{\pi'}/f_{\pi'}$, $r_{\pi'}$, and $r_{V'}$, the production cross section scales as $m_{A'}^{-2}$ (see \cref{app:production}).
This scaling is cancelled with $m_{A'}$ dependence of the probability $\mathcal{P}_\mathrm{dec} \propto m_{A'}^2$ [see \cref{eq:prob_bottomline}].

Within the sensitivity reach, the dark vector mesons have the proper decay length of $\mathcal{O}(1)\,\mathrm{[m]}$.
The dark photon decays into the SM particles with the proper decay length of $\mathcal{O}(1)\,\mathrm{[mm]}$ or below with the very same dark-photon parameters $(m_{A'},\epsilon)$.
The future sensitivity of the LHC forward searches to the dark vector meson is compatible with that of some visible decay searches of dark photons: in particular, the prompt decay searches at Belle-II and LHCb, the displaced vertex searches at LHCb, and the long-lived particle searches at the fixed-target experiments.

We choose different mass spectrum: $r_{\pi'} = m_{\pi'}/m_{A'} =1.2$, $r_{V'} = m_{V'}/m_{A'} =2$ in the top panels, and $r_{\pi'} = 1.25$, $r_{V'} = 1.5$ in the bottom panels. 
The mass difference between $V'$ and $\pi'$ in the former case is larger than that in the latter case.
The LHC forward searches are more sensitive to large kinetic mixing in the range of $10^{-4} \lesssim \epsilon \lesssim 10^{-2}$ in the bottom panels compared to the top panels since the dark-charged vector mesons tend to be long-lived due to the small mass difference.

\begin{figure}[t]
\begin{minipage}[t]{0.5\linewidth}
    \centering
    \includegraphics[width=1\textwidth]{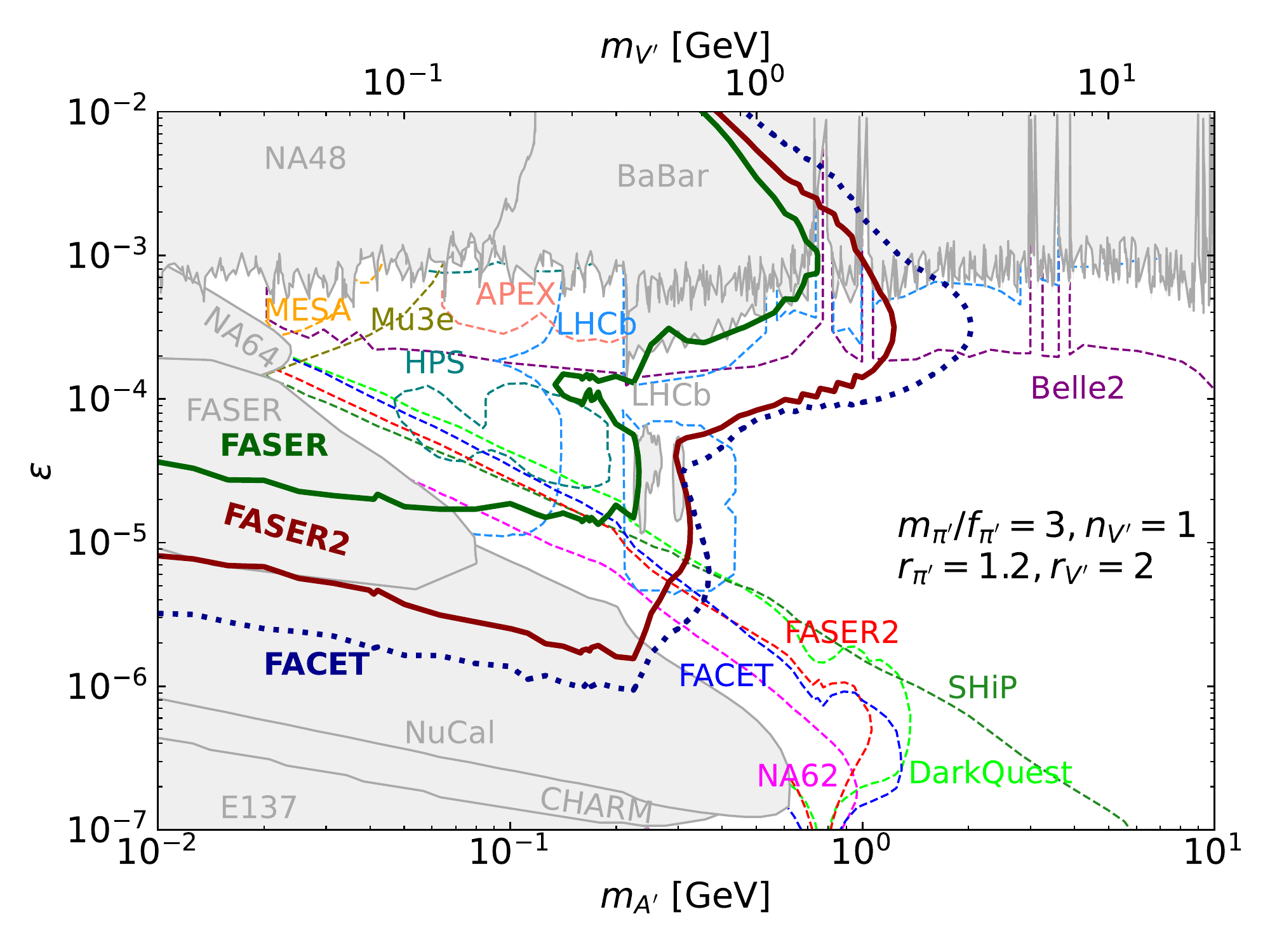}
\end{minipage}
\begin{minipage}[t]{0.5\linewidth}
    \centering
    \includegraphics[width=1\textwidth]{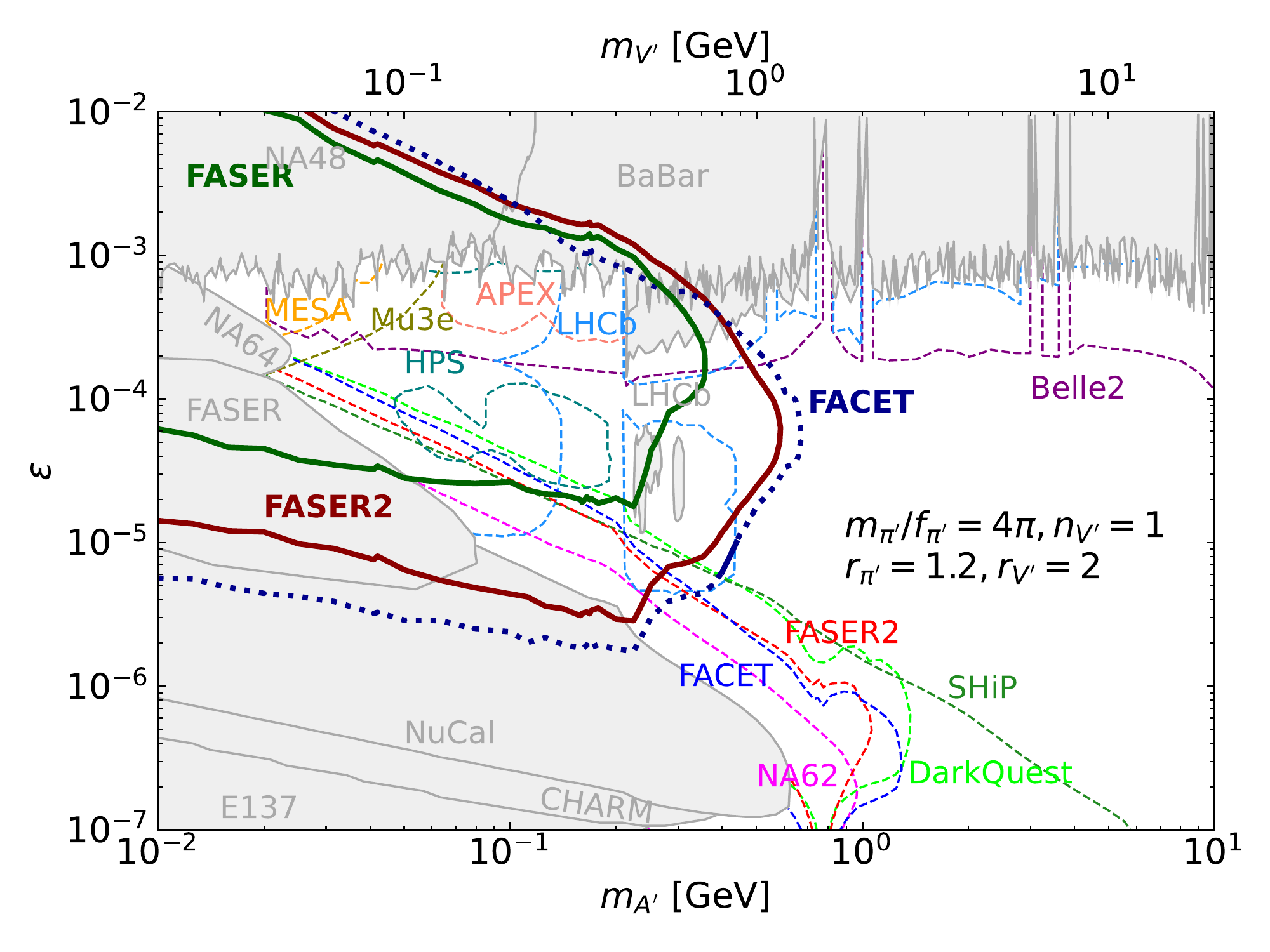}
\end{minipage}

\begin{minipage}[t]{0.5\linewidth}
    \centering
    \includegraphics[width=1\textwidth]{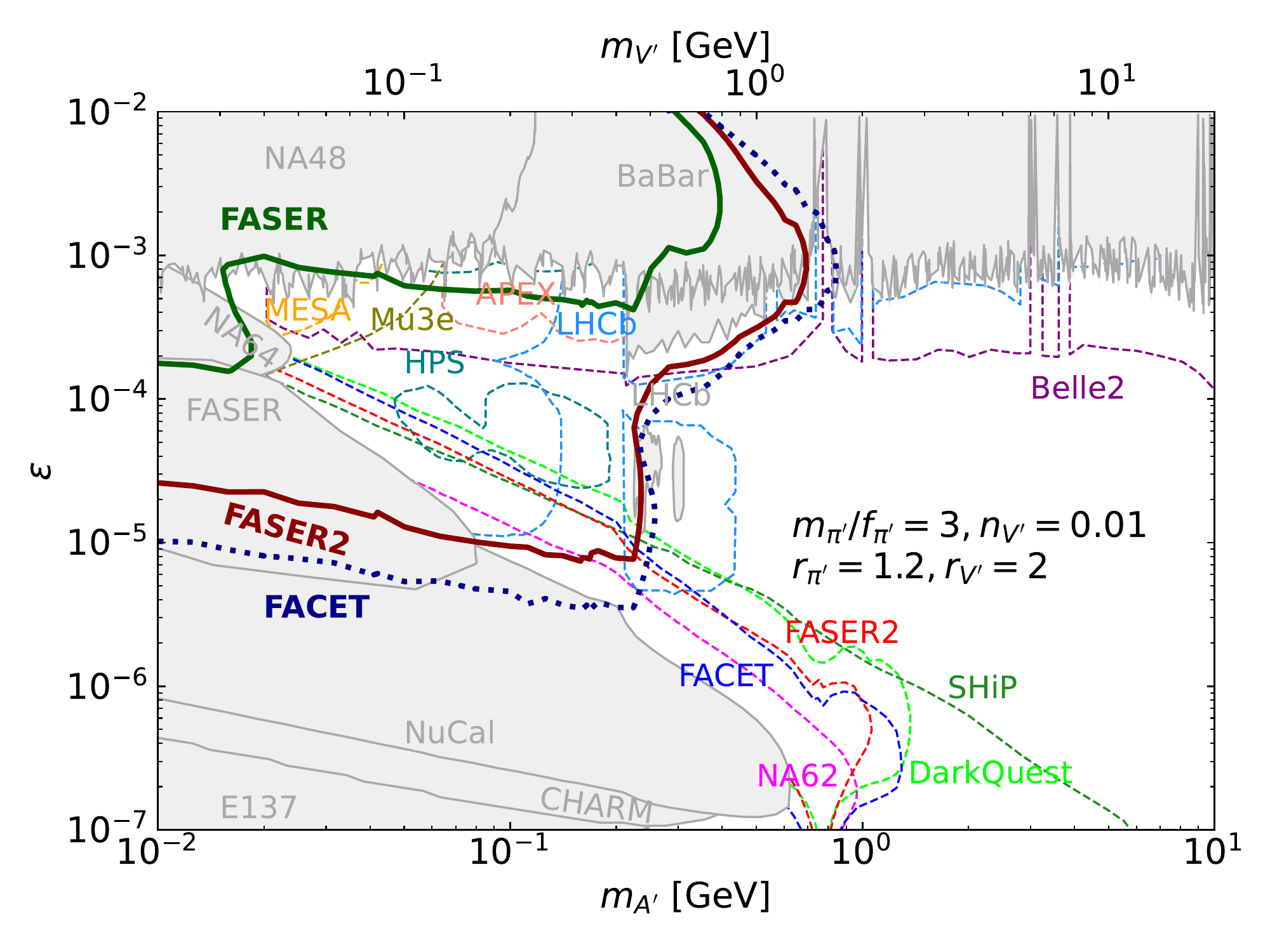}
\end{minipage}
\begin{minipage}[t]{0.5\linewidth}
    \centering
    \includegraphics[width=1\textwidth]{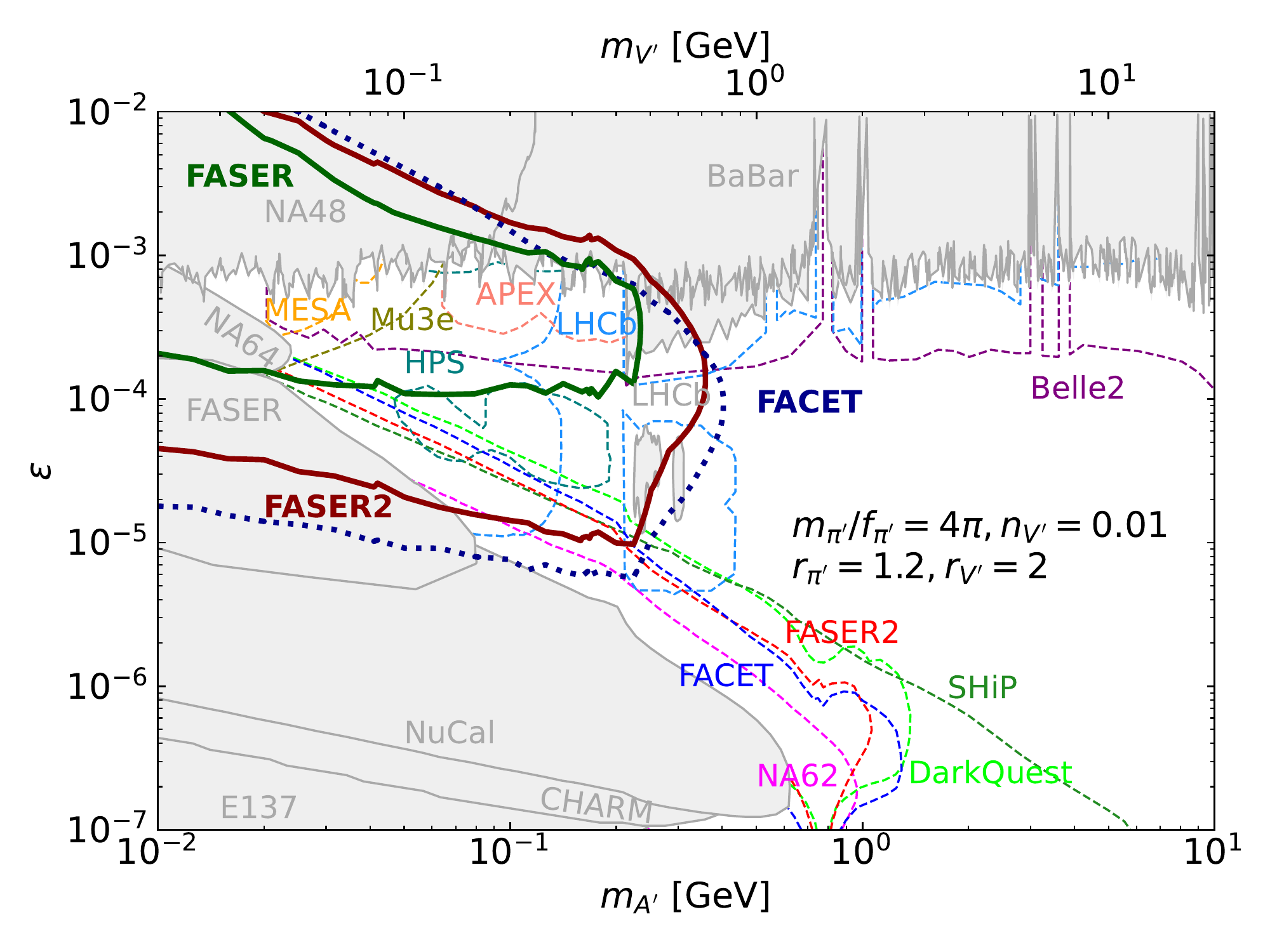}
\end{minipage}
\caption{
  Existing constraints on dark photon and sensitivity of FASER/FASER2 and FACET to dark vector meson produced through off-shell dark photon and naive hadronizaiton. 
  The result is evaluated with parameters as $r_\pi = m_{\pi}/m_{A'}=1.2$, $r_V = m_{V}/m_{A'}=2$, and $m_{\pi}/f_{\pi}=3(4\pi)$.
  }
\label{fig:adm_ofs_hadron}
\end{figure}

The dark hadrons might be produced via dark hadronization since the dark vector meson mass is close to the dynamical scale of dark QCD. 
Using the differential cross section given by \cref{eq:diffxsec_had}, we obtain the sensitivity reach of FASER(dark green), FASER2(dark red) and FACET(dark blue) in \cref{fig:adm_ofs_hadron}.
In the figure, we assume the constant multiplicity to be $n_{V'} = 1.0 \, (0.01) $ in the top (bottom) panels.
Compared to the production via the direct coupling (\ref{eq:ApiVcoupling}), the future sensitivity does not significantly change as $n_{V'} = 1.0$.

\section{Summary\label{sec:summary}}

The confining gauge dynamics in the dark sector is a natural realization of DM with the mass in the range of sub-GeV to GeV; the DM in the mass range has recently attracted much attention in several DM scenarios, such as SIMP, ADM, and so on.
Vector mesons appear as a natural consequence of the confining gauge dynamics, and they would play important roles in the DM scenarios.
When the dark sector is connected with the SM sector via the dark photon portal, dark vector mesons can be produced at accelerator-based experiments via the dark photon, and their decay via the dark photon leaves visible signals for specific mass spectra of dark mesons at the long-lived particle searches.
In this paper, we have studied the prospected sensitivities of the LHC forward detectors to search for dark vector mesons, in particular, in two different scenarios: dark-pion DM scenarios and composite ADM scenarios (dark baryon DM). 
There are two forward-detector experiments for exploring long-lived particles during the HL-LHC era: FASER2 near the ATLAS detector and FACET near the CMS detector.

Depending on scenarios, dark photons play different roles, and the required mass spectrum is also different: $m_{\pi'} \lesssim m_{A'}$ in the dark-pion DM models, and $m_{A'} \lesssim m_{\pi'}, m_{N'}$ in the composite ADM models.
This difference leads to the different production of dark vector mesons at the collider experiments: one is the production through the on-shell dark photon decay (in dark-pion DM scenarios), and another is the production by mediating the off-shell dark photon (in composite ADM scenarios).
We have found that, in the on-shell production case, the LHC forward-detector searches of dark vector mesons are comparable to the fixed-target experiments, DarkQuest, which have already been studied in the literature. 
It is worthwhile to note that FACET is possible to explore slightly more broad parameter space than FASER2 thanks to the detector design and the event selection criteria.
In the case of the dark photon being lightest in the dark sector, the future sensitivity of the LHC forward-detector searches covers the parameter space that will be explored by the visible decay searches of dark photons at the fixed-target experiments and the collider experiments, such as HPS, LHCb, and Belle-II.

\subsection*{Acknowledgement}
We thank Qing-Hong Cao for the fruitful discussion.
We also thank Felix Kling for the useful discussion about the package \texttt{FORESEE}, and Burak Hacisahinoglu for the information on simulation regarding the FACET experiment.
T. K. also thank Ayuki Kamada for the useful discussion.
The work is supported in part by the National Science Foundation of China under Grant Nos. 11675002, 11635001, 11725520, 12235001 (T.K. and S.Y.), and 12250410248 (T.K.).

\appendix
\section{Dark Vector Mesons\label{app:vector_meson}}

We list the relevant interactions among dark sector particles, dark vector mesons, dark pions, and dark photons in this appendix.
We first show the relevant Lagrangian, and we will give brief explanations of each terms in the following subsections. 
The relevant Lagrangian is decomposed into five parts as follows.
\eqs{
  \mathcal{L} = \mathcal{L}_{\chi\mathrm{PT}} + \mathcal{L}_\mathrm{WZW} + \mathcal{L}_\mathrm{HLS} + \mathcal{L}_\mathrm{portal} + \mathcal{L}_\mathrm{SM} \,.
}
$\mathcal{L}_\mathrm{SM}$ is the SM Lagrangian, and the chiral Lagrangian $\mathcal{L}_{\chi\mathrm{PT}}$ is given by \cref{eq:chiralLagrangian}.
The gauged WZW Lagrangian $\mathcal{L}_\mathrm{WZW}$ including both the dark photon and the dark vector mesons has a complicated structure, and the interactions with the small numbers of pions are given by 
\eqs{
  \mathcal{L}_\mathrm{WZW} & \supset 
  \frac{2N_C}{15 \pi^2 f_\pi^5} \epsilon^{\mu\nu\rho\sigma} \mathrm{Tr} (\pi' \partial_\mu \pi' \partial_\nu \pi' \partial_\rho \pi' \partial_\sigma \pi') 
  + \frac{N_C e'^2}{16 \pi^2 f_\pi} \mathrm{Tr} \left( Q^2 T^a\right) \epsilon^{\mu\nu\rho\sigma} A'_{\mu \nu} A'_{\rho\sigma} \pi'^a  \\
  & \qquad + \frac{i N_C}{3\pi^2 f_\pi^3} A'_\mu \epsilon^{\mu\nu\rho\sigma} \mathrm{Tr} \left(
    Q \partial_\nu \pi' \partial_\rho \pi' \partial_\sigma \pi'
  \right) 
  - \frac{N_C e' g}{8 \pi^2 f_\pi }\epsilon^{\mu\nu\rho\sigma} \partial_\mu A'_\nu \mathrm{Tr} \left( 
    \{Q, V'_\rho\} \partial_\sigma \pi' 
  \right) \\
  & \qquad 
  - \frac{N_C g^2 }{8 \pi^2 f_\pi }\epsilon^{\mu\nu\rho\sigma} \mathrm{Tr} \left( 
    \{\partial_\mu V'_\nu, V'_\rho\} \partial_\sigma \pi' 
  \right)
  \,.
	\label{eq:piAVinteraction}
}
In particular, the dark-charged vector mesons can decay into dark pion and dark photons through the fourth term.
$\mathcal{L}_\mathrm{portal}$ denotes the Lagrangian for the dark photon, which is given by
\eqs{
  \mathcal{L}_\mathrm{portal} 
  = - \frac14 F'_{\mu\nu} F^{'\mu\nu} 
  - \frac{\epsilon}{2} F_{\mu\nu} A^{'\mu\nu} 
  + \frac12  m_{A'}^2 A_\mu' A^{' \mu} \,.
}
Here, the second term is the kinetic mixing term between the dark photon and our QED, and $\epsilon$ is the kinetic mixing. 

The interactions with the dark vector mesons are described by the HLS.
The HLS Lagrangian is given by 
\eqs{
  \mathcal{L}_\mathrm{HLS} & =
  - \frac{1}{2} \mathrm{Tr} (V'_{\mu\nu} V'^{\mu\nu})
  - \frac{e'}{g} \mathrm{Tr} (Q V'^{\mu\nu}) F'_{\mu\nu} \\
  & \qquad + \frac12 m_{V'}^2 V'^a_\mu V'^{a \mu} 
  + g_{V'\pi'\pi'} f^{abc} V'^a_\mu \pi'^b \partial^\mu \pi'^c 
  + \cdots \,.
}
The second term provides the kinetic mixing between the dark photons and the dark-neutral vector mesons.
Therefore, there are two contributions to give the $\pi A'V'$ processes: one is the direct coupling from the fourth term of \cref{eq:piAVinteraction}, and another is a combination the $A'$-$V'$ mixing and $V'V'\pi'$ coupling [from the fifth term of \cref{eq:piAVinteraction}].

\subsection{Vector Mesons}
We describe vector mesons as the dynamical gauge bosons of hidden local gauge symmetry in the chiral effective theory~\cite{Bando:1984ej}.
In this section, we briefly introduce the vector mesons following Ref.~\cite{Bando:1984ej}.
The global flavor symmetry $G$ is broken down to $H$, and the pion fields $\pi^a$ correspond to the coordinate of the coset space $G/H$.
The chiral Lagrangian describes the effective theory of pions and is given in the text, see \cref{eq:chiralLagrangian}.
The normalization of the decay constant given by \cref{eq:chiralLagrangian} follows $f_\pi = 93\,\mathrm{MeV}$ in the SM.

Now, let us consider the simplest case, $G = SU(2)_L \otimes SU(2)_R$ and $H = SU(2)_V$.
In the UV theory, quarks transform as $q_L \to U_L q_L$ and $q_R \to U_R q_R$ under the flavor symmetry where $U_L$ and $U_R$ are unitary matrices of $SU(2)_L \times SU(2)_R$.
$\Sigma$ is the matrix-valued field, corresponding to $\Sigma_{ij} \sim \overline q_{R j} q_{L i}$ with the subscripts being the flavor indices.
Hence, $\Sigma$ transforms as $\Sigma \to U_L \Sigma U_R^\dag$ under the flavor symmetry.
It is straightforward to extend the following analysis to the case of $SU(N_f)_L \otimes SU(N_f)_R/SU(N_f)_V$.

Besides the global symmetry $G = SU(2)_L \otimes SU(2)_R$, the chiral Lagrangian possesses a local symmetry $SU(2)_V$, which is hidden in terms of $\Sigma$.
We introduce new fields $\xi_L \,, \xi_R$ and the massive Yang-Mills field.
\eqs{
  \Sigma(x) \equiv \xi_L^\dag (x) \xi_R(x) \,, \qquad 
  \widehat V'_\mu(x) = \widehat V'_\mu(x)^a T^a \,.
}
Under the symmetry, $[SU(2)_L \otimes SU(2)_R]_\mathrm{global}\otimes[SU(2)_V]_\mathrm{local}$, these fields are transformed as follows.
\eqs{s
  \xi_L(x) & \to h(x) \xi_L(x) U_L^\dag\,, \qquad
  \xi_R(x) \to h(x) \xi_R(x) U_R^\dag\,, \\
  \widehat V'_\mu(x) & \to i h(x) \partial_\mu h^\dag(x) +  h(x) \widehat V'_\mu(x) h^\dag(x) \,.
}
Here, $h(x)$ denotes the gauge group transformation of $[SU(2)_V]_\mathrm{local}$.
There are two invariants under this transformation and the parity transformation of $\xi_L$ and $\xi_R$.
The invariant Lagrangian is given by
\eqs{
  \mathcal{L} & = \mathcal{L}_A + a \mathcal{L}_V \,, \\
  \mathcal{L}_V & = - \frac{f_{\pi'}^2}{4} \mathrm{Tr}[(D_\mu \xi_L)\xi_L^\dag + (D_\mu \xi_R)\xi_R^\dag]^2 \,, \\
  \mathcal{L}_A & = - \frac{f_{\pi'}^2}{4} \mathrm{Tr}[(D_\mu \xi_L)\xi_L^\dag - (D_\mu \xi_R)\xi_R^\dag]^2 \,.
  \label{eq:invLag_xiLR}
}
Here, the covariant derivative is $D_\mu \xi_{L,R} = \partial_\mu \xi_{L,R} - i g \widehat V'_\mu(x) \xi_{L,R}$ with $g$ being the gauge coupling of the hidden local gauge dynamics.
This Lagrangian should be equivalent to the original Lagrangian \cref{eq:chiralLagrangian}. 
The parameter $a$ is undetermined, which we assume to be $a = 2$ in the text to follow the KSRF relation. 
The Lagrangian (\ref{eq:invLag_xiLR}) contains the mass term for $V_\mu$.
We have new phase degrees of freedom in addition to the pion degrees of freedom, which are eaten by the vector mesons: 
\eqs{
  \xi_L(x) = e^{i \Phi/f_{\pi'}} \hat \xi_L(x)\,, \qquad
  \xi_R(x) = e^{i \Phi/f_{\pi'}} \hat \xi_R(x)\,.
}
The Lagrangian also contains the two-point interaction of the vector mesons and the unphysical pNG bosons.
We can remove the unphysical pNG bosons by choosing the gauge fixing to be the unitary gauge, which is given by
\eqs{
  \xi_L^\dag = \xi_R \equiv \xi = \exp \left( \frac{i \pi'^a T^a}{f_{\pi'}}\right)\,.
}

The kinetic term of $\widehat V'$ is not included in the Lagrangian, but it is expected to arise from the non-perturbative effect in QCD.
Taking the unitary gauge, the Lagrangian is rewritten in terms of $\xi$ as follows:
\eqs{
  \mathcal{L} & =
  \frac{f_{\pi'}^2}{4} \mathrm{Tr}(\partial_\mu \Sigma \partial^\mu \Sigma^\dag)
  - \frac{1}{4} \widehat V'^a_{\mu\nu} \widehat V'^{a\mu\nu}
  + a g^2 f_{\pi'}^2 \mathrm{Tr} \left[ \widehat V'_\mu - \frac{1}{2ig} (\partial_\mu \xi^\dag \cdot \xi + \partial_\mu \xi \cdot \xi^\dag) \right]^2 \,.
  \label{eq:invLag_xi}
}
Here, $\Sigma = \xi^2$ in this gauge, and $\widehat V'_{\mu\nu}$ is the field strength of the dynamical gauge fields, $\widehat V'_{\mu\nu} \equiv \partial_\mu \widehat V'_\nu - \partial_\nu \widehat V'_\mu - i g [\widehat V'_\mu, \widehat V'_\nu]$.
The Lagrangian is expanded up to $\mathcal{O}(\pi'^2)$ as follows.
\eqs{
  \mathcal{L} & =
  \frac12 (\partial_\mu \pi'^a)^2
  - \frac{1}{4} \widehat V'^a_{\mu\nu} \widehat V'^{a\mu\nu}
  + \frac12 ag^2 f_\pi^2 \widehat V'^a_\mu \widehat V'^{a \mu} 
  + \frac12 a g f^{abc} \widehat V'^a_\mu \pi'^b \partial^\mu \pi'^c + \cdots \,.
  \label{eq:Lag_piV}
}
The vector meson mass is universal in the same flavor multiplet, and is determined by the pion decay constant.
Meanwhile, the $V'\pi'\pi'$ coupling is determined by the gauge coupling of the hidden local symmetry:
\eqs{
  m_{V'}^2 = a g^2 f_{\pi'}^2 \,, \qquad 
  g_{V'\pi'\pi'} = \frac{a}{2} g \,.
  \label{eq:app_KSRF}
}
We obtain the KSRF relation given in \cref{eq:KSRF_relation} when we set $a = 2$.

We modify the covariant derivative by introducing the $U(1)$ electromagnetic interaction as follows:
\eqs{
  D_\mu \xi_{L,R} = \partial_\mu \xi_{L,R} - i g \widehat V'_\mu(x) \xi_{L,R} + i e_0 B_\mu \xi_{L,R} Q_{L,R} \,.
}
Here, $B_\mu$ denotes the $U(1)$ gauge field, and the action of the charge matrices is defined as the action on $\Sigma$: $D_\mu\Sigma = \partial_\mu \Sigma - i e_0 B_\mu (Q_L^\dag \Sigma - \Sigma Q_R)$%
\footnote{
  We always have an uncertainty by the unit matrix to define the charge matrices of $\xi$ from those of quarks. 
}.
As far as the mass basis of quarks is identified with the charge basis, the charge matrices are identical $Q_R = Q_L \equiv Q$ and hermite ($Q^\dag = Q$).
We consider the identical case in the following.
In the presence of the $U(1)$ gauge field, the Lagrangian in the unitary gauge given by \cref{eq:invLag_xiLR} takes the form:
\eqs{
  \mathcal{L}_V & = \frac{f_{\pi'}^2}{4} \mathrm{Tr}[D_\mu \Sigma D^\mu \Sigma^\dag] \,, \\
  a \mathcal{L}_A & = a f_{\pi'}^2 \mathrm{Tr} \left[ \widehat V'_\mu - \frac{e_0}{2g} B_\mu (\xi^\dag Q \xi + \xi Q \xi^\dag) - \frac{1}{2ig^2} (\partial_\mu \xi^\dag \cdot \xi + \partial_\mu \xi \cdot \xi^\dag) \right]^2 \,.
}
The second term contains the mass mixing between the vector meson and the $U(1)$ gauge boson, and hence we redefine the vector meson field as the mass eigenstates.
This mass mixing is absorbed by a shift of the vector mesons, and then, in turn, the kinetic mixing appears.
The newly defined vector meson fields and its field strength tensors are 
\eqs{
  V'_\mu & = \widehat V_\mu - \frac{e_0}{g} B_\mu Q \,, \qquad
  V'_{\mu \nu} = \widehat V_{\mu \nu} - \frac{e_0}{g} B_{\mu\nu} Q \,, \\
  V'_{\mu \nu} & \equiv D_\mu V'_\nu - D_\nu V'_\mu - i g [V'_\mu, V'_\nu] \,, \qquad 
  D_\mu V'_\nu = \partial_\mu V'_\nu - i e_0 B_\mu [Q,V'_\nu]\,.
}
Up to $\mathcal{O}(\pi'^2)$, the Lagrangian is expanded as follows:
\eqs{
  \mathcal{L} & =
  \frac12 (\partial_\mu \pi'^a)^2
  - \frac{1}{2} \mathrm{Tr} (V'_{\mu\nu} V'^{\mu\nu})
  - \frac{e_0}{g} \mathrm{Tr} (Q V'^{\mu\nu}) B_{\mu\nu}
  - \frac{1}{4} \left( 1 + \frac{e_0^2}{g^2} \right) B_{\mu\nu} B^{\mu\nu} \\
  & \qquad 
  + \frac12 m_{V'}^2 V'^a_\mu V'^{a \mu} 
  + g_{V'\pi'\pi'} f^{abc} V'^a_\mu \pi'^b \partial^\mu \pi'^c 
  - 2 i e_0 B_\mu \mathrm{Tr}(Q [\pi', \partial^\mu \pi']) 
  + \cdots \,.
}
Here, $m_{V'}^2$ and $g_{V'\pi'\pi'}$ are defined in \cref{eq:app_KSRF}.
The kinetic term for $B_\mu$ is not canonically normalized, and hence we define the canonically normalized field and the electromagnetic coupling as follows.
\eqs{
  A'_\mu = \sqrt{1 + \frac{e_0^2}{g^2}} B_\mu \,, \qquad 
  e' = \frac{g}{\sqrt{g^2+e_0^2}} e_0 \,.
}
The coupling to the dynamical gauge boson is typically strong, it is expected $g \gg e_0$, which follows $e' \simeq e_0$.
The Lagrangian at the leading order is rewritten as follows.
\eqs{
  \mathcal{L} & =
  \frac12 (\partial_\mu \pi^a)^2
  - \frac{1}{2} \mathrm{Tr} (V'_{\mu\nu} V'^{\mu\nu})
  - \frac{e'}{g} \mathrm{Tr} (Q V'^{\mu\nu}) F'_{\mu\nu}
  - \frac{1}{4} F'_{\mu\nu} F'^{\mu\nu} \\
  & \qquad 
  + \frac12 m_{V'}^2 V'^a_\mu V'^{a \mu} 
  + g_{V'\pi'\pi'} f^{abc} V'^a_\mu \pi'^b \partial^\mu \pi'^c 
  - 2 i e' A_\mu \mathrm{Tr}(Q [\pi', \partial^\mu \pi']) 
  + \cdots \,.
  \label{eq:gaugedWZW}
}
Here, $F'_{\mu\nu}$ is the field strength tensor of $A'_\mu$.

\subsection{Wess-Zumino-Witten term}

Next, we introduce the gauged WZW term~\cite{Wess:1971yu,Witten:1983tw} of pions to discuss the interactions among dark photon, dark vector meson, and dark pion.
The non-linear sigma model field $\Sigma$ is a mapping of the four-dimensional spacetime $M$ into $G/H$. 
We can find a topological action on five-dimensional spacetime, which is given by a unique Lagrangian consisting only of $\Sigma$. 
The Lagrangian is written as the total derivative at the leading order of $\pi'$ not $\Sigma$, and hence the volume integral of the Lagrangian gives the surface integral corresponding to the WZW term.
For $G = SU(N_f)_L \times SU(N_f)_R$ and $H = SU(N_f)_V$, the action takes the form
\eqs{
  i N \Gamma(\Sigma) = \frac{2N i}{15 \pi^2 f_{\pi'}^5} \int_M d^4 x \epsilon^{\mu\nu\rho\sigma} \mathrm{Tr} (\pi' \partial_\mu \pi' \partial_\nu \pi' \partial_\rho \pi' \partial_\sigma \pi') \,.
}
As we will see below, $N = N_C$ for $SU(N_C)$ gauge theory.
The chiral anomaly term is reproduced by incorporating the coupling to electromagnetism into the WZW term. 
$\Gamma$ is invariant under global infinitesimal transformation $\Sigma \to \Sigma + i \epsilon [Q, \Sigma]$.
However, the global symmetry cannot be gauged just by replacing the derivative to the covariant derivative since the global symmetry is non-linearly realized in the four-dimensional expression for $\Gamma$.
We construct the gauge invariant functional by finding the Noether current $J^\mu$ and adding extra terms to maintain the gauge invariance.
\eqs{
  & \overline \Gamma(\Sigma, A_\mu) 
  = \Gamma(\Sigma) 
  - e' \int d^4x A'_\mu J^\mu \\
  & \quad + \frac{i e'^2}{24 \pi^2} \int d^4x \epsilon^{\mu\nu\rho\sigma} \partial_\mu A'_\nu A'_\rho \mathrm{Tr} 
    Q^2 \left[(\partial_\sigma \Sigma) \Sigma^{-1} 
    + \Sigma^{-1} (\partial_\sigma \Sigma) \right] \\
  & \quad + \frac{i e'^2}{24 \pi^2} \int d^4x \epsilon^{\mu\nu\rho\sigma} \partial_\mu A'_\nu A'_\rho \mathrm{Tr}
  Q \Sigma Q \Sigma^{-1} (\partial_\sigma \Sigma) \Sigma^{-1} \,.
  \label{eq:WZWactionU1}
}
Here, the Noether current is given by
\eqs{
  J^\mu = \frac{1}{48\pi^2} \epsilon^{\mu\nu\rho\sigma} \mathrm{Tr} Q \left[ 
    (\partial_\nu \Sigma \Sigma^{-1}) (\partial_\rho \Sigma \Sigma^{-1}) (\partial_\sigma \Sigma \Sigma^{-1}) 
    + (\Sigma^{-1} \partial_\nu \Sigma) (\Sigma^{-1}\partial_\rho \Sigma) (\Sigma^{-1} \partial_\sigma \Sigma) 
  \right] \,.
}
We find the last term reproduces the chiral anomaly term as $N=N_C$.

The WZW Lagrangian \cref{eq:WZWactionU1} is not still invariant under the HLS.
We can incorporate the vector mesons in the WZW Lagrangian by gauging the global flavor symmetry \cite{Kaymakcalan:1983qq,Fujiwara:1984mp}.
We show the relevant $V' V' \pi'$ processes that arise from the gauge invariant 4-form.
\eqs{
  \mathcal{L}_\mathrm{WZW} & \supset 
  \frac{i N_C g^2}{8 \pi^2}\epsilon^{\mu\nu\rho\sigma} \mathrm{Tr} \left[ 
    \partial_\mu V'_\nu (V'_\rho \partial_\sigma \pi - \partial_\rho \pi V'_\sigma)
  \right] 
  \,.
} 
Even though we use interaction terms with assuming the vector meson dominance (for a review, see \cite{Harada:2003jx}), we describe the effective vertices by replacing $V'$ with $A'$ through the kinetic mixing term in \cref{eq:gaugedWZW}.
We obtain the effective terms for $\pi' V' A'$ interactions in \cref{eq:piAVinteraction} by the replacement.

\section{Decay Rate \label{app:Decay_Rate}}

Let us summarize the decays of dark particles in this appendix. 
First, we focus on the decay of dark photons. 
The dark photon predominantly decays into the SM particles through the kinetic mixing $\epsilon$ when $A'$ is lightest among dark-sector particles. 
\eqs{
  \Gamma(A' \to \ell^+ \ell^-)
  = \frac{\alpha \epsilon^2}{3} m_{A'} \sqrt{1 - \frac{4 m_\ell^2}{m_{A'}^2}} \left( 1 + \frac{2 m_\ell^2}{m_{A'}^2} \right) \,.
}
Here, $m_\ell$ denotes the mass of a lepton $\ell$.
When the dark photon is heavier than dark pions and dark vector mesons, the dark photon predominantly decays into dark hadrons.
When the mass of dark photon is heavier than dark hadrons, the dominant decay is given by $A' \to \pi' \pi'$ and $A' \to V' \pi'$. 
Each decay rate is given by 
\eqs{
  \Gamma_{A' \to \pi'^a + \pi'^b}
  & = \frac{2 \alpha' m_{A'}}{3 \pi^4} \left[ \mathrm{Tr}(\{Q, T^a\} T^b) \right]^2 \left( 1 - \frac{4 m_{\pi'}^2}{m_{A'}^2} \right)^{3/2} \left( 1 - \frac{m_{A'}^2}{m_{V'}^2} \right)^{-2} \,.
}
and 
\eqs{
  \Gamma_{A' \to V'^a + \pi'^b}
  & = \frac{3}{128 \pi^4} \frac{g^2 \alpha'}{f_{\pi'}^2}
  \left[ \mathrm{Tr}(\{Q, T^a\} T^b) \right]^2 \\
  & \qquad \times 
  \frac{1}{m_{A'}^3}[m_{A'}^2 - (m_{V'}+m_{\pi'})^2]^{3/2}[m_{A'}^2 - (m_{V'}-m_{\pi'})^2]^{3/2} \,.
}

Next, we discuss the decay of dark vector mesons. 
The dark-neutral vector meson decays through the kinetic mixing, and the decay rate is given by 
\eqs{
  \Gamma(V'^{a} \to \ell^+ \ell^-) = \frac{16 \pi \alpha' \alpha \epsilon^2}{3 g^2} [\mathrm{Tr}(Q T^a)]^2 \frac{m_{V'}^5}{m_{A'}^4} \left(\frac{m_{A'}^2}{m_{A'}^2 - m_{V'}^2}\right)^2 \sqrt{1 - \frac{4 m_\ell^2}{m_{V'}^2}} \left( 1 + \frac{2 m_\ell^2}{m_{V'}^2} \right) \,.
}
The hadronic decay of the dark vector mesons is obtained by multiplying the $R$-ratio at the collision energy of $\sqrt{s} = m_{V'}$ to $\Gamma(V'^{a} \to \mu^+ \mu^-)$.
As for the three-body decay of the dark-charged vector mesons, we use an integral form of the decay rate in order to incorporate the hadronic decay. 
The process is divided into two pieces at the intermediate off-shell dark photon: $V'^a \to \pi'^a + A'^\ast(q)$ and $A'^\ast(q) \to \overline f f$, where $q$ denotes the four-momentum of the intermediate dark photon.
Inserting unity to the differential decay rate, we can divide it into two parts. 
\eqs{
  1 & = \int \frac{d^4 q}{(2\pi)^4} \int \frac{d m_\ast^2}{2\pi} (2\pi)^4 \delta^4(q - p_3 - p_4) \theta(q^0) 2 \pi \delta(m_\ast^2 - q^2) \\
  & = \int \frac{d m_\ast^2}{2\pi} \int \frac{d^3 q}{(2\pi)^3 2 E_q} (2\pi)^4 \delta^4(q - p_3 - p_4) \,,
}
where $E_q^2 = m_\ast^2 + \vec{q}^{\;2}$.
Now, we consider the rest frame of initial particle. 
$m_\ast^2$ is kinematically allowed in the range of $4 m_f^2 \leq m_\ast^2 \leq (m_{V'} - m_{\pi'})^2$.

Using the decomposition of the decay rate, the total decay rate of $V'$ is given by
\eqs{
  d \Gamma (V' \to \pi' f \overline f) 
  & = \frac{d m_\ast^2}{\pi} \Gamma_{V' \to \pi' + A'^\ast}(m_{A'}=m_\ast)
  \frac{m_\ast \Gamma_{A'^\ast \to f \overline f}(m_{A'}=m_\ast)}{(m_\ast^2 - m_{A'}^2)^2 + (m_{A'} \Gamma_{A'})^2} \,.
}
Here, the decay rate of $V'$ is given by exchanging masses of $\Gamma_{A' \to V'+\pi'^b}$, and takes the form:
\eqs{
  \Gamma_{V'^a \to \pi'^b + A'} (m_{A'})
  & = \frac{3}{128 \pi^4} \frac{g^2 \alpha'}{f_{\pi'}^2}
  \left[ \mathrm{Tr}(\{Q, T^a\} T^b) \right]^2 \\
  & \qquad \times 
  \frac{1}{m_{V'}^3}[m_{V'}^2 - (m_{A'}+m_{\pi'})^2]^{3/2}[m_{V'}^2 - (m_{A'}-m_{\pi'})^2]^{3/2} \,.
}

When the momentum transfer $q^2 = m_\ast^2$ is small to open the decay into the dark mesons ($A'^\ast \to \pi' \pi'$), the (off-shell) dark photon predominantly decays into the SM fermions (and hadrons).
The decay rate into the SM particle, $\Gamma(A'^\ast \to f(p_3)+\overline{f}(p_4))$, is 
\eqs{
  \Gamma_{A'^\ast \to f \overline f}(m_{A'}=m_\ast)
  & = \frac{\epsilon^2 \alpha m_\ast}{3} \left( 1 + \frac{2 m_f^2}{m_\ast^2} \right) \sqrt{1 - \frac{4 m_f^2}{m_\ast^2}} \,, \\
  \Gamma_{A'^\ast \to \text{hadrons}}(m_{A'}=m_\ast)
  & = R(\sqrt{s} = m_\ast) \Gamma_{A'^\ast \to f \overline f}(m_{A'}=m_\ast) \,.
}
Here, $R(\sqrt{s}) \equiv  \sigma(e^+e^- \to \text{hadrons}) / \sigma(e^+e^- \to \mu^+ \mu^-)$ taken from Ref.~\cite{Zyla:2020zbs}.

\section{Production \label{app:production}}
We discuss the off-shell production of the dark vector mesons.

\subsection{Bremsstrahlung \label{app:brems}}

We start from the dark photon production from proton bremsstrahlung following the discussion of Ref.~\cite{Feng:2017uoz}.
The production cross section of dark photon is given by
\eqs{
  d \sigma_\mathrm{Brem} = w_{A' p}(z,p_T^2) F_{1,p}^2(m_{A'}^2) d z d p_T^2 \sigma_{pp}(s') \,.
}
Here, $z \equiv p_{A',z}/p$ is the ratio of the dark photon momentum along the beamline and the beam particle momentum, $F_{1,p}(m_{A'}^2)$ is the timelike form factor of the proton, and $p_T$ denotes the transverse momentum of the produced dark photon.
The center-of-mass energy after emitting the dark photon is $s' = 2 m_p (E_p - E_{A'})$ where $E_p$ and $E_{A'}$ the energies of the beam proton and the dark photon, respectively. 
$\sigma_{pp}(s') \equiv \sigma_\mathrm{tot}(s') - \sigma_\mathrm{ela}(s')$ denotes the inelastic scattering cross section, and we use the experimental data for the cross section.
We parametrize the proton-proton total cross section following Refs.~\cite{Cudell:2001pn,ParticleDataGroup:2016lqr}, and the elastic cross section following Ref.~\cite{TOTEM:2017asr}.
\eqs{
  \sigma_\mathrm{tot} & = 34.41 + 13.07\left( \frac{s}{s_0} \right)^{-0.4473} + 7.394\left( \frac{s}{s_0} \right)^{-0.5486} + 0.272\left(\ln{ \frac{s}{s_0}}\right)^{2} \,, \\
  \sigma_\mathrm{ela} & = 11.84 - 1.617\ln{\frac{s}{s_1}} + 0.1359\left(\ln{ \frac{s}{s_1}}\right)^{2} \,, \\
}
where $s_0 = (4\,\mathrm{GeV})^2$ and $s_1 = (1\,\mathrm{GeV})^2$, and the cross section is in unit of $\mathrm{mb}$.
$w_{A' p}(z,p_T^2)$ is the factorizing weight-function that is computed using the FWW method in which the full process is divided into two subprocesses, a part with emitting dark photon and a part with inelastic scattering of hadrons~\cite{Fermi:1924tc,vonWeizsacker:1934nji,Williams:1934ad,Williams:1935dka}.
\eqs{
  w_{A' p}(z,p_T^2) & = 
  - \frac{\epsilon^2 \alpha}{2 \pi} \frac{1}{z U} 
  \left[ \frac{1+(1-z)^2}{z} 
  + 2 (1-z) \left( \frac{2 m_p^2 + m_{A'}^2}{U} + \frac{2 z m_p^4}{U^2} \right) 
  \right. \\
  & \qquad \left. + 2 (1-z) \frac{1 + (1-z)^2}{z} \frac{m_p^2 m_{A'}^2}{U^2} 
  + \frac{(1-z)^2}{z} \frac{2 m_{A'}^4}{U^2}
  \right] \,.
}
Here, $U$ measures the virtuality of the internal proton after emitting the dark photon and is defined by 
\eqs{
  U = - z m_p^2 - \frac{1-z}{z} m_{A'}^2 - \frac{p_T^2}{z} \,.
}
As far as the virtuality is quite small, the approximation using the FWW method is plausible. 
The FWW approximation can be valid as far as the following condition is satisfied (limited to the massive vector boson production)~\cite{Blumlein:2013cua,deNiverville:2016rqh}.
\eqs{
  E_p \,, E_{A'} \,, E_p - E_{A'} \gg m_p \,, m_{A'} \,, p_T \,.
  \label{eq:criteria_FWW}
}
The beam particle is more energetic than the produced particles, and hence, dark photons are softly and collinearly produced. 
Most of energy is still carried by the beam particle even after emitting dark photons. 

For the dark vector meson production through off-shell dark photon from proton bremsstrahlung, the production can be divided into two parts, as explained in the text [see \cref{eq:offshell_productionxsec}].
The dark vector meson production cross section is a convolution of these two parts.
\eqs{
  d \sigma 
  & = \frac{d m_{A'}^{\ast 2}}{\pi} \frac{m_{A'}^{\ast} \Gamma_{A' \to V' \pi'}}{(m_{A'}^{\ast 2} - m_{A'}^2)^2 + (m_{A'} \Gamma_{A'})^2} d \sigma_\mathrm{Brem} \\
  & = \frac{d m_{A'}^{\ast 2}}{\pi} \frac{m_{A'}^{\ast} \Gamma_{A' \to V' \pi'}}{(m_{A'}^{\ast 2} - m_{A'}^2)^2 + (m_{A'} \Gamma_{A'})^2} w(z,p_T^2) F_{1,p}^2(m_{A'}^{\ast 2}) d z d p_T^2 \sigma_{pp}(s') \,.
}
We use the differential cross section integrated over $m_{A'}^{\ast 2}$ and $z$ in order to depict the kinematic distribution of the dark vector mesons in \cref{fig:spectrum_ADM_set2}.

\subsection{Meson Decay}

We consider the $V'$ production via the off-shell dark photon through the meson decays.
For the three-body decay process $\varphi(p_0) \to \gamma(p_1) + A'^\ast(q) \to \gamma(p_1) + \pi'(p_2) + V'(p_3)$, the phase-space integral of the final states is written as
\eqs{
  \int \frac{d q^2}{2 \pi} \frac{|\vec{p}_1|}{16 \pi^2 m_\varphi} \int d \cos \theta_{1q} d \varphi_{1q}
  \frac{|\vec{p}_3|}{16 \pi^2 \sqrt{q^2}} \int d \cos \theta_{23} d \varphi_{23} \,.
}
Here, $q$ denotes the four momentum of the intermediate dark photon. 
$\theta_{1q}$ and $\varphi_{1q}$ denote the angle variables of $p_1$ with respect to the beam axis in the center-of-mass frame of $p_1$ and $q$ system. 
Meanwhile, $\theta_{23}$ and $\varphi_{23}$ denote the angle variables of $p_3$ with respect to the $q$ direction in the center-of-mass frame of $p_2$ and $p_3$ system.
The amplitude square for $\varphi \to \gamma V' \pi'$ is
\eqs{
  |\mathcal{M}|^2
  & = \left( \frac{3 \alpha' m_{V'}^2}{8 \pi^2 f_{\pi'}^4} \right)
  \left( \frac{\alpha \epsilon}{\pi f_\varphi} \right)^2
  \left[ q^4 \left(p_1 \cdot p_3\right)^2
  -m_{V'}^2 q^2 \left(p_1 \cdot q\right)^2
  +2 \left(p_1 \cdot q\right)^2 \left(p_3 \cdot q\right)^2\right. \\
  & \qquad \left.-2 q^2 \left(p_1\cdot p_3\right) \left(p_1 \cdot q\right) \left(p_3 \cdot q\right)
  \right]
  \frac{1}{\left(q^2 - m_{A'}^2\right)^2} \,.
}
where the coefficient is specific for the case that the final dark vector mesons are $\eta$ and $\rho^{0}$.
The differential decay width respect to $q^2$ and $\cos \theta_{23}$ is
\eqs{
  \frac{d \Gamma_{\varphi \to \gamma V'\pi'}}{d q^2 d \cos \theta_{23}}
  & = \frac{1}{2 \pi} \frac{1}{2 m_\varphi} \frac{|\vec{p}_1|}{4 \pi m_\varphi} \frac{|\vec{p}_3|}{8 \pi \sqrt{q^2}} |\mathcal{M}|^2 \,.
}

\subsection{Drell-Yan Process}

We simulate the dark photon production through the Drell-Yan process using the Hidden-Valley module in \texttt{PYTHIA 8}~\cite{Sjostrand:2014zea}.
The encoded hard process is at the LO of $\alpha_s$, and the initial state radiation can partly mimic the next-to-leading order (NLO) effect, which endows dark-photon transverse momentum.
Thus, the (polar-)angle distribution of dark photon highly depends on the initial state radiation, and hence the choice of the PDF may influence the angle distribution.
As a consequence, the distribution of the dark vector mesons also depends on the PDF sets.
Following \texttt{FORESEE}, we simulate the dark photon from Drell-Yan process using default parameters in \texttt{PYTHIA 8} with the \texttt{NNPDF2.3} PDF set, QCD+QED at the LO, and $\alpha_{s}(M_Z)=0.130$.

\begin{figure}[!htbp]
\begin{minipage}[t]{0.5\linewidth}
    \centering
    \includegraphics[width=1\textwidth]{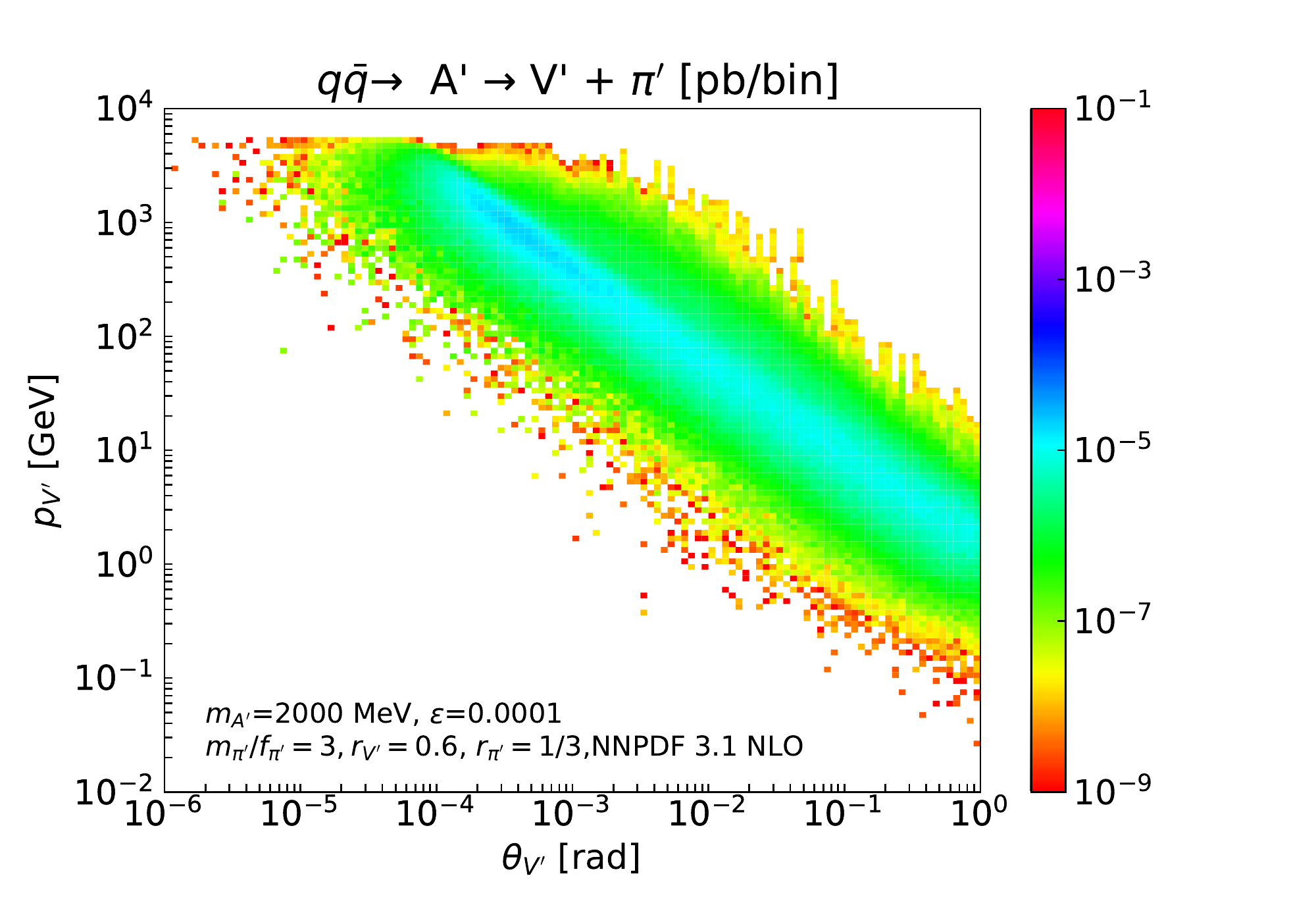}
\end{minipage}
\begin{minipage}[t]{0.5\linewidth}
    \centering
    \includegraphics[width=1\textwidth]{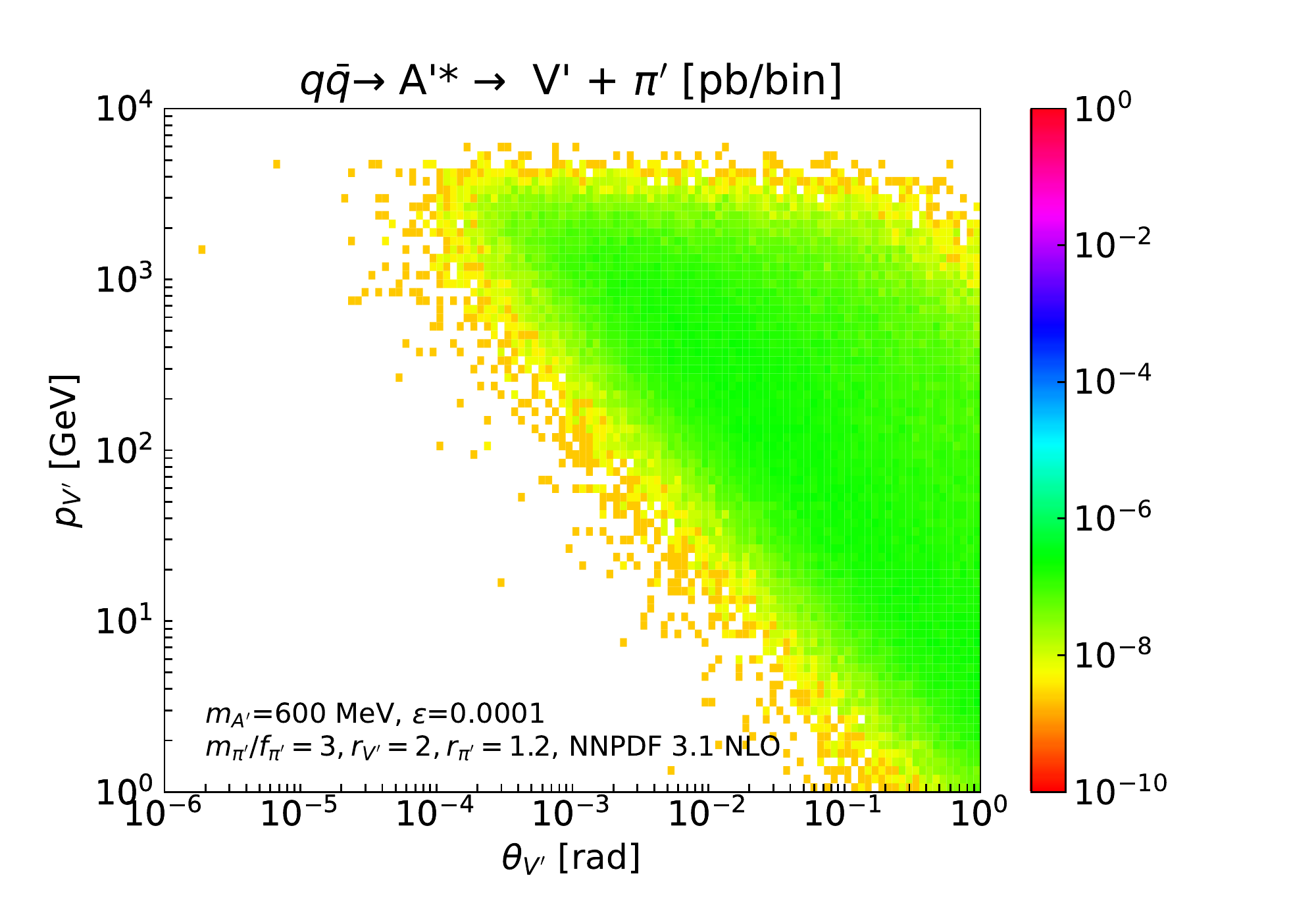}
\end{minipage}
\caption{
  Dark vector meson distribution in the $(\theta_{V'},p_{V'})$ plane. The dark vector mesons are produced through off-shell (left) and on-shell (right) dark photon from Drell-Yan process.
  We use different PDF option compared to \cref{fig:spectrum_SIMP,fig:spectrum_ADM_set2}.
  }
\label{fig:spectrum_DY_nlo}
\end{figure}

We compare the momentum distributions based on \texttt{NNPDF 3.1} NLO with those in \cref{fig:spectrum_SIMP,fig:spectrum_ADM_set2}. 
Compared to the corresponding result in \cref{fig:spectrum_SIMP,fig:spectrum_ADM_set2}, the distribution does not significantly change, but the cross section is slightly changed in the large $\theta_{V'}$ and small $p_{V'}$.
The final sensitivity contour will not change drastically since the forward detectors only cover the small angle of $\theta \lesssim 10^{-3}\,\mathrm{rad}$.
This is because, in both cases, the angle distribution is predominantly controlled by the production angle via $A'\text{--}\pi'\text{--}V'$ interaction.
In other words, the angle distribution of dark vector mesons is hardly sensitive to the PDF sets.

\section{Invisible Decay Signals \label{app:invisibleNA64}}

In \cref{fig:simp_ons_detect}, we include the existing constraints by the invisible signal searches at NA64~\cite{Banerjee:2019pds}. 
The original study constraints only on the dark photon mediator model, where $A'$ predominantly decays into a pair of DM particles with the branching fraction of $\mathrm{Br}(A' \to \overline\chi \chi) \simeq 1$.
In the dark-pion DM models, there are two dominant decay mode of $A'$, $A' \to \pi'\pi'$ and $A' \to \pi' V'$.
However, the latter would leave visible signals as we have discussed in the text for sufficiently small $\epsilon$.
The decay gives visible signals when the decay occurs at the detector area, and the decay signal is vetoed by the detection criteria. 
We can reinterpret the constraint by NA64 on the minimal model, the upper bound denoted by $\epsilon_\mathrm{NA64}$, into that on the dark-pion DM models, the bound denoted by $\epsilon_\mathrm{SIMP}$, by the following relation.
\eqs{
  \epsilon^2_\mathrm{NA64} = \epsilon^2_\mathrm{SIMP} \left[ \mathrm{Br}(A' \to \pi'\pi') + \mathrm{Br}(A' \to \pi' V') (1-\mathcal{P}_\mathrm{dec})\theta(c\tau_{V'}\beta\gamma-L_\mathrm{min}) \right]\,.
}
Here, $\mathcal{P}_\mathrm{dec}$ denotes the decay probability of $V'$ inside the detector volume, defined in \cref{eq:signalnumber}, and the step function represents the exclusion of $V'$ with decaying before the detector area.

\bibliographystyle{utphys}
\bibliography{ref}

\providecommand{\href}[2]{#2}\begingroup\raggedright\begin{thebibliography}{100}

\bibitem{Lee:1977ua}
B.~W. Lee and S.~Weinberg, ``{Cosmological Lower Bound on Heavy Neutrino
  Masses},'' \href{http://dx.doi.org/10.1103/PhysRevLett.39.165}{{\em Phys.
  Rev. Lett.} {\bfseries 39} (1977) 165--168}.

\bibitem{Gudnason:2006yj}
S.~B. Gudnason, C.~Kouvaris, and F.~Sannino, ``{Dark Matter from new
  Technicolor Theories},''
  \href{http://dx.doi.org/10.1103/PhysRevD.74.095008}{{\em Phys. Rev.}
  {\bfseries D74} (2006) 095008},
\href{http://arxiv.org/abs/hep-ph/0608055}{{\ttfamily arXiv:hep-ph/0608055
  [hep-ph]}}.

\bibitem{Dietrich:2006cm}
D.~D. Dietrich and F.~Sannino, ``{Conformal window of SU(N) gauge theories with
  fermions in higher dimensional representations},''
  \href{http://dx.doi.org/10.1103/PhysRevD.75.085018}{{\em Phys. Rev.}
  {\bfseries D75} (2007) 085018},
\href{http://arxiv.org/abs/hep-ph/0611341}{{\ttfamily arXiv:hep-ph/0611341
  [hep-ph]}}.

\bibitem{Khlopov:2007ic}
M.~{\relax Yu}. Khlopov and C.~Kouvaris, ``{Strong Interactive Massive
  Particles from a Strong Coupled Theory},''
  \href{http://dx.doi.org/10.1103/PhysRevD.77.065002}{{\em Phys. Rev.}
  {\bfseries D77} (2008) 065002},
\href{http://arxiv.org/abs/0710.2189}{{\ttfamily arXiv:0710.2189 [astro-ph]}}.

\bibitem{Khlopov:2008ty}
M.~{\relax Yu}. Khlopov and C.~Kouvaris, ``{Composite dark matter from a model
  with composite Higgs boson},''
  \href{http://dx.doi.org/10.1103/PhysRevD.78.065040}{{\em Phys. Rev.}
  {\bfseries D78} (2008) 065040},
\href{http://arxiv.org/abs/0806.1191}{{\ttfamily arXiv:0806.1191 [astro-ph]}}.

\bibitem{Foadi:2008qv}
R.~Foadi, M.~T. Frandsen, and F.~Sannino, ``{Technicolor Dark Matter},''
  \href{http://dx.doi.org/10.1103/PhysRevD.80.037702}{{\em Phys. Rev.}
  {\bfseries D80} (2009) 037702},
\href{http://arxiv.org/abs/0812.3406}{{\ttfamily arXiv:0812.3406 [hep-ph]}}.

\bibitem{Mardon:2009gw}
J.~Mardon, Y.~Nomura, and J.~Thaler, ``{Cosmic Signals from the Hidden
  Sector},'' \href{http://dx.doi.org/10.1103/PhysRevD.80.035013}{{\em Phys.
  Rev.} {\bfseries D80} (2009) 035013},
\href{http://arxiv.org/abs/0905.3749}{{\ttfamily arXiv:0905.3749 [hep-ph]}}.

\bibitem{Kribs:2009fy}
G.~D. Kribs, T.~S. Roy, J.~Terning, and K.~M. Zurek, ``{Quirky Composite Dark
  Matter},'' \href{http://dx.doi.org/10.1103/PhysRevD.81.095001}{{\em Phys.
  Rev.} {\bfseries D81} (2010) 095001},
\href{http://arxiv.org/abs/0909.2034}{{\ttfamily arXiv:0909.2034 [hep-ph]}}.

\bibitem{Barbieri:2010mn}
R.~Barbieri, S.~Rychkov, and R.~Torre, ``{Signals of composite
  electroweak-neutral Dark Matter: LHC/Direct Detection interplay},''
  \href{http://dx.doi.org/10.1016/j.physletb.2010.04.010}{{\em Phys. Lett.}
  {\bfseries B688} (2010) 212--215},
\href{http://arxiv.org/abs/1001.3149}{{\ttfamily arXiv:1001.3149 [hep-ph]}}.

\bibitem{Blennow:2010qp}
M.~Blennow, B.~Dasgupta, E.~Fernandez-Martinez, and N.~Rius, ``{Aidnogenesis
  via Leptogenesis and Dark Sphalerons},''
  \href{http://dx.doi.org/10.1007/JHEP03(2011)014}{{\em JHEP} {\bfseries 03}
  (2011) 014},
\href{http://arxiv.org/abs/1009.3159}{{\ttfamily arXiv:1009.3159 [hep-ph]}}.

\bibitem{Lewis:2011zb}
R.~Lewis, C.~Pica, and F.~Sannino, ``{Light Asymmetric Dark Matter on the
  Lattice: SU(2) Technicolor with Two Fundamental Flavors},''
  \href{http://dx.doi.org/10.1103/PhysRevD.85.014504}{{\em Phys. Rev.}
  {\bfseries D85} (2012) 014504},
\href{http://arxiv.org/abs/1109.3513}{{\ttfamily arXiv:1109.3513 [hep-ph]}}.

\bibitem{Appelquist:2013ms}
{\bfseries Lattice Strong Dynamics (LSD)} Collaboration, T.~Appelquist {\em
  et~al.}, ``{Lattice calculation of composite dark matter form factors},''
  \href{http://dx.doi.org/10.1103/PhysRevD.88.014502}{{\em Phys. Rev.}
  {\bfseries D88} no.~1, (2013) 014502},
\href{http://arxiv.org/abs/1301.1693}{{\ttfamily arXiv:1301.1693 [hep-ph]}}.

\bibitem{Hietanen:2013fya}
A.~Hietanen, R.~Lewis, C.~Pica, and F.~Sannino, ``{Composite Goldstone Dark
  Matter: Experimental Predictions from the Lattice},''
  \href{http://dx.doi.org/10.1007/JHEP12(2014)130}{{\em JHEP} {\bfseries 12}
  (2014) 130},
\href{http://arxiv.org/abs/1308.4130}{{\ttfamily arXiv:1308.4130 [hep-ph]}}.

\bibitem{Cline:2013zca}
J.~M. Cline, Z.~Liu, G.~Moore, and W.~Xue, ``{Composite strongly interacting
  dark matter},'' \href{http://dx.doi.org/10.1103/PhysRevD.90.015023}{{\em
  Phys. Rev.} {\bfseries D90} no.~1, (2014) 015023},
\href{http://arxiv.org/abs/1312.3325}{{\ttfamily arXiv:1312.3325 [hep-ph]}}.

\bibitem{Appelquist:2014jch}
{\bfseries Lattice Strong Dynamics (LSD)} Collaboration, T.~Appelquist {\em
  et~al.}, ``{Composite bosonic baryon dark matter on the lattice: SU(4) baryon
  spectrum and the effective Higgs interaction},''
  \href{http://dx.doi.org/10.1103/PhysRevD.89.094508}{{\em Phys. Rev.}
  {\bfseries D89} no.~9, (2014) 094508},
\href{http://arxiv.org/abs/1402.6656}{{\ttfamily arXiv:1402.6656 [hep-lat]}}.

\bibitem{Hietanen:2014xca}
A.~Hietanen, R.~Lewis, C.~Pica, and F.~Sannino, ``{Fundamental Composite Higgs
  Dynamics on the Lattice: SU(2) with Two Flavors},''
  \href{http://dx.doi.org/10.1007/JHEP07(2014)116}{{\em JHEP} {\bfseries 07}
  (2014) 116},
\href{http://arxiv.org/abs/1404.2794}{{\ttfamily arXiv:1404.2794 [hep-lat]}}.

\bibitem{Krnjaic:2014xza}
G.~Krnjaic and K.~Sigurdson, ``{Big Bang Darkleosynthesis},''
  \href{http://dx.doi.org/10.1016/j.physletb.2015.11.001}{{\em Phys. Lett.}
  {\bfseries B751} (2015) 464--468},
\href{http://arxiv.org/abs/1406.1171}{{\ttfamily arXiv:1406.1171 [hep-ph]}}.

\bibitem{Detmold:2014qqa}
W.~Detmold, M.~McCullough, and A.~Pochinsky, ``{Dark Nuclei I: Cosmology and
  Indirect Detection},''
  \href{http://dx.doi.org/10.1103/PhysRevD.90.115013}{{\em Phys. Rev.}
  {\bfseries D90} no.~11, (2014) 115013},
\href{http://arxiv.org/abs/1406.2276}{{\ttfamily arXiv:1406.2276 [hep-ph]}}.

\bibitem{Detmold:2014kba}
W.~Detmold, M.~McCullough, and A.~Pochinsky, ``{Dark nuclei. II. Nuclear
  spectroscopy in two-color QCD},''
  \href{http://dx.doi.org/10.1103/PhysRevD.90.114506}{{\em Phys. Rev.}
  {\bfseries D90} no.~11, (2014) 114506},
\href{http://arxiv.org/abs/1406.4116}{{\ttfamily arXiv:1406.4116 [hep-lat]}}.

\bibitem{Asano:2014wra}
M.~Asano and R.~Kitano, ``{Partially Composite Dark Matter},''
  \href{http://dx.doi.org/10.1007/JHEP09(2014)171}{{\em JHEP} {\bfseries 09}
  (2014) 171},
\href{http://arxiv.org/abs/1406.6374}{{\ttfamily arXiv:1406.6374 [hep-ph]}}.

\bibitem{Brod:2014loa}
J.~Brod, J.~Drobnak, A.~L. Kagan, E.~Stamou, and J.~Zupan, ``{Stealth QCD-like
  strong interactions and the $t \bar {t}$ asymmetry},''
  \href{http://dx.doi.org/10.1103/PhysRevD.91.095009}{{\em Phys. Rev.}
  {\bfseries D91} no.~9, (2015) 095009},
\href{http://arxiv.org/abs/1407.8188}{{\ttfamily arXiv:1407.8188 [hep-ph]}}.

\bibitem{Antipin:2014qva}
O.~Antipin, M.~Redi, and A.~Strumia, ``{Dynamical generation of the weak and
  Dark Matter scales from strong interactions},''
  \href{http://dx.doi.org/10.1007/JHEP01(2015)157}{{\em JHEP} {\bfseries 01}
  (2015) 157},
\href{http://arxiv.org/abs/1410.1817}{{\ttfamily arXiv:1410.1817 [hep-ph]}}.

\bibitem{Hardy:2014mqa}
E.~Hardy, R.~Lasenby, J.~March-Russell, and S.~M. West, ``{Big Bang Synthesis
  of Nuclear Dark Matter},''
  \href{http://dx.doi.org/10.1007/JHEP06(2015)011}{{\em JHEP} {\bfseries 06}
  (2015) 011},
\href{http://arxiv.org/abs/1411.3739}{{\ttfamily arXiv:1411.3739 [hep-ph]}}.

\bibitem{Appelquist:2015yfa}
{\bfseries Lattice Strong Dynamics (LSD)} Collaboration, T.~Appelquist {\em
  et~al.}, ``{Stealth Dark Matter: Dark scalar baryons through the Higgs
  portal},'' \href{http://dx.doi.org/10.1103/PhysRevD.92.075030}{{\em Phys.
  Rev.} {\bfseries D92} no.~7, (2015) 075030},
\href{http://arxiv.org/abs/1503.04203}{{\ttfamily arXiv:1503.04203 [hep-ph]}}.

\bibitem{Appelquist:2015zfa}
{\bfseries Lattice Strong Dynamics (LSD)} Collaboration, T.~Appelquist {\em
  et~al.}, ``{Detecting Stealth Dark Matter Directly through Electromagnetic
  Polarizability},''
  \href{http://dx.doi.org/10.1103/PhysRevLett.115.171803}{{\em Phys. Rev.
  Lett.} {\bfseries 115} no.~17, (2015) 171803},
\href{http://arxiv.org/abs/1503.04205}{{\ttfamily arXiv:1503.04205 [hep-ph]}}.

\bibitem{Antipin:2015xia}
O.~Antipin, M.~Redi, A.~Strumia, and E.~Vigiani, ``{Accidental Composite Dark
  Matter},'' \href{http://dx.doi.org/10.1007/JHEP07(2015)039}{{\em JHEP}
  {\bfseries 07} (2015) 039},
\href{http://arxiv.org/abs/1503.08749}{{\ttfamily arXiv:1503.08749 [hep-ph]}}.

\bibitem{Hardy:2015boa}
E.~Hardy, R.~Lasenby, J.~March-Russell, and S.~M. West, ``{Signatures of Large
  Composite Dark Matter States},''
  \href{http://dx.doi.org/10.1007/JHEP07(2015)133}{{\em JHEP} {\bfseries 07}
  (2015) 133},
\href{http://arxiv.org/abs/1504.05419}{{\ttfamily arXiv:1504.05419 [hep-ph]}}.

\bibitem{Co:2016akw}
R.~T. Co, K.~Harigaya, and Y.~Nomura, ``{Chiral Dark Sector},''
  \href{http://dx.doi.org/10.1103/PhysRevLett.118.101801}{{\em Phys. Rev.
  Lett.} {\bfseries 118} no.~10, (2017) 101801},
\href{http://arxiv.org/abs/1610.03848}{{\ttfamily arXiv:1610.03848 [hep-ph]}}.

\bibitem{Dienes:2016vei}
K.~R. Dienes, F.~Huang, S.~Su, and B.~Thomas, ``{Dynamical Dark Matter from
  Strongly-Coupled Dark Sectors},''
  \href{http://dx.doi.org/10.1103/PhysRevD.95.043526}{{\em Phys. Rev.}
  {\bfseries D95} no.~4, (2017) 043526},
\href{http://arxiv.org/abs/1610.04112}{{\ttfamily arXiv:1610.04112 [hep-ph]}}.

\bibitem{Ishida:2016fbp}
H.~Ishida, S.~Matsuzaki, and Y.~Yamaguchi, ``{Bosonic-Seesaw Portal Dark
  Matter},'' \href{http://dx.doi.org/10.1093/ptep/ptx132}{{\em PTEP} {\bfseries
  2017} no.~10, (2017) 103B01},
\href{http://arxiv.org/abs/1610.07137}{{\ttfamily arXiv:1610.07137 [hep-ph]}}.

\bibitem{Lonsdale:2017mzg}
S.~J. Lonsdale, M.~Schroor, and R.~R. Volkas, ``{Asymmetric Dark Matter and the
  hadronic spectra of hidden QCD},''
  \href{http://dx.doi.org/10.1103/PhysRevD.96.055027}{{\em Phys. Rev.}
  {\bfseries D96} no.~5, (2017) 055027},
\href{http://arxiv.org/abs/1704.05213}{{\ttfamily arXiv:1704.05213 [hep-ph]}}.

\bibitem{Berryman:2017twh}
J.~M. Berryman, A.~de~Gouv{\^e}a, K.~J. Kelly, and Y.~Zhang, ``{Dark Matter and
  Neutrino Mass from the Smallest Non-Abelian Chiral Dark Sector},''
  \href{http://dx.doi.org/10.1103/PhysRevD.96.075010}{{\em Phys. Rev.}
  {\bfseries D96} no.~7, (2017) 075010},
\href{http://arxiv.org/abs/1706.02722}{{\ttfamily arXiv:1706.02722 [hep-ph]}}.

\bibitem{Gresham:2017zqi}
M.~I. Gresham, H.~K. Lou, and K.~M. Zurek, ``{Nuclear Structure of Bound States
  of Asymmetric Dark Matter},''
  \href{http://dx.doi.org/10.1103/PhysRevD.96.096012}{{\em Phys. Rev.}
  {\bfseries D96} no.~9, (2017) 096012},
\href{http://arxiv.org/abs/1707.02313}{{\ttfamily arXiv:1707.02313 [hep-ph]}}.

\bibitem{Gresham:2017cvl}
M.~I. Gresham, H.~K. Lou, and K.~M. Zurek, ``{Early Universe synthesis of
  asymmetric dark matter nuggets},''
  \href{http://dx.doi.org/10.1103/PhysRevD.97.036003}{{\em Phys. Rev.}
  {\bfseries D97} no.~3, (2018) 036003},
\href{http://arxiv.org/abs/1707.02316}{{\ttfamily arXiv:1707.02316 [hep-ph]}}.

\bibitem{Mitridate:2017oky}
A.~Mitridate, M.~Redi, J.~Smirnov, and A.~Strumia, ``{Dark Matter as a weakly
  coupled Dark Baryon},'' \href{http://dx.doi.org/10.1007/JHEP10(2017)210}{{\em
  JHEP} {\bfseries 10} (2017) 210},
\href{http://arxiv.org/abs/1707.05380}{{\ttfamily arXiv:1707.05380 [hep-ph]}}.

\bibitem{Gresham:2018anj}
M.~I. Gresham, H.~K. Lou, and K.~M. Zurek, ``{Astrophysical Signatures of
  Asymmetric Dark Matter Bound States},''
  \href{http://dx.doi.org/10.1103/PhysRevD.98.096001}{{\em Phys. Rev.}
  {\bfseries D98} no.~9, (2018) 096001},
\href{http://arxiv.org/abs/1805.04512}{{\ttfamily arXiv:1805.04512 [hep-ph]}}.

\bibitem{Ibe:2018juk}
M.~Ibe, A.~Kamada, S.~Kobayashi, and W.~Nakano, ``{Composite Asymmetric Dark
  Matter with a Dark Photon Portal},''
  \href{http://dx.doi.org/10.1007/JHEP11(2018)203}{{\em JHEP} {\bfseries 11}
  (2018) 203}, \href{http://arxiv.org/abs/1805.06876}{{\ttfamily
  arXiv:1805.06876 [hep-ph]}}.

\bibitem{Braaten:2018xuw}
E.~Braaten, D.~Kang, and R.~Laha, ``{Production of dark-matter bound states in
  the early universe by three-body recombination},''
  \href{http://dx.doi.org/10.1007/JHEP11(2018)084}{{\em JHEP} {\bfseries 11}
  (2018) 084},
\href{http://arxiv.org/abs/1806.00609}{{\ttfamily arXiv:1806.00609 [hep-ph]}}.

\bibitem{Francis:2018xjd}
A.~Francis, R.~J. Hudspith, R.~Lewis, and S.~Tulin, ``{Dark Matter from Strong
  Dynamics: The Minimal Theory of Dark Baryons},''
\href{http://arxiv.org/abs/1809.09117}{{\ttfamily arXiv:1809.09117 [hep-ph]}}.

\bibitem{Bai:2018dxf}
Y.~Bai, A.~J. Long, and S.~Lu, ``{Dark Quark Nuggets},''
  \href{http://dx.doi.org/10.1103/PhysRevD.99.055047}{{\em Phys. Rev. D}
  {\bfseries 99} no.~5, (2019) 055047},
  \href{http://arxiv.org/abs/1810.04360}{{\ttfamily arXiv:1810.04360
  [hep-ph]}}.

\bibitem{Chu:2018faw}
X.~Chu, C.~Garcia-Cely, and H.~Murayama, ``{Finite-size dark matter and its
  effect on small-scale structure},''
  \href{http://dx.doi.org/10.1103/PhysRevLett.124.041101}{{\em Phys. Rev.
  Lett.} {\bfseries 124} no.~4, (2020) 041101},
  \href{http://arxiv.org/abs/1901.00075}{{\ttfamily arXiv:1901.00075
  [hep-ph]}}.

\bibitem{Hall:2019rld}
E.~Hall, T.~Konstandin, R.~McGehee, and H.~Murayama, ``{Asymmetric Matters from
  a Dark First-Order Phase Transition},''
  \href{http://arxiv.org/abs/1911.12342}{{\ttfamily arXiv:1911.12342
  [hep-ph]}}.

\bibitem{Tsai:2020vpi}
Y.-D. Tsai, R.~McGehee, and H.~Murayama, ``{Resonant Self-Interacting Dark
  Matter from Dark QCD},''
  \href{http://dx.doi.org/10.1103/PhysRevLett.128.172001}{{\em Phys. Rev.
  Lett.} {\bfseries 128} no.~17, (2022) 172001},
  \href{http://arxiv.org/abs/2008.08608}{{\ttfamily arXiv:2008.08608
  [hep-ph]}}.

\bibitem{Asadi:2021yml}
P.~Asadi, E.~D. Kramer, E.~Kuflik, G.~W. Ridgway, T.~R. Slatyer, and
  J.~Smirnov, ``{Accidentally Asymmetric Dark Matter},''
  \href{http://dx.doi.org/10.1103/PhysRevLett.127.211101}{{\em Phys. Rev.
  Lett.} {\bfseries 127} no.~21, (2021) 211101},
  \href{http://arxiv.org/abs/2103.09822}{{\ttfamily arXiv:2103.09822
  [hep-ph]}}.

\bibitem{Asadi:2021pwo}
P.~Asadi, E.~D. Kramer, E.~Kuflik, G.~W. Ridgway, T.~R. Slatyer, and
  J.~Smirnov, ``{Thermal squeezeout of dark matter},''
  \href{http://dx.doi.org/10.1103/PhysRevD.104.095013}{{\em Phys. Rev. D}
  {\bfseries 104} no.~9, (2021) 095013},
  \href{http://arxiv.org/abs/2103.09827}{{\ttfamily arXiv:2103.09827
  [hep-ph]}}.

\bibitem{Zhang:2021orr}
M.~Zhang, ``{Leptophilic composite asymmetric dark matter and its detection},''
  \href{http://dx.doi.org/10.1103/PhysRevD.104.055008}{{\em Phys. Rev. D}
  {\bfseries 104} no.~5, (2021) 055008},
  \href{http://arxiv.org/abs/2104.06988}{{\ttfamily arXiv:2104.06988
  [hep-ph]}}.

\bibitem{Bottaro:2021aal}
S.~Bottaro, M.~Costa, and O.~Popov, ``{Asymmetric accidental composite dark
  matter},'' \href{http://dx.doi.org/10.1007/JHEP11(2021)055}{{\em JHEP}
  {\bfseries 11} (2021) 055}, \href{http://arxiv.org/abs/2104.14244}{{\ttfamily
  arXiv:2104.14244 [hep-ph]}}.

\bibitem{Ibe:2021gil}
M.~Ibe, S.~Kobayashi, and K.~Watanabe, ``{Chiral composite asymmetric dark
  matter},'' \href{http://dx.doi.org/10.1007/JHEP07(2021)220}{{\em JHEP}
  {\bfseries 07} (2021) 220}, \href{http://arxiv.org/abs/2105.07642}{{\ttfamily
  arXiv:2105.07642 [hep-ph]}}.

\bibitem{Hall:2021zsk}
E.~Hall, R.~McGehee, H.~Murayama, and B.~Suter, ``{Asymmetric Dark Matter May
  Not Be Light},'' \href{http://arxiv.org/abs/2107.03398}{{\ttfamily
  arXiv:2107.03398 [hep-ph]}}.

\bibitem{Asadi:2022vkc}
P.~Asadi, E.~D. Kramer, E.~Kuflik, T.~R. Slatyer, and J.~Smirnov, ``{Glueballs
  in a thermal squeezeout model},''
  \href{http://dx.doi.org/10.1007/JHEP07(2022)006}{{\em JHEP} {\bfseries 07}
  (2022) 006}, \href{http://arxiv.org/abs/2203.15813}{{\ttfamily
  arXiv:2203.15813 [hep-ph]}}.

\bibitem{Kobzarev:1966qya}
I.~{\relax Yu}. Kobzarev, L.~B. Okun, and I.~{\relax Ya}. Pomeranchuk, ``{On
  the possibility of experimental observation of mirror particles},'' {\em Sov.
  J. Nucl. Phys.} {\bfseries 3} no.~6, (1966) 837--841.
[Yad. Fiz.3,1154(1966)].

\bibitem{Blinnikov:1983gh}
S.~I. Blinnikov and M.~Khlopov, ``{Possible astronomical effects of mirror
  particles},'' {\em Sov. Astron.} {\bfseries 27} (1983) 371--375.

\bibitem{Kolb:1985bf}
E.~W. Kolb, D.~Seckel, and M.~S. Turner, ``{The shadow world of superstring
  theories},''
\href{http://dx.doi.org/10.1038/314415a0}{{\em Nature} {\bfseries 314} (1985)
  415--419}.

\bibitem{Khlopov:1989fj}
M.~Y. Khlopov, G.~M. Beskin, N.~E. Bochkarev, L.~A. Pustylnik, and S.~A.
  Pustylnik, ``{Observational Physics of Mirror World},'' {\em Sov. Astron.}
  {\bfseries 35} (1991) 21.

\bibitem{Hodges:1993yb}
H.~M. Hodges, ``{Mirror baryons as the dark matter},''
\href{http://dx.doi.org/10.1103/PhysRevD.47.456}{{\em Phys. Rev.} {\bfseries
  D47} (1993) 456--459}.

\bibitem{Foot:2004pa}
R.~Foot, ``{Mirror matter-type dark matter},''
  \href{http://dx.doi.org/10.1142/S0218271804006449}{{\em Int. J. Mod. Phys.}
  {\bfseries D13} (2004) 2161--2192},
\href{http://arxiv.org/abs/astro-ph/0407623}{{\ttfamily arXiv:astro-ph/0407623
  [astro-ph]}}.

\bibitem{Kribs:2016cew}
G.~D. Kribs and E.~T. Neil, ``{Review of strongly-coupled composite dark matter
  models and lattice simulations},''
  \href{http://dx.doi.org/10.1142/S0217751X16430041}{{\em Int. J. Mod. Phys.}
  {\bfseries A31} no.~22, (2016) 1643004},
\href{http://arxiv.org/abs/1604.04627}{{\ttfamily arXiv:1604.04627 [hep-ph]}}.

\bibitem{Beylin:2020bsz}
V.~Beylin, M.~Khlopov, V.~Kuksa, and N.~Volchanskiy, ``{New physics of strong
  interaction and Dark Universe},''
  \href{http://dx.doi.org/10.3390/universe6110196}{{\em Universe} {\bfseries 6}
  no.~11, (2020) 196}, \href{http://arxiv.org/abs/2010.13678}{{\ttfamily
  arXiv:2010.13678 [hep-ph]}}.

\bibitem{Hochberg:2014dra}
Y.~Hochberg, E.~Kuflik, T.~Volansky, and J.~G. Wacker, ``{Mechanism for Thermal
  Relic Dark Matter of Strongly Interacting Massive Particles},''
  \href{http://dx.doi.org/10.1103/PhysRevLett.113.171301}{{\em Phys. Rev.
  Lett.} {\bfseries 113} (2014) 171301},
  \href{http://arxiv.org/abs/1402.5143}{{\ttfamily arXiv:1402.5143 [hep-ph]}}.

\bibitem{Hochberg:2014kqa}
Y.~Hochberg, E.~Kuflik, H.~Murayama, T.~Volansky, and J.~G. Wacker, ``{Model
  for Thermal Relic Dark Matter of Strongly Interacting Massive Particles},''
  \href{http://dx.doi.org/10.1103/PhysRevLett.115.021301}{{\em Phys. Rev.
  Lett.} {\bfseries 115} no.~2, (2015) 021301},
  \href{http://arxiv.org/abs/1411.3727}{{\ttfamily arXiv:1411.3727 [hep-ph]}}.

\bibitem{Wess:1971yu}
J.~Wess and B.~Zumino, ``{Consequences of anomalous Ward identities},''
  \href{http://dx.doi.org/10.1016/0370-2693(71)90582-X}{{\em Phys. Lett. B}
  {\bfseries 37} (1971) 95--97}.

\bibitem{Witten:1983tw}
E.~Witten, ``{Global Aspects of Current Algebra},''
  \href{http://dx.doi.org/10.1016/0550-3213(83)90063-9}{{\em Nucl. Phys. B}
  {\bfseries 223} (1983) 422--432}.

\bibitem{Spergel:1999mh}
D.~N. Spergel and P.~J. Steinhardt, ``{Observational Evidence for
  Self-Interacting Cold Dark Matter},''
  \href{http://dx.doi.org/10.1103/PhysRevLett.84.3760}{{\em Phys. Rev. Lett.}
  {\bfseries 84} (2000) 3760--3763},
  \href{http://arxiv.org/abs/astro-ph/9909386}{{\ttfamily
  arXiv:astro-ph/9909386}}.

\bibitem{Tulin:2017ara}
S.~Tulin and H.-B. Yu, ``{Dark Matter Self-interactions and Small Scale
  Structure},'' \href{http://dx.doi.org/10.1016/j.physrep.2017.11.004}{{\em
  Phys. Rept.} {\bfseries 730} (2018) 1--57},
  \href{http://arxiv.org/abs/1705.02358}{{\ttfamily arXiv:1705.02358
  [hep-ph]}}.

\bibitem{Choi:2018iit}
S.-M. Choi, H.~M. Lee, P.~Ko, and A.~Natale, ``{Resolving phenomenological
  problems with strongly-interacting-massive-particle models with dark vector
  resonances},'' \href{http://dx.doi.org/10.1103/PhysRevD.98.015034}{{\em Phys.
  Rev. D} {\bfseries 98} no.~1, (2018) 015034},
  \href{http://arxiv.org/abs/1801.07726}{{\ttfamily arXiv:1801.07726
  [hep-ph]}}.

\bibitem{Berlin:2018tvf}
A.~Berlin, N.~Blinov, S.~Gori, P.~Schuster, and N.~Toro, ``{Cosmology and
  Accelerator Tests of Strongly Interacting Dark Matter},''
  \href{http://dx.doi.org/10.1103/PhysRevD.97.055033}{{\em Phys. Rev. D}
  {\bfseries 97} no.~5, (2018) 055033},
  \href{http://arxiv.org/abs/1801.05805}{{\ttfamily arXiv:1801.05805
  [hep-ph]}}.

\bibitem{Kaplinghat:2015aga}
M.~Kaplinghat, S.~Tulin, and H.-B. Yu, ``{Dark Matter Halos as Particle
  Colliders: Unified Solution to Small-Scale Structure Puzzles from Dwarfs to
  Clusters},'' \href{http://dx.doi.org/10.1103/PhysRevLett.116.041302}{{\em
  Phys. Rev. Lett.} {\bfseries 116} no.~4, (2016) 041302},
  \href{http://arxiv.org/abs/1508.03339}{{\ttfamily arXiv:1508.03339
  [astro-ph.CO]}}.

\bibitem{Bjorken:1988as}
J.~D. Bjorken, S.~Ecklund, W.~R. Nelson, A.~Abashian, C.~Church, B.~Lu, L.~W.
  Mo, T.~A. Nunamaker, and P.~Rassmann, ``{Search for Neutral Metastable
  Penetrating Particles Produced in the SLAC Beam Dump},''
  \href{http://dx.doi.org/10.1103/PhysRevD.38.3375}{{\em Phys. Rev. D}
  {\bfseries 38} (1988) 3375}.

\bibitem{Celentano:2014wya}
{\bfseries HPS} Collaboration, A.~Celentano, ``{The Heavy Photon Search
  experiment at Jefferson Laboratory},''
  \href{http://dx.doi.org/10.1088/1742-6596/556/1/012064}{{\em J. Phys. Conf.
  Ser.} {\bfseries 556} no.~1, (2014) 012064},
  \href{http://arxiv.org/abs/1505.02025}{{\ttfamily arXiv:1505.02025
  [physics.ins-det]}}.

\bibitem{Izaguirre:2014bca}
E.~Izaguirre, G.~Krnjaic, P.~Schuster, and N.~Toro, ``{Testing GeV-Scale Dark
  Matter with Fixed-Target Missing Momentum Experiments},''
  \href{http://dx.doi.org/10.1103/PhysRevD.91.094026}{{\em Phys. Rev. D}
  {\bfseries 91} no.~9, (2015) 094026},
  \href{http://arxiv.org/abs/1411.1404}{{\ttfamily arXiv:1411.1404 [hep-ph]}}.

\bibitem{SeaQuest:2017kjt}
{\bfseries SeaQuest} Collaboration, C.~A. Aidala {\em et~al.}, ``{The SeaQuest
  Spectrometer at Fermilab},''
  \href{http://dx.doi.org/10.1016/j.nima.2019.03.039}{{\em Nucl. Instrum. Meth.
  A} {\bfseries 930} (2019) 49--63},
  \href{http://arxiv.org/abs/1706.09990}{{\ttfamily arXiv:1706.09990
  [physics.ins-det]}}.

\bibitem{Feng:2017uoz}
J.~L. Feng, I.~Galon, F.~Kling, and S.~Trojanowski, ``{ForwArd Search
  ExpeRiment at the LHC},''
  \href{http://dx.doi.org/10.1103/PhysRevD.97.035001}{{\em Phys. Rev. D}
  {\bfseries 97} no.~3, (2018) 035001},
  \href{http://arxiv.org/abs/1708.09389}{{\ttfamily arXiv:1708.09389
  [hep-ph]}}.

\bibitem{Cerci:2021nlb}
S.~Cerci {\em et~al.}, ``{FACET: A new long-lived particle detector in the very
  forward region of the CMS experiment},''
  \href{http://dx.doi.org/10.1007/JHEP06(2022)110}{{\em JHEP} {\bfseries 2022}
  no.~06, (2022) 110}, \href{http://arxiv.org/abs/2201.00019}{{\ttfamily
  arXiv:2201.00019 [hep-ex]}}.

\bibitem{Gligorov:2017nwh}
V.~V. Gligorov, S.~Knapen, M.~Papucci, and D.~J. Robinson, ``{Searching for
  Long-lived Particles: A Compact Detector for Exotics at LHCb},''
  \href{http://dx.doi.org/10.1103/PhysRevD.97.015023}{{\em Phys. Rev. D}
  {\bfseries 97} no.~1, (2018) 015023},
  \href{http://arxiv.org/abs/1708.09395}{{\ttfamily arXiv:1708.09395
  [hep-ph]}}.

\bibitem{Chou:2016lxi}
J.~P. Chou, D.~Curtin, and H.~J. Lubatti, ``{New Detectors to Explore the
  Lifetime Frontier},''
  \href{http://dx.doi.org/10.1016/j.physletb.2017.01.043}{{\em Phys. Lett. B}
  {\bfseries 767} (2017) 29--36},
  \href{http://arxiv.org/abs/1606.06298}{{\ttfamily arXiv:1606.06298
  [hep-ph]}}.

\bibitem{Scherer:2002tk}
S.~Scherer, ``{Introduction to chiral perturbation theory},'' {\em Adv. Nucl.
  Phys.} {\bfseries 27} (2003) 277,
  \href{http://arxiv.org/abs/hep-ph/0210398}{{\ttfamily arXiv:hep-ph/0210398}}.

\bibitem{Ecker:1989yg}
G.~Ecker, J.~Gasser, H.~Leutwyler, A.~Pich, and E.~de~Rafael, ``{Chiral
  Lagrangians for Massive Spin 1 Fields},''
  \href{http://dx.doi.org/10.1016/0370-2693(89)91627-4}{{\em Phys. Lett. B}
  {\bfseries 223} (1989) 425--432}.

\bibitem{Schwinger:1967tc}
J.~S. Schwinger, ``{Chiral dynamics},''
  \href{http://dx.doi.org/10.1016/0370-2693(67)90277-8}{{\em Phys. Lett. B}
  {\bfseries 24} (1967) 473--476}.

\bibitem{Gasiorowicz:1969kn}
S.~Gasiorowicz and D.~A. Geffen, ``{Effective Lagrangians and field algebras
  with chiral symmetry},''
  \href{http://dx.doi.org/10.1103/RevModPhys.41.531}{{\em Rev. Mod. Phys.}
  {\bfseries 41} (1969) 531--573}.

\bibitem{Kaymakcalan:1983qq}
O.~Kaymakcalan, S.~Rajeev, and J.~Schechter, ``{Nonabelian Anomaly and Vector
  Meson Decays},'' \href{http://dx.doi.org/10.1103/PhysRevD.30.594}{{\em Phys.
  Rev. D} {\bfseries 30} (1984) 594}.

\bibitem{Meissner:1987ge}
U.~G. Meissner, ``{Low-Energy Hadron Physics from Effective Chiral Lagrangians
  with Vector Mesons},''
  \href{http://dx.doi.org/10.1016/0370-1573(88)90090-7}{{\em Phys. Rept.}
  {\bfseries 161} (1988) 213}.

\bibitem{Gasser:1983yg}
J.~Gasser and H.~Leutwyler, ``{Chiral Perturbation Theory to One Loop},''
  \href{http://dx.doi.org/10.1016/0003-4916(84)90242-2}{{\em Annals Phys.}
  {\bfseries 158} (1984) 142}.

\bibitem{Bando:1984ej}
M.~Bando, T.~Kugo, S.~Uehara, K.~Yamawaki, and T.~Yanagida, ``{Is rho Meson a
  Dynamical Gauge Boson of Hidden Local Symmetry?},''
  \href{http://dx.doi.org/10.1103/PhysRevLett.54.1215}{{\em Phys. Rev. Lett.}
  {\bfseries 54} (1985) 1215}.

\bibitem{Harada:2003jx}
M.~Harada and K.~Yamawaki, ``{Hidden local symmetry at loop: A New perspective
  of composite gauge boson and chiral phase transition},''
  \href{http://dx.doi.org/10.1016/S0370-1573(03)00139-X}{{\em Phys. Rept.}
  {\bfseries 381} (2003) 1--233},
  \href{http://arxiv.org/abs/hep-ph/0302103}{{\ttfamily arXiv:hep-ph/0302103}}.

\bibitem{Kawarabayashi:1966kd}
K.~Kawarabayashi and M.~Suzuki, ``{Partially conserved axial vector current and
  the decays of vector mesons},''
  \href{http://dx.doi.org/10.1103/PhysRevLett.16.255}{{\em Phys. Rev. Lett.}
  {\bfseries 16} (1966) 255}.

\bibitem{Riazuddin:1966sw}
Riazuddin and Fayyazuddin, ``{Algebra of current components and decay widths of
  rho and K* mesons},'' \href{http://dx.doi.org/10.1103/PhysRev.147.1071}{{\em
  Phys. Rev.} {\bfseries 147} (1966) 1071--1073}.

\bibitem{Hochberg:2015vrg}
Y.~Hochberg, E.~Kuflik, and H.~Murayama, ``{SIMP Spectroscopy},''
  \href{http://dx.doi.org/10.1007/JHEP05(2016)090}{{\em JHEP} {\bfseries 05}
  (2016) 090}, \href{http://arxiv.org/abs/1512.07917}{{\ttfamily
  arXiv:1512.07917 [hep-ph]}}.

\bibitem{Manohar:1983md}
A.~Manohar and H.~Georgi, ``{Chiral Quarks and the Nonrelativistic Quark
  Model},'' \href{http://dx.doi.org/10.1016/0550-3213(84)90231-1}{{\em Nucl.
  Phys. B} {\bfseries 234} (1984) 189--212}.

\bibitem{Georgi:1992dw}
H.~Georgi, ``{Generalized dimensional analysis},''
  \href{http://dx.doi.org/10.1016/0370-2693(93)91728-6}{{\em Phys. Lett. B}
  {\bfseries 298} (1993) 187--189},
  \href{http://arxiv.org/abs/hep-ph/9207278}{{\ttfamily arXiv:hep-ph/9207278}}.

\bibitem{Katz:2020ywn}
A.~Katz, E.~Salvioni, and B.~Shakya, ``{Split SIMPs with Decays},''
  \href{http://dx.doi.org/10.1007/JHEP10(2020)049}{{\em JHEP} {\bfseries 10}
  (2020) 049}, \href{http://arxiv.org/abs/2006.15148}{{\ttfamily
  arXiv:2006.15148 [hep-ph]}}.

\bibitem{Clowe:2003tk}
D.~Clowe, A.~Gonzalez, and M.~Markevitch, ``{Weak lensing mass reconstruction
  of the interacting cluster 1E0657-558: Direct evidence for the existence of
  dark matter},'' \href{http://dx.doi.org/10.1086/381970}{{\em Astrophys. J.}
  {\bfseries 604} (2004) 596--603},
  \href{http://arxiv.org/abs/astro-ph/0312273}{{\ttfamily
  arXiv:astro-ph/0312273}}.

\bibitem{Markevitch:2003at}
M.~Markevitch, A.~H. Gonzalez, D.~Clowe, A.~Vikhlinin, L.~David, W.~Forman,
  C.~Jones, S.~Murray, and W.~Tucker, ``{Direct constraints on the dark matter
  self-interaction cross-section from the merging galaxy cluster 1E0657-56},''
  \href{http://dx.doi.org/10.1086/383178}{{\em Astrophys. J.} {\bfseries 606}
  (2004) 819--824}, \href{http://arxiv.org/abs/astro-ph/0309303}{{\ttfamily
  arXiv:astro-ph/0309303}}.

\bibitem{Randall:2008ppe}
S.~W. Randall, M.~Markevitch, D.~Clowe, A.~H. Gonzalez, and M.~Bradac,
  ``{Constraints on the Self-Interaction Cross-Section of Dark Matter from
  Numerical Simulations of the Merging Galaxy Cluster 1E 0657-56},''
  \href{http://dx.doi.org/10.1086/587859}{{\em Astrophys. J.} {\bfseries 679}
  (2008) 1173--1180}, \href{http://arxiv.org/abs/0704.0261}{{\ttfamily
  arXiv:0704.0261 [astro-ph]}}.

\bibitem{tHooft:1973alw}
G.~'t~Hooft, ``{A Planar Diagram Theory for Strong Interactions},''
  \href{http://dx.doi.org/10.1016/0550-3213(74)90154-0}{{\em Nucl. Phys. B}
  {\bfseries 72} (1974) 461}.

\bibitem{tHooft:1974pnl}
G.~'t~Hooft, ``{A Two-Dimensional Model for Mesons},''
  \href{http://dx.doi.org/10.1016/0550-3213(74)90088-1}{{\em Nucl. Phys. B}
  {\bfseries 75} (1974) 461--470}.

\bibitem{Witten:1979kh}
E.~Witten, ``{Baryons in the 1/n Expansion},''
  \href{http://dx.doi.org/10.1016/0550-3213(79)90232-3}{{\em Nucl. Phys. B}
  {\bfseries 160} (1979) 57--115}.

\bibitem{Witten:1979vv}
E.~Witten, ``{Current Algebra Theorems for the U(1) Goldstone Boson},''
  \href{http://dx.doi.org/10.1016/0550-3213(79)90031-2}{{\em Nucl. Phys. B}
  {\bfseries 156} (1979) 269--283}.

\bibitem{Coleman:1980mx}
S.~R. Coleman and E.~Witten, ``{Chiral Symmetry Breakdown in Large N
  Chromodynamics},'' \href{http://dx.doi.org/10.1103/PhysRevLett.45.100}{{\em
  Phys. Rev. Lett.} {\bfseries 45} (1980) 100}.

\bibitem{Witten:1980sp}
E.~Witten, ``{Large N Chiral Dynamics},''
  \href{http://dx.doi.org/10.1016/0003-4916(80)90325-5}{{\em Annals Phys.}
  {\bfseries 128} (1980) 363}.

\bibitem{Lonsdale:2018xwd}
S.~J. Lonsdale and R.~R. Volkas, ``{Comprehensive asymmetric dark matter
  model},'' \href{http://dx.doi.org/10.1103/PhysRevD.97.103510}{{\em Phys.
  Rev.} {\bfseries D97} no.~10, (2018) 103510},
\href{http://arxiv.org/abs/1801.05561}{{\ttfamily arXiv:1801.05561 [hep-ph]}}.

\bibitem{Ibe:2018tex}
M.~Ibe, A.~Kamada, S.~Kobayashi, T.~Kuwahara, and W.~Nakano, ``{Ultraviolet
  Completion of a Composite Asymmetric Dark Matter Model with a Dark Photon
  Portal},'' \href{http://dx.doi.org/10.1007/JHEP03(2019)173}{{\em JHEP}
  {\bfseries 03} (2019) 173}, \href{http://arxiv.org/abs/1811.10232}{{\ttfamily
  arXiv:1811.10232 [hep-ph]}}.

\bibitem{Ibe:2019ena}
M.~Ibe, A.~Kamada, S.~Kobayashi, T.~Kuwahara, and W.~Nakano, ``{Baryon-Dark
  Matter Coincidence in Mirrored Unification},''
  \href{http://dx.doi.org/10.1103/PhysRevD.100.075022}{{\em Phys. Rev. D}
  {\bfseries 100} no.~7, (2019) 075022},
  \href{http://arxiv.org/abs/1907.03404}{{\ttfamily arXiv:1907.03404
  [hep-ph]}}.

\bibitem{Blennow:2012de}
M.~Blennow, E.~Fernandez-Martinez, O.~Mena, J.~Redondo, and P.~Serra,
  ``{Asymmetric Dark Matter and Dark Radiation},''
  \href{http://dx.doi.org/10.1088/1475-7516/2012/07/022}{{\em JCAP} {\bfseries
  07} (2012) 022}, \href{http://arxiv.org/abs/1203.5803}{{\ttfamily
  arXiv:1203.5803 [hep-ph]}}.

\bibitem{BaBar:2017tiz}
{\bfseries BaBar} Collaboration, J.~P. Lees {\em et~al.}, ``{Search for
  Invisible Decays of a Dark Photon Produced in ${e}^{+}{e}^{-}$ Collisions at
  BaBar},'' \href{http://dx.doi.org/10.1103/PhysRevLett.119.131804}{{\em Phys.
  Rev. Lett.} {\bfseries 119} no.~13, (2017) 131804},
  \href{http://arxiv.org/abs/1702.03327}{{\ttfamily arXiv:1702.03327
  [hep-ex]}}.

\bibitem{Belle-II:2018jsg}
{\bfseries Belle-II} Collaboration, W.~Altmannshofer {\em et~al.}, ``{The Belle
  II Physics Book},'' \href{http://dx.doi.org/10.1093/ptep/ptz106}{{\em PTEP}
  {\bfseries 2019} no.~12, (2019) 123C01},
  \href{http://arxiv.org/abs/1808.10567}{{\ttfamily arXiv:1808.10567
  [hep-ex]}}. [Erratum: PTEP 2020, 029201 (2020)].

\bibitem{Banerjee:2019pds}
D.~Banerjee {\em et~al.}, ``{Dark matter search in missing energy events with
  NA64},'' \href{http://dx.doi.org/10.1103/PhysRevLett.123.121801}{{\em Phys.
  Rev. Lett.} {\bfseries 123} no.~12, (2019) 121801},
  \href{http://arxiv.org/abs/1906.00176}{{\ttfamily arXiv:1906.00176
  [hep-ex]}}.

\bibitem{LDMX:2018cma}
{\bfseries LDMX} Collaboration, T.~\r{A}kesson {\em et~al.}, ``{Light Dark
  Matter eXperiment (LDMX)},''
  \href{http://arxiv.org/abs/1808.05219}{{\ttfamily arXiv:1808.05219
  [hep-ex]}}.

\bibitem{Riordan:1987aw}
E.~M. Riordan {\em et~al.}, ``{A Search for Short Lived Axions in an Electron
  Beam Dump Experiment},''
  \href{http://dx.doi.org/10.1103/PhysRevLett.59.755}{{\em Phys. Rev. Lett.}
  {\bfseries 59} (1987) 755}.

\bibitem{Bross:1989mp}
A.~Bross, M.~Crisler, S.~H. Pordes, J.~Volk, S.~Errede, and J.~Wrbanek, ``{A
  Search for Shortlived Particles Produced in an Electron Beam Dump},''
  \href{http://dx.doi.org/10.1103/PhysRevLett.67.2942}{{\em Phys. Rev. Lett.}
  {\bfseries 67} (1991) 2942--2945}.

\bibitem{Davier:1989wz}
M.~Davier and H.~Nguyen~Ngoc, ``{An Unambiguous Search for a Light Higgs
  Boson},'' \href{http://dx.doi.org/10.1016/0370-2693(89)90174-3}{{\em Phys.
  Lett. B} {\bfseries 229} (1989) 150--155}.

\bibitem{Konaka:1986cb}
A.~Konaka {\em et~al.}, ``{Search for Neutral Particles in Electron Beam Dump
  Experiment},'' \href{http://dx.doi.org/10.1103/PhysRevLett.57.659}{{\em Phys.
  Rev. Lett.} {\bfseries 57} (1986) 659}.

\bibitem{NA64:2018lsq}
{\bfseries NA64} Collaboration, D.~Banerjee {\em et~al.}, ``{Search for a
  Hypothetical 16.7 MeV Gauge Boson and Dark Photons in the NA64 Experiment at
  CERN},'' \href{http://dx.doi.org/10.1103/PhysRevLett.120.231802}{{\em Phys.
  Rev. Lett.} {\bfseries 120} no.~23, (2018) 231802},
  \href{http://arxiv.org/abs/1803.07748}{{\ttfamily arXiv:1803.07748
  [hep-ex]}}.

\bibitem{CHARM:1985nku}
{\bfseries CHARM} Collaboration, F.~Bergsma {\em et~al.}, ``{A Search for
  Decays of Heavy Neutrinos in the Mass Range 0.5-{GeV} to 2.8-{GeV}},''
  \href{http://dx.doi.org/10.1016/0370-2693(86)91601-1}{{\em Phys. Lett. B}
  {\bfseries 166} (1986) 473--478}.

\bibitem{Gninenko:2012eq}
S.~N. Gninenko, ``{Constraints on sub-GeV hidden sector gauge bosons from a
  search for heavy neutrino decays},''
  \href{http://dx.doi.org/10.1016/j.physletb.2012.06.002}{{\em Phys. Lett. B}
  {\bfseries 713} (2012) 244--248},
  \href{http://arxiv.org/abs/1204.3583}{{\ttfamily arXiv:1204.3583 [hep-ph]}}.

\bibitem{NA482:2015wmo}
{\bfseries NA48/2} Collaboration, J.~R. Batley {\em et~al.}, ``{Search for the
  dark photon in $\pi^0$ decays},''
  \href{http://dx.doi.org/10.1016/j.physletb.2015.04.068}{{\em Phys. Lett. B}
  {\bfseries 746} (2015) 178--185},
  \href{http://arxiv.org/abs/1504.00607}{{\ttfamily arXiv:1504.00607
  [hep-ex]}}.

\bibitem{LSND:1997vqj}
{\bfseries LSND} Collaboration, C.~Athanassopoulos {\em et~al.}, ``{Evidence
  for muon-neutrino ---\ensuremath{>} electron-neutrino oscillations from pion
  decay in flight neutrinos},''
  \href{http://dx.doi.org/10.1103/PhysRevC.58.2489}{{\em Phys. Rev. C}
  {\bfseries 58} (1998) 2489--2511},
  \href{http://arxiv.org/abs/nucl-ex/9706006}{{\ttfamily
  arXiv:nucl-ex/9706006}}.

\bibitem{Batell:2009di}
B.~Batell, M.~Pospelov, and A.~Ritz, ``{Exploring Portals to a Hidden Sector
  Through Fixed Targets},''
  \href{http://dx.doi.org/10.1103/PhysRevD.80.095024}{{\em Phys. Rev. D}
  {\bfseries 80} (2009) 095024},
  \href{http://arxiv.org/abs/0906.5614}{{\ttfamily arXiv:0906.5614 [hep-ph]}}.

\bibitem{Blumlein:2011mv}
J.~Bl\"umlein and J.~Brunner, ``{New Exclusion Limits for Dark Gauge Forces
  from Beam-Dump Data},''
  \href{http://dx.doi.org/10.1016/j.physletb.2011.05.046}{{\em Phys. Lett. B}
  {\bfseries 701} (2011) 155--159},
  \href{http://arxiv.org/abs/1104.2747}{{\ttfamily arXiv:1104.2747 [hep-ex]}}.

\bibitem{Blumlein:2013cua}
J.~Bl\"umlein and J.~Brunner, ``{New Exclusion Limits on Dark Gauge Forces from
  Proton Bremsstrahlung in Beam-Dump Data},''
  \href{http://dx.doi.org/10.1016/j.physletb.2014.02.029}{{\em Phys. Lett. B}
  {\bfseries 731} (2014) 320--326},
  \href{http://arxiv.org/abs/1311.3870}{{\ttfamily arXiv:1311.3870 [hep-ph]}}.

\bibitem{Gardner:2015wea}
S.~Gardner, R.~J. Holt, and A.~S. Tadepalli, ``{New Prospects in Fixed Target
  Searches for Dark Forces with the SeaQuest Experiment at Fermilab},''
  \href{http://dx.doi.org/10.1103/PhysRevD.93.115015}{{\em Phys. Rev. D}
  {\bfseries 93} no.~11, (2016) 115015},
  \href{http://arxiv.org/abs/1509.00050}{{\ttfamily arXiv:1509.00050
  [hep-ph]}}.

\bibitem{Berlin:2018pwi}
A.~Berlin, S.~Gori, P.~Schuster, and N.~Toro, ``{Dark Sectors at the Fermilab
  SeaQuest Experiment},''
  \href{http://dx.doi.org/10.1103/PhysRevD.98.035011}{{\em Phys. Rev. D}
  {\bfseries 98} no.~3, (2018) 035011},
  \href{http://arxiv.org/abs/1804.00661}{{\ttfamily arXiv:1804.00661
  [hep-ph]}}.

\bibitem{Apyan:2022tsd}
A.~Apyan {\em et~al.}, ``{DarkQuest: A dark sector upgrade to SpinQuest at the
  120 GeV Fermilab Main Injector},'' in {\em {2022 Snowmass Summer Study}}.
\newblock 3, 2022.
\newblock \href{http://arxiv.org/abs/2203.08322}{{\ttfamily arXiv:2203.08322
  [hep-ex]}}.

\bibitem{Zyla:2020zbs}
{\bfseries Particle Data Group} Collaboration, P.~Zyla {\em et~al.}, ``{Review
  of Particle Physics},'' \href{http://dx.doi.org/10.1093/ptep/ptaa104}{{\em
  PTEP} {\bfseries 2020} no.~8, (2020) 083C01}.

\bibitem{Poulin:2016anj}
V.~Poulin, J.~Lesgourgues, and P.~D. Serpico, ``{Cosmological constraints on
  exotic injection of electromagnetic energy},''
  \href{http://dx.doi.org/10.1088/1475-7516/2017/03/043}{{\em JCAP} {\bfseries
  03} (2017) 043}, \href{http://arxiv.org/abs/1610.10051}{{\ttfamily
  arXiv:1610.10051 [astro-ph.CO]}}.

\bibitem{Kamada:2021cow}
A.~Kamada and T.~Kuwahara, ``{LHC lifetime frontier and visible decay searches
  in composite asymmetric dark matter models},''
  \href{http://dx.doi.org/10.1007/JHEP03(2022)176}{{\em JHEP} {\bfseries 03}
  (2022) 176}, \href{http://arxiv.org/abs/2112.01202}{{\ttfamily
  arXiv:2112.01202 [hep-ph]}}.

\bibitem{FASER:2018eoc}
{\bfseries FASER} Collaboration, A.~Ariga {\em et~al.},
  ``{FASER\textquoteright{}s physics reach for long-lived particles},''
  \href{http://dx.doi.org/10.1103/PhysRevD.99.095011}{{\em Phys. Rev. D}
  {\bfseries 99} no.~9, (2019) 095011},
  \href{http://arxiv.org/abs/1811.12522}{{\ttfamily arXiv:1811.12522
  [hep-ph]}}.

\bibitem{Berlin:2018jbm}
A.~Berlin and F.~Kling, ``{Inelastic Dark Matter at the LHC Lifetime Frontier:
  ATLAS, CMS, LHCb, CODEX-b, FASER, and MATHUSLA},''
  \href{http://dx.doi.org/10.1103/PhysRevD.99.015021}{{\em Phys. Rev. D}
  {\bfseries 99} no.~1, (2019) 015021},
  \href{http://arxiv.org/abs/1810.01879}{{\ttfamily arXiv:1810.01879
  [hep-ph]}}.

\bibitem{MoriondFASER:2023}
C.~Gwilliam, ``{First Physics Results of the FASER Experiment},'' 2023.
\newblock
  \url{https://moriond.in2p3.fr/QCD/2023/WednesdayMorning/Gwilliam.pdf}.

\bibitem{Petersen:2023hgm}
{\bfseries FASER} Collaboration, B.~Petersen, ``{First Physics Results from the
  FASER Experiment},''
\newblock 5, 2023.
\newblock \href{http://arxiv.org/abs/2305.08665}{{\ttfamily arXiv:2305.08665
  [hep-ex]}}.

\bibitem{Kling:2021fwx}
F.~Kling and S.~Trojanowski, ``{Forward experiment sensitivity estimator for
  the LHC and future hadron colliders},''
  \href{http://dx.doi.org/10.1103/PhysRevD.104.035012}{{\em Phys. Rev. D}
  {\bfseries 104} no.~3, (2021) 035012},
  \href{http://arxiv.org/abs/2105.07077}{{\ttfamily arXiv:2105.07077
  [hep-ph]}}.

\bibitem{Pierog:2013ria}
T.~Pierog, I.~Karpenko, J.~M. Katzy, E.~Yatsenko, and K.~Werner, ``{EPOS LHC:
  Test of collective hadronization with data measured at the CERN Large Hadron
  Collider},'' \href{http://dx.doi.org/10.1103/PhysRevC.92.034906}{{\em Phys.
  Rev. C} {\bfseries 92} no.~3, (2015) 034906},
  \href{http://arxiv.org/abs/1306.0121}{{\ttfamily arXiv:1306.0121 [hep-ph]}}.

\bibitem{Alloul:2013bka}
A.~Alloul, N.~D. Christensen, C.~Degrande, C.~Duhr, and B.~Fuks, ``{FeynRules
  2.0 - A complete toolbox for tree-level phenomenology},''
  \href{http://dx.doi.org/10.1016/j.cpc.2014.04.012}{{\em Comput. Phys.
  Commun.} {\bfseries 185} (2014) 2250--2300},
  \href{http://arxiv.org/abs/1310.1921}{{\ttfamily arXiv:1310.1921 [hep-ph]}}.

\bibitem{Alwall:2014hca}
J.~Alwall, R.~Frederix, S.~Frixione, V.~Hirschi, F.~Maltoni, O.~Mattelaer,
  H.~S. Shao, T.~Stelzer, P.~Torrielli, and M.~Zaro, ``{The automated
  computation of tree-level and next-to-leading order differential cross
  sections, and their matching to parton shower simulations},''
  \href{http://dx.doi.org/10.1007/JHEP07(2014)079}{{\em JHEP} {\bfseries 07}
  (2014) 079}, \href{http://arxiv.org/abs/1405.0301}{{\ttfamily arXiv:1405.0301
  [hep-ph]}}.

\bibitem{Sjostrand:2014zea}
T.~Sj\"ostrand, S.~Ask, J.~R. Christiansen, R.~Corke, N.~Desai, P.~Ilten,
  S.~Mrenna, S.~Prestel, C.~O. Rasmussen, and P.~Z. Skands, ``{An introduction
  to PYTHIA 8.2},'' \href{http://dx.doi.org/10.1016/j.cpc.2015.01.024}{{\em
  Comput. Phys. Commun.} {\bfseries 191} (2015) 159--177},
  \href{http://arxiv.org/abs/1410.3012}{{\ttfamily arXiv:1410.3012 [hep-ph]}}.

\bibitem{NNPDF:2017mvq}
{\bfseries NNPDF} Collaboration, R.~D. Ball {\em et~al.}, ``{Parton
  distributions from high-precision collider data},''
  \href{http://dx.doi.org/10.1140/epjc/s10052-017-5199-5}{{\em Eur. Phys. J. C}
  {\bfseries 77} no.~10, (2017) 663},
  \href{http://arxiv.org/abs/1706.00428}{{\ttfamily arXiv:1706.00428
  [hep-ph]}}.

\bibitem{Kim:1972gw}
K.~J. Kim and Y.-S. Tsai, ``{AN IMPROVED WEIZSACKER-WILLIAMS METHOD AND
  PHOTOPRODUCTION OF LEPTON PAIRS},''
  \href{http://dx.doi.org/10.1016/0370-2693(72)90622-3}{{\em Phys. Lett. B}
  {\bfseries 40} (1972) 665--670}.

\bibitem{Kim:1973he}
K.~J. Kim and Y.-S. Tsai, ``{IMPROVED WEIZSACKER-WILLIAMS METHOD AND ITS
  APPLICATION TO LEPTON AND W BOSON PAIR PRODUCTION},''
  \href{http://dx.doi.org/10.1103/PhysRevD.8.3109}{{\em Phys. Rev. D}
  {\bfseries 8} (1973) 3109}.

\bibitem{Foroughi-Abari:2021zbm}
S.~Foroughi-Abari and A.~Ritz, ``{Dark sector production via proton
  bremsstrahlung},'' \href{http://dx.doi.org/10.1103/PhysRevD.105.095045}{{\em
  Phys. Rev. D} {\bfseries 105} no.~9, (2022) 095045},
  \href{http://arxiv.org/abs/2108.05900}{{\ttfamily arXiv:2108.05900
  [hep-ph]}}.

\bibitem{Bjorken:2009mm}
J.~D. Bjorken, R.~Essig, P.~Schuster, and N.~Toro, ``{New Fixed-Target
  Experiments to Search for Dark Gauge Forces},''
  \href{http://dx.doi.org/10.1103/PhysRevD.80.075018}{{\em Phys. Rev. D}
  {\bfseries 80} (2009) 075018},
  \href{http://arxiv.org/abs/0906.0580}{{\ttfamily arXiv:0906.0580 [hep-ph]}}.

\bibitem{Andreas:2012mt}
S.~Andreas, C.~Niebuhr, and A.~Ringwald, ``{New Limits on Hidden Photons from
  Past Electron Beam Dumps},''
  \href{http://dx.doi.org/10.1103/PhysRevD.86.095019}{{\em Phys. Rev. D}
  {\bfseries 86} (2012) 095019},
  \href{http://arxiv.org/abs/1209.6083}{{\ttfamily arXiv:1209.6083 [hep-ph]}}.

\bibitem{deNiverville:2011it}
P.~deNiverville, M.~Pospelov, and A.~Ritz, ``{Observing a light dark matter
  beam with neutrino experiments},''
  \href{http://dx.doi.org/10.1103/PhysRevD.84.075020}{{\em Phys. Rev. D}
  {\bfseries 84} (2011) 075020},
  \href{http://arxiv.org/abs/1107.4580}{{\ttfamily arXiv:1107.4580 [hep-ph]}}.

\bibitem{Batell:2014mga}
B.~Batell, R.~Essig, and Z.~Surujon, ``{Strong Constraints on Sub-GeV Dark
  Sectors from SLAC Beam Dump E137},''
  \href{http://dx.doi.org/10.1103/PhysRevLett.113.171802}{{\em Phys. Rev.
  Lett.} {\bfseries 113} no.~17, (2014) 171802},
  \href{http://arxiv.org/abs/1406.2698}{{\ttfamily arXiv:1406.2698 [hep-ph]}}.

\bibitem{Aguilar-Arevalo:2017mqx}
{\bfseries MiniBooNE} Collaboration, A.~A. Aguilar-Arevalo {\em et~al.},
  ``{Dark Matter Search in a Proton Beam Dump with MiniBooNE},''
  \href{http://dx.doi.org/10.1103/PhysRevLett.118.221803}{{\em Phys. Rev.
  Lett.} {\bfseries 118} no.~22, (2017) 221803},
  \href{http://arxiv.org/abs/1702.02688}{{\ttfamily arXiv:1702.02688
  [hep-ex]}}.

\bibitem{BaBar:2014zli}
{\bfseries BaBar} Collaboration, J.~P. Lees {\em et~al.}, ``{Search for a Dark
  Photon in $e^+e^-$ Collisions at BaBar},''
  \href{http://dx.doi.org/10.1103/PhysRevLett.113.201801}{{\em Phys. Rev.
  Lett.} {\bfseries 113} no.~20, (2014) 201801},
  \href{http://arxiv.org/abs/1406.2980}{{\ttfamily arXiv:1406.2980 [hep-ex]}}.

\bibitem{LHCb:2019vmc}
{\bfseries LHCb} Collaboration, R.~Aaij {\em et~al.}, ``{Search for
  $A'\to\mu^+\mu^-$ Decays},''
  \href{http://dx.doi.org/10.1103/PhysRevLett.124.041801}{{\em Phys. Rev.
  Lett.} {\bfseries 124} no.~4, (2020) 041801},
  \href{http://arxiv.org/abs/1910.06926}{{\ttfamily arXiv:1910.06926
  [hep-ex]}}.

\bibitem{Doria:2018sfx}
L.~Doria, P.~Achenbach, M.~Christmann, A.~Denig, P.~G\"ulker, and H.~Merkel,
  ``{Search for light dark matter with the MESA accelerator},'' in {\em {13th
  Conference on the Intersections of Particle and Nuclear Physics}}.
\newblock 9, 2018.
\newblock \href{http://arxiv.org/abs/1809.07168}{{\ttfamily arXiv:1809.07168
  [hep-ex]}}.

\bibitem{Doria:2019sux}
L.~Doria, P.~Achenbach, M.~Christmann, A.~Denig, and H.~Merkel, ``{Dark Matter
  at the Intensity Frontier: the new MESA electron accelerator facility},''
  \href{http://dx.doi.org/10.22323/1.360.0022}{{\em PoS} {\bfseries ALPS2019}
  (2020) 022}, \href{http://arxiv.org/abs/1908.07921}{{\ttfamily
  arXiv:1908.07921 [hep-ex]}}.

\bibitem{Echenard:2014lma}
B.~Echenard, R.~Essig, and Y.-M. Zhong, ``{Projections for Dark Photon Searches
  at Mu3e},'' \href{http://dx.doi.org/10.1007/JHEP01(2015)113}{{\em JHEP}
  {\bfseries 01} (2015) 113}, \href{http://arxiv.org/abs/1411.1770}{{\ttfamily
  arXiv:1411.1770 [hep-ph]}}.

\bibitem{APEX:2011dww}
{\bfseries APEX} Collaboration, S.~Abrahamyan {\em et~al.}, ``{Search for a New
  Gauge Boson in Electron-Nucleus Fixed-Target Scattering by the APEX
  Experiment},'' \href{http://dx.doi.org/10.1103/PhysRevLett.107.191804}{{\em
  Phys. Rev. Lett.} {\bfseries 107} (2011) 191804},
  \href{http://arxiv.org/abs/1108.2750}{{\ttfamily arXiv:1108.2750 [hep-ex]}}.

\bibitem{SHiP:2015vad}
{\bfseries SHiP} Collaboration, M.~Anelli {\em et~al.}, ``{A facility to Search
  for Hidden Particles (SHiP) at the CERN SPS},''
  \href{http://arxiv.org/abs/1504.04956}{{\ttfamily arXiv:1504.04956
  [physics.ins-det]}}.

\bibitem{Alekhin:2015byh}
S.~Alekhin {\em et~al.}, ``{A facility to Search for Hidden Particles at the
  CERN SPS: the SHiP physics case},''
  \href{http://dx.doi.org/10.1088/0034-4885/79/12/124201}{{\em Rept. Prog.
  Phys.} {\bfseries 79} no.~12, (2016) 124201},
  \href{http://arxiv.org/abs/1504.04855}{{\ttfamily arXiv:1504.04855
  [hep-ph]}}.

\bibitem{Ilten:2015hya}
P.~Ilten, J.~Thaler, M.~Williams, and W.~Xue, ``{Dark photons from charm mesons
  at LHCb},'' \href{http://dx.doi.org/10.1103/PhysRevD.92.115017}{{\em Phys.
  Rev. D} {\bfseries 92} no.~11, (2015) 115017},
  \href{http://arxiv.org/abs/1509.06765}{{\ttfamily arXiv:1509.06765
  [hep-ph]}}.

\bibitem{Ilten:2016tkc}
P.~Ilten, Y.~Soreq, J.~Thaler, M.~Williams, and W.~Xue, ``{Proposed Inclusive
  Dark Photon Search at LHCb},''
  \href{http://dx.doi.org/10.1103/PhysRevLett.116.251803}{{\em Phys. Rev.
  Lett.} {\bfseries 116} no.~25, (2016) 251803},
  \href{http://arxiv.org/abs/1603.08926}{{\ttfamily arXiv:1603.08926
  [hep-ph]}}.

\bibitem{LHCb:2017trq}
{\bfseries LHCb} Collaboration, R.~Aaij {\em et~al.}, ``{Search for Dark
  Photons Produced in 13 TeV $pp$ Collisions},''
  \href{http://dx.doi.org/10.1103/PhysRevLett.120.061801}{{\em Phys. Rev.
  Lett.} {\bfseries 120} no.~6, (2018) 061801},
  \href{http://arxiv.org/abs/1710.02867}{{\ttfamily arXiv:1710.02867
  [hep-ex]}}.

\bibitem{Fujiwara:1984mp}
T.~Fujiwara, T.~Kugo, H.~Terao, S.~Uehara, and K.~Yamawaki, ``{Nonabelian
  Anomaly and Vector Mesons as Dynamical Gauge Bosons of Hidden Local
  Symmetries},'' \href{http://dx.doi.org/10.1143/PTP.73.926}{{\em Prog. Theor.
  Phys.} {\bfseries 73} (1985) 926}.

\bibitem{Cudell:2001pn}
J.~R. Cudell, V.~Ezhela, P.~Gauron, K.~Kang, Y.~V. Kuyanov, S.~Lugovsky,
  B.~Nicolescu, and N.~Tkachenko, ``{Hadronic scattering amplitudes:
  Medium-energy constraints on asymptotic behavior},''
  \href{http://dx.doi.org/10.1103/PhysRevD.65.074024}{{\em Phys. Rev. D}
  {\bfseries 65} (2002) 074024},
  \href{http://arxiv.org/abs/hep-ph/0107219}{{\ttfamily arXiv:hep-ph/0107219}}.

\bibitem{ParticleDataGroup:2016lqr}
{\bfseries Particle Data Group} Collaboration, C.~Patrignani {\em et~al.},
  ``{Review of Particle Physics},''
  \href{http://dx.doi.org/10.1088/1674-1137/40/10/100001}{{\em Chin. Phys. C}
  {\bfseries 40} no.~10, (2016) 100001}.

\bibitem{TOTEM:2017asr}
{\bfseries TOTEM} Collaboration, G.~Antchev {\em et~al.}, ``{First measurement
  of elastic, inelastic and total cross-section at $\sqrt{s}=13$ TeV by TOTEM
  and overview of cross-section data at LHC energies},''
  \href{http://dx.doi.org/10.1140/epjc/s10052-019-6567-0}{{\em Eur. Phys. J. C}
  {\bfseries 79} no.~2, (2019) 103},
  \href{http://arxiv.org/abs/1712.06153}{{\ttfamily arXiv:1712.06153
  [hep-ex]}}.

\bibitem{Fermi:1924tc}
E.~Fermi, ``{On the Theory of the impact between atoms and electrically charged
  particles},'' \href{http://dx.doi.org/10.1007/BF03184853}{{\em Z. Phys.}
  {\bfseries 29} (1924) 315--327}.

\bibitem{vonWeizsacker:1934nji}
C.~F. von Weizsacker, ``{Radiation emitted in collisions of very fast
  electrons},'' \href{http://dx.doi.org/10.1007/BF01333110}{{\em Z. Phys.}
  {\bfseries 88} (1934) 612--625}.

\bibitem{Williams:1934ad}
E.~J. Williams, ``{Nature of the high-energy particles of penetrating radiation
  and status of ionization and radiation formulae},''
  \href{http://dx.doi.org/10.1103/PhysRev.45.729}{{\em Phys. Rev.} {\bfseries
  45} (1934) 729--730}.

\bibitem{Williams:1935dka}
E.~J. Williams, ``{Correlation of certain collision problems with radiation
  theory},'' {\em Kong. Dan. Vid. Sel. Mat. Fys. Med.} {\bfseries 13N4} no.~4,
  (1935) 1--50.

\bibitem{deNiverville:2016rqh}
P.~deNiverville, C.-Y. Chen, M.~Pospelov, and A.~Ritz, ``{Light dark matter in
  neutrino beams: production modelling and scattering signatures at MiniBooNE,
  T2K and SHiP},'' \href{http://dx.doi.org/10.1103/PhysRevD.95.035006}{{\em
  Phys. Rev. D} {\bfseries 95} no.~3, (2017) 035006},
  \href{http://arxiv.org/abs/1609.01770}{{\ttfamily arXiv:1609.01770
  [hep-ph]}}.

\end{thebibliography}\endgroup

\end{document}